%% file: nuSTORM-EoI-arXiv.tex
\documentclass{99-Styles/IDS-NF}
\input 99-Styles/GetPackages

\usepackage{bibentry}
\nobibliography*

\begin{document}

\makeatletter

\input 99-Styles/nuSTORM-EoI-defs

\parindent 10pt
\pagenumbering{roman}
\setcounter{page}{1}
\pagestyle{plain}

\input 20-arXiv/20-00-Top-matter
\input 20-arXiv/20-02b-AuthorList-for-arXiv
\clearpage
\input 00b-Executive-summary/00b-Executive-summary

\clearpage
\pagenumbering{arabic}                   
\setcounter{page}{1}

\input 01-Introduction/01-Introduction

\input 02-Motivation/02-Motivation

\input 03-nuSTORM-facility/03-nuSTORM-facility

\input 04-Implementing-nuSTORM/04-Implementing-nuSTORM

\input 05-Proposal/05-Proposal

\clearpage
\bibliographystyle{99-Styles/utphys}
\bibliography{Concatenated-bibliography}

\clearpage
\appendix
\input 10A-HiResPhys/10A-HiResPhys

\end{document}

%% file: 99-Styles/GetPackages.tex
\usepackage{times}
\usepackage{bm}         
\usepackage{amsmath,amssymb,bm,slashed} 
\usepackage{amssymb}    
\usepackage{graphicx}   
\usepackage{verbatim}   
\usepackage{color}      
\usepackage{subfigure}  
\usepackage{hyperref}   
\usepackage{enumerate}

\usepackage{fmtcount}   

\usepackage[square,comma,numbers,sort&compress]{natbib}

%% file: 99-Styles/nuSTORM-EoI-defs.tex
\newcommand{\bra}[1]{\ensuremath{\langle #1 |}}   
\newcommand{\ket}[1]{\ensuremath{| #1 \rangle}}   
\newcommand{\bigbra}[1]{\ensuremath{\big\langle #1 \big|}}   
\newcommand{\bigket}[1]{\ensuremath{\big| #1 \big\rangle}}   
\newcommand{\amp}[3]{\ensuremath{\left\langle #1 \,\left|\, #2%
                     \,\right|\, #3 \right\rangle}}  
\newcommand{\sprod}[2]{\ensuremath{\left\langle #1 |%
                     #2 \right\rangle}}  
\newcommand{\ev}[1]{\ensuremath{\left\langle #1 %
                     \right\rangle}} 
\newcommand{\ds}[1]{\ensuremath{\! \frac{d^3#1}{(2\pi)^3 %
                     \sqrt{2 E_\vec{#1}}} \,}} 
\newcommand{\dst}[1]{\ensuremath{\! %
                     \frac{d^4#1}{(2\pi)^4} \,}} 
\newcommand{\tr}{\text{tr}}
\newcommand{\sgn}{\text{sgn}}
\newcommand{\diag}{\text{diag}}
\newcommand{\BR}{\text{BR}}

\renewcommand{\vec}[1]{{\mathbf{#1}}}
\renewcommand{\Re}{{\text{Re}}}
\renewcommand{\Im}{{\text{Im}}}
\newcommand{\iso}[2]{{\ensuremath{{}^{#2}}\ensuremath{\rm #1}}}
\newcommand{\eps}{{\ensuremath{\epsilon}}}
\newcommand{\draftnote}[1]{{\bf\color{red} \MakeUppercase{#1}}}
\newcommand{\panm}[1]{{\color{blue} #1}}
\providecommand{\abs}[1]{\lvert#1\rvert}
\providecommand{\norm}[1]{\lVert#1\rVert}

\def\parenbar{\mathpalette\p@renb@r}
\def\p@renb@r#1#2{\vbox{%
  \ifx#1\scriptscriptstyle \dimen@.7em\dimen@ii.2em\else
  \ifx#1\scriptstyle \dimen@.8em\dimen@ii.25em\else
  \dimen@1em\dimen@ii.4em\fi\fi \offinterlineskip
  \ialign{\hfill##\hfill\cr
    \vbox{\hrule width\dimen@ii}\cr
    \noalign{\vskip-.3ex}%
    \hbox to\dimen@{$\mathchar300\hfil\mathchar301$}\cr
    \noalign{\vskip-.3ex}%
    $#1#2$\cr}}}

%
\providecommand{\anmne}{\mbox{$\bar\nu_{\mu} \rightarrow \bar\nu_e$}} 
\providecommand{\nmne}{\mbox{$\nu_{\mu}\rightarrow\nu_e$}} 
\providecommand{\anm}{\mbox{$\bar\nu_\mu$}} 
\providecommand{\nm}{\mbox{$\nu_\mu$}}
\providecommand{\nue}{\mbox{$\nu_e$}} 
\providecommand{\ane}{\mbox{$\bar\nu_e$}} 
\providecommand{\enu}{\mbox{$E_\nu$}}
\providecommand{\piz}{\mbox{$\pi^0 $}}
\providecommand{\pip}{\mbox{$\pi^+$}} 
\providecommand{\pim}{\mbox{$\pi^-$}}

%% file: 20-arXiv/20-00-Top-matter.tex
\thispagestyle{empty}
\begin{tabular}{p{5cm} p{5cm} p{6cm}}
  \setlength{\baselineskip}{0.25\baselineskip}
    \leftline{\today}                                         &
    \centering{          }                                    &
    \rightline{$\nu$STORM EoI}
    \vspace{-0.4cm}\\
    \hline
\end{tabular}

\begin{center}
  {\bf\LARGE 
    Neutrinos from Stored Muons (\boldmath{$\nu$}STORM):      \\
    Expression of Interest
    \footnote{
             Submitted to the SPSC at CERN on the $5^{th}$ April
             2013 \cite{Adey:1537983}
             }
  }
  \vspace{-0.5cm}
\end{center}

%% file: 20-arXiv/20-02b-AuthorList-for-arXiv.tex
\vspace{1.0cm}
\noindent
D.~Adey$^1$, S.K.~Agarwalla$^4$, C.M.~Ankenbrandt$^{2,1}$, 
R.~Asfandiyarov$^5$, J.J.~Back$^6$, G.~Barker$^6$, E.~Baussan$^7$,
R.~Bayes$^8$, S.~Bhadra$^9$, V.~Blackmore$^{11}$
% ...

\vspace{1.0cm}
{\noindent\tiny
\begin{flushleft}
$^1$   Fermilab, P.O. Box 500, Batavia, IL 60510-5011, USA \\
$^2$   Muons Inc., 552 N. Batavia Avenue, Batavia, IL 60510, USA \\
% ...
$^4$   Institute of Physics, Sachivalaya Marg, Sainik School Post, 
       Bhubaneswar 751005, Orissa, India \\
$^5$   University de Geneve, 24, Quai Ernest-Ansermet, 1211 Geneva 4,
       Suisse \\
$^6$   Department of Physics, University of Warwick, Coventry,
       CV4 7AL, UK \\
$^7$   IPHC, Universit\'e de Strasbourg, CNRS/IN2P3, F-67037 Strasbourg, 
       France \\
$^8$   School of Physics and Astronomy, Kelvin Building, University of
       Glasgow, Glasgow G12 8QQ, Scotland, UK \\
$^9$   Department of Physics and Astronomy, York University, 4700
       Keele Street, Toronto, Ontario, M3J 1P3, Canada \\
% ...
$^{11}$ Particle Physics Department, The Denys Wilkinson Building,
       Keble Road, Oxford, OX1 3RH, UK \\
% ...
\end{flushleft}
}

%% file: 00b-Executive-summary/00b-Executive-summary.tex
\section*{Executive summary}

The $\nu$STORM facility has been designed to deliver beams of
$\parenbar{\nu}_e$ and $\parenbar{\nu}_\mu$ from the decay of a
stored $\mu^\pm$ beam with a central momentum of 3.8\,GeV/c and
a momentum spread of 10\% \cite{Kyberd:2012iz}.
The facility is unique in that it will:
\begin{itemize}
  \item Serve the future long- and short-baseline neutrino-oscillation
    programmes by providing definitive measurements of
    $\parenbar{\nu}_e N$ and $\parenbar{\nu}_\mu N$ scattering cross
    sections with percent-level precision;
  \item Allow searches for sterile neutrinos of exquisite sensitivity
    to be carried out; and
  \item Constitute the essential first step in the incremental
    development of muon accelerators as a powerful new
    technique for particle physics.
\end{itemize}
 
The race to discover CP-invariance violation in the lepton sector and
to determine the neutrino mass-hierarchy has begun with the recent
discovery that $\theta_{13} \ne 0$ 
\cite{An:2012eh,Ahn:2012nd,Abe:2011fz,Abe:2011sj,Adamson:2011qu}.
The measured value of $\theta_{13}$ is large 
($\sin^2 2 \theta_{13} \sim 0.1$) so measurements of oscillation
probabilities with uncertainties at the percent level are required.
For the next generation of long-baseline experiments to reach the
requisite precision requires that the $\parenbar{\nu}_e N$ and
the $\parenbar{\nu}_\mu N$ cross sections are known precisely for
neutrino energies ($E_\nu$) in the range $0.5 < E_\nu < 3$\,GeV.  
At $\nu$STORM, the flavour composition of the beam and the
neutrino-energy spectrum are both precisely known.
The storage-ring instrumentation combined with measurements at a
near detector will allow the neutrino flux to be determined to a
precision of 1\% or better.
$\nu$STORM is therefore unique as it makes it possible to measure the
$\parenbar{\nu}_e N$ and the $\parenbar{\nu}_\mu N$ cross 
sections with a precision $\simeq 1$\% over the required
neutrino-energy range.

A number of results have been reported that can be interpreted as
hints for oscillations involving sterile neutrinos
\cite{Aguilar:2001ty,AguilarArevalo:2007it,AguilarArevalo:2010wv,Mueller:2011nm,Huber:2011wv,Mention:2011rk,Anselmann:1994ar,Hampel:1997fc,Abdurashitov:1996dp,Abdurashitov:1998ne,Abdurashitov:2005tb}
(for a recent review see \cite{Abazajian:2012ys}).
Taken together, these hints warrant a systematically different and
definitive search for sterile neutrinos.
A magnetised iron neutrino detector at a distance of 
$\simeq 1\,500$\,m from the storage ring combined with a near
detector, identical but with a fiducial mass one tenth that of the far
detector, placed at 20--50\,m, will allow searches for active/sterile
neutrino oscillations in both the appearance and disappearance
channels.
Simulations of the $\nu_e \rightarrow \nu_\mu$ appearance channel show
that the presently allowed region can be excluded at the $10\sigma$
level while in the $\nu_e$ disappearance channel, $\nu$STORM has the
statistical power to exclude the presently allowed parameter space.
Furthermore, the definitive studies of $\parenbar{\nu}_e N$ 
($\parenbar{\nu}_\mu N$) scattering that can be done at $\nu$STORM
will allow backgrounds to be quantified precisely.

The European Strategy for Particle Physics provides for the
development of a vibrant neutrino-physics programme in Europe in which
CERN plays an essential enabling role \cite{Nakada:2013ESG}.
$\nu$STORM is ideally matched to the development of such a programme
combining first-rate discovery potential with a unique
neutrino-nucleus scattering programme.
$\nu$STORM could be developed in the North Area at CERN as part of the
CERN Neutrino Facility (CENF) \cite{CERN:EDMS1233951}.
Furthermore, $\nu$STORM is capable of providing the technology
test-bed that is needed to prove the techniques required by the
Neutrino Factory and, eventually, the Muon Collider.
$\nu$STORM is therefore the critical first step in establishing a
revolutionary new technique for particle physics.

Of the world's proton-accelerator laboratories, only CERN and FNAL
have the infrastructure required to mount $\nu$STORM.
In view of the fact that no siting decision has yet been taken, the
purpose of this Expression of Interest (EoI) is to request the
resources required to:
\begin{itemize}
  \item Investigate in detail how $\nu$STORM could be implemented at
    CERN; and 
  \item Develop options for decisive European contributions to the
    $\nu$STORM facility and experimental programme wherever the 
    facility is sited.
\end{itemize}
The EoI defines a two-year programme culminating in the delivery of
a Technical Design Report.

%% file: 01-Introduction/01-Introduction.tex
\section{Introduction}
\label{Sect:Intro}

\input 01-Introduction/01-01-Overview/01-01-Overview
\input 01-Introduction/01-02-Fit-to-CERN/01-02-Fit-to-CERN

%% file: 01-Introduction/01-01-Overview/01-01-Overview.tex
\subsection{Overview}
\label{SubSect:Overview}

Muon accelerators have been proposed as sources of intense,
high-energy electron- and muon-neutrino beams at the Neutrino Factory
\cite{Bandyopadhyay:2007kx,Choubey:2011zzq} and as the basis for
multi-TeV $l^+ l^-$ collisions at the Muon Collider
\cite{:1900cvd,Holmes:2010zz}.
An incremental approach to the development of the facility has been
outlined in \cite{Kaplan:2012zzb}.
At $\nu$STORM, a stored muon beam with a central momentum of
3.8\,GeV/c and a momentum spread of 10\% will:
\begin{itemize}
  \item Serve a first-rate neutrino-physics programme will encompass:
    \begin{itemize}
      \item Detailed and precise studies of electron- and
        muon-neutrino-nucleus scattering over the energy range
        required by the future long- and short-baseline neutrino
        oscillation programme; and
      \item Exquisitely sensitive searches for sterile neutrinos in
        both appearance and disappearance modes; and
    \end{itemize}
  \item Provide the technology test-bed required to carry-out the R\&D
    critical to the implementation of the next increment in the
    muon-accelerator based particle-physics programme.
\end{itemize}
$\nu$STORM is, therefore, the essential first step in the incremental
development of muon accelerators as a new technique for particle
physics.

Neutrino oscillations are readily described in terms of three
neutrino-mass eigenstates and a unitary mixing matrix that relates the
mass states to the flavour states (the Standard Neutrino Model,
S$\nu$M)
\cite{Pontecorvo:1957cp,Pontecorvo:1957qd,Maki:1962mu,Bilenky:2001rz}.
The three-neutrino-mixing paradigm is able to give an accurate
description of the observed fluxes of neutrinos produced in the
sun, by cosmic ray interactions in the atmosphere, by high-energy
particle accelerators and \mbox{anti-}neutrinos produced by nuclear
reactors \cite{Amsler20081}.
However, a number of results can not be described by the S$\nu$M.
First, the LSND collaboration reported evidence for
$\bar{\nu}_\mu \rightarrow \bar{\nu}_e$
oscillations corresponding to a mass-squared difference of 
$\sim 1$\,eV$^2$ \cite{Aguilar:2001ty}; a value which is much larger
than the two mass-squared differences of the S$\nu$M.
A third mass-squared difference, if confirmed, would imply a fourth
neutrino-mass state and hence the existence of a sterile neutrino.
The MiniBooNE experiment observed an effect consistent
with the LSND result
\cite{AguilarArevalo:2007it,AguilarArevalo:2010wv}. 
A further hint for the existence of sterile neutrinos may be provided
by the discrepancy between the measured reactor-neutrino flux and that
obtained in calculations of the expected flux
\cite{Mueller:2011nm,Huber:2011wv,Mention:2011rk}.
Finally, the GALLEX and SAGE experiments reported anomalies in the
rate of neutrinos observed from the sources used to calibrate their 
radio-chemical detection techniques
\cite{Anselmann:1994ar,Hampel:1997fc,Abdurashitov:1996dp,Abdurashitov:1998ne,Abdurashitov:2005tb}.
A detailed review of the relevant data may be found in
\cite{Abazajian:2012ys}.

Now that the small mixing angle $\theta_{13}$ is known
\cite{An:2012eh,Ahn:2012nd,Abe:2011fz,Abe:2011sj,Adamson:2011qu}, the
emphasis of the study of the S$\nu$M has shifted to the determination
of the mass hierarchy and the search for CP-invariance violation.
In a conventional super-beam experiment, both of these objectives
requires the measurement of $\nu_e$ ($\bar{\nu}_e$) appearance in a 
$\nu_\mu$ ($\bar{\nu}_\mu$) beam.
With a sufficiently large data sample, the measurement of the mass
hierarchy is relatively insensitive to systematic uncertainties.
By contrast, the sensitivity to CP-invariance violation depends
critically on systematic effects in general and on the knowledge of
the $\nu_e N$ ($\bar{\nu}_e N$) cross sections in particular
\cite{Huber:2007em,Coloma:2012ji}.
The $\nu$STORM facility described in this Expression of Interest (EoI)
is unique in that it is capable of serving a near detector (or suite
of near detectors)  that will be able to measure $\nu_e N$
($\bar{\nu}_e N$) and $\nu_\mu N$ ($\bar{\nu}_\mu N$) cross sections
at the percent level and of studying the hadronic final states.

Unambiguous evidence for the existence of one or more sterile
neutrinos would revolutionise the field.
$\nu$STORM is capable of making the measurements required to confirm
or refute the evidence for sterile neutrinos summarised above using a
technique that is both qualitatively and quantitatively new
\cite{Kyberd:2012iz}.
The $\nu$STORM facility has been designed to deliver beams of $\nu_e$
($\bar{\nu}_e$) and $\bar{\nu}_\mu$ ($\nu_\mu$).
A detector located at a distance $\sim 1\,500$\,m from the end of one
of the straight sections will be able to make sensitive searches for
the existence of sterile neutrinos.
If no appearance 
($\bar{\nu}_\mu \rightarrow \bar{\nu}_e$)
signal is observed, the LSND allowed region can be ruled out at the
$\sim 10 \sigma$ level.
Instrumenting the $\nu$STORM neutrino beam with a near detector at a
distance of $\sim 50$\,m makes it possible to search for sterile
neutrinos in the disappearance $\nu_e \rightarrow \nu_X$ and
$\nu_\mu \rightarrow \nu_X$ channels.
In the disappearance search, the absence of a signal would permit the 
presently allowed region to be excluded at the 90\% confidence level
\cite{Winter:2012sk}.

By providing the ideal technology test-bed, the $\nu$STORM facility
will play a pivotal role in the development of neutrino detectors,
accelerator systems and instrumentation techniques.
It is capable of providing a high-intensity, high-emittance,
low-energy muon beam for studies of ionisation cooling and of
supporting the development of the high-resolution, totally-active,
magnetised neutrino detectors.
The development of the $\nu$STORM ring, together with the
instrumentation required for the $\nu N$-scattering and
sterile-neutrino-search programmes will allow the next step in the
development of muon accelerators for particle physics to be defined.
Just as the Cambridge Electron Accelerator \cite{Oasis:Lib:Harvard},
built by Harvard and MIT at the end of the '50s, was the first in a
series of electron synchrotrons that culminated in LEP, $\nu$STORM has
the potential to establish a new technique for particle physics that
can be developed to deliver the high-energy $\nu_e$ ($\bar{\nu}_e$) 
beams required to elucidate the physics of flavour at the Neutrino
Factory \cite{Choubey:2011zzq} and to provide the basis for multi-TeV
lepton-antilepton collisions at the Muon Collider
\cite{Holmes:2010zz}.  

%% file: 01-Introduction/01-02-Fit-to-CERN/01-02-Fit-to-CERN.tex
\subsection{\boldmath{$\nu$}STORM and the emerging
            CERN neutrino programme}
\label{SubSect:FitToCERN}

\subsubsection{Short-baseline neutrino facility in the North Area}

\noindent
It has been proposed to develop the North Area at CERN to host a
portfolio of neutrino experiments \cite{CERN:EDMS1233951}.
In the short term, it has been proposed that a search for sterile
neutrinos be carried out by the ICARUS and NESSiE collaborations
\cite{Antonello:2012hf,Antonello:2012qx}.
These experiments will be served by a conventional neutrino beam
generated by the fast extraction of protons from the SPS at 100\,GeV.
For these experiments to take sufficient data before the second long
shutdown of the LHC in 2017 requires that the beam and experiments be
implemented such that data taking can start early in 2016.
$\nu$STORM requires a primary proton beam similar to that which is
being prepared for ICARUS/NESSiE but with a smaller transverse and
longitudinal emittance.
A beam with the appropriate properties will be available once LINAC4
becomes operational after the 2017 long shutdown 
\cite{Arnaudon:2006jt}.
The near and far source--detector distances required by $\nu$STORM
closely match those specified for ICARUS/NESSiE.

The concept for the implementation of the $\nu$STORM facility at CERN
presented in this EoI is self-consistent and is capable of delivering
searches for sterile neutrinos with exquisite sensitivity and serving
a unique and detailed $\nu_{e,\mu} N$ ($\bar{\nu}_{e,\mu} N$)
scattering programme.
Given the technical synergies, it is natural to consider how the
$\nu$STORM facility could be developed first to enhance and then to
take forward the short-baseline neutrino programme at CERN.

\subsubsection{A step on the way to the Neutrino Factory}

To go beyond the sensitivity offered by the next generation super-beam
experiments requires the development of novel techniques for the
production of neutrino beams and novel detector systems.
Pure $\nu_e$ ($\bar{\nu}_e$) beams may be generated from the decay of
radioactive ions at a ``beta-beam'' facility \cite{Benedikt:2011za}.
The low charge-to-mass ratio of the ions places a practical limit of
$\sim 1$\,GeV on the neutrino energies that can be produced in this
way.
Alternatively, high-energy electron- and muon-neutrino beams of
precisely known flux may be generated from the decay of stored muon
beams at the Neutrino Factory \cite{Choubey:2011zzq}.

The Neutrino Factory has been shown to offer a sensitivity to
CP-invariance violation superior to that which can be achieved at any
other proposed facility \cite{Choubey:2011zzq,Bandyopadhyay:2007kx}.
The EURO$\nu$ consortium demonstrated that the CERN baseline
($\gamma=100$) beta-beam becomes competitive only if it is combined
with the CERN-Frejus super-beam, or a super-beam of comparable
performance \cite{EUROnu2012}.
Detailed and precise measurements of neutrino oscillations will be
required for the physics of flavour to be elucidated.
The challenge to the experimental community is to establish a
programme capable of delivering measurements of the neutrino-mixing
parameters with a precision approaching that with which the quark
mixing parameters are known.
Only the Neutrino Factory offers such precision.

It is conceivable that the Neutrino Factory can be implemented in a
series of increments or stages---each increment offering a first-rate
neutrino-science programme and being capable of delivering the R\&D
required for the development of the subsequent increment.
The International Design Study for the Neutrino Factory (IDS-NF)
collaboration will include a discussion of the incremental
implementation of the facility in its Reference Design Report that
will be published in the autumn of 2013.
The $\nu$STORM facility, by proving the feasibility of using stored
muon beams to provide neutrino beams for physics, will be the
essential first increment.

\subsubsection{Long-baseline neutrino oscillation physics}

The present generation of long-baseline neutrino-oscillation
experiments (MINOS \cite{Ables:1995wq}, T2K \cite{Abe:2011ks},
NO$\nu$A \cite{Ayres:2004js}) will continue to refine the
measurements of the mixing parameters.
Their data, taken together with that obtained in atmospheric-neutrino
experiments, may constrain the neutrino mass hierarchy at the 
$2\sigma$---$3\sigma$ confidence level.
However, even in combination with all oscillation data, the present
generation of experiments will be essentially insensitive to leptonic
CP-invariance violation.

High-power conventional neutrino beams serving very large detectors
have been proposed to determine the mass hierarchy.
Such ``super-beam'' experiments fall into two broad categories:
narrow-band beams, in which a low-energy ($E_\nu \le 1$\,GeV) beam is
used to illuminate a detector 100\,km---300\,km from the source; and
wide-band beams in which neutrinos with energies spanning the range
$\sim 1$\,GeV to $10$\,GeV illuminate a detector at a distance of
between 700\,km and 2\,300\,km.

The opportunities for CERN to host a next-generation super-beam has
been studied by the EURO$\nu$ Framework Programme 7 (FP7) Design Study 
consortium \cite{EUROnu}. 
EURO$\nu$ studied a narrow-band beam generated using the 5\,GeV, 4\,MW
Superconducting Proton Linac (SPL) at CERN illuminating the MEMPHYS,
450\,kT water Cherenkov detector located in the Laboratoire Souterrain
de Modane (LSM) at a distance of 130\,km (this option is referred to
as CERN-Frejus since the LSM is located in the Frejus tunnel)
\cite{Baussan:2012wf}.

The study of super-beam experiments at CERN is now being taken forward
by the LAGUNA-LBNO FP7 Design Study consortium \cite{LAGUNA-LBNO}.  
In LAGUNA-LBNO, the CERN-Frejus narrow-band beam continues to be
developed and a new wide-band beam option, the Long-Baseline Neutrino 
Observatory (LBNO), is being considered \cite{Rubbia:LBNO}.
LBNO calls for a high-energy, wide-band neutrino beam to be created
using protons from the SPS.
The beam would serve a suite of detectors in the Pyh\"asalmi mine in
Finland, at a distance of 2\,300\,km from CERN.
The long baseline, coupled with the wide-band beam makes
CERN-Pyh\"asalmi a powerful option since it would allow LBNO to
determine the mass hierarchy at a confidence level in excess of
$5\sigma$ no matter what the value of the CP phase.
Alternative proposals for next generation super-beam experiments have
been brought forward in Japan (the Tokai to Hyper-Kamiokande, T2HK,
experiment \cite{Abe:2011ts}) and in the US (the Long-Baseline
Neutrino Experiment, LBNE \cite{Akiri:2011dv}).

Each of the super-beam experiments outlined above exploits the
sub-leading $\nu_\mu \rightarrow \nu_e$ 
($\bar{\nu}_\mu \rightarrow \bar{\nu}_e$)
oscillation to determine the mass hierarchy and to search for leptonic
CP-invariance violation.
At present, data on neutrino-nucleus scattering in the energy range of
interest is limited to relatively sparse $\nu_\mu N$ 
($\bar{\nu}_\mu N$) measurements; $\nu_e N$ ($\bar{\nu}_e N$)
cross sections being inferred from the $\nu_\mu N$ ($\bar{\nu}_\mu N$)
measurements.
As a result, uncertainties in oscillation measurements made using
conventional beams suffer from systematic uncertainties arising from
the absence of reliable electron-neutrino-nucleus (and
muon-neutrino-nucleus) scattering cross sections.
Moreover, the lack of knowledge of the relevant cross sections gives
rise to correlated uncertainties in the estimate of the neutrino-beam
flux.

$\nu$STORM will make detailed studies of both 
$\nu_e N$ ($\bar{\nu}_e N$) and $\nu_\mu N$ ($\bar{\nu}_\mu N$)
scattering. 
As discussed in this EoI, an appropriately designed suite of near
detectors will be able to determine the scattering cross sections and
provide detailed information on the hadronic final states.
The latter will be of first importance not only in the long-baseline
oscillation programme, but will allow the systematic study of the
sources of background that currently affect sterile-neutrino
searches.
The cross-section measurements that $\nu$STORM will provide will
therefore be an essential part of the emerging CERN neutrino
programme.

%% file: 02-Motivation/02-Motivation.tex
\section{Motivation}
\label{Sect:Motivation}

The case for the $\nu$STORM facility rests on three themes.
First, the uniquely well-known neutrino beam generated in muon decay
may be exploited to make detailed studies of neutrino-nucleus
scattering over the neutrino-energy range of interest to present and
future long- and short-baseline neutrino oscillation experiment from
In long-baseline experiments, these measurements are required to break
the correlation between the cross-section and flux uncertainties and
to reduce the overall systematic uncertainty to a level that justifies
the investment in high-power conventional super-beam experiments such
as T2HK, LBNE, LBNO and SPL-Frejus.
The $\nu$STORM $\parenbar{\nu} N$ scattering programme is no less
important for the next generation of short-baseline experiments for
which uncertainties in the magnitude and shape of backgrounds to the
sterile-neutrino searches will be come critically important.
Second, the $\nu$STORM neutrino beam, instrumented with a pair of
magnetised detectors near and far, will allow searches for sterile
neutrinos of exquisite sensitivity to be carried out.
The signal to background ratio for this combination is of order ten
and is much larger than other accelerator-based projects.
Thirdly, the storage ring itself, and the muon beam it contains, can
be used to carry out the R\&D programme required to implement the next
step in the incremental development of muon accelerators for particle
physics.
The muon accelerator programme has the potential to elucidation of the
physics of flavour at the Neutrino Factory and to provide multi-TeV
$l^+ l^-$ collisions at the Muon Collider.
Just as the three legs of a tripod make it a uniquely stable platform,
the three individually-compelling themes that make up the case for
$\nu$STORM constitute a uniquely robust case for a facility that will
be at once immensely productive scientifically and seminal in the
creation of a new technique for particle physics. 

\input 02-Motivation/02-02-Neutrino-scattering/02-02-Neutrino-scattering
\input 02-Motivation/02-01-Steriles/02-01-Steriles
\input 02-Motivation/02-03-RnD/02-03-RnD

%% file: 02-Motivation/02-02-Neutrino-scattering/02-02-Neutrino-scattering.tex
\subsection{Neutrino-nucleus scattering}
\label{SubSect:nuNScat}

\subsubsection{Introduction}

To date, neutrino oscillations \cite{Beringer:1900zz} remain the only
observed and confirmed phenomenon not described by the Standard Model
(SM) of particle physics.
Neutrino-oscillation data, combined with searches for kinematic
effects of neutrino mass in tritium-decay experiments, very clearly
indicate that, in the S$\nu$M, the mass of the heaviest neutrino must
be smaller than $\sim\!1$\,eV.
This mass is too small to be explained naturally by the Higgs
mechanism, making it necessary to invoke physics beyond the SM to
explain neutrino mass and mixing.
The detailed exploration of the neutrino sector is one of the most
important goals for the next decade in particle physics research.
The neutrino community is converging on the conclusion that a
wide-band long-baseline (LBL) accelerator-based neutrino experiment is
an important part of this research programme
\cite{Akiri:2011dv,Rubbia:LBNO,Abe:2011ts}.
The principal goals of the next-generation LBL experiments are the
determination of the neutrino mass-hierarchy and the search for
CP-invariance violation.
Recent observations that the value of the third neutrino mixing angle,
$\theta_{13}$, is large \cite{An:2012eh,Ahn:2012nd,Abe:2011sj} mean
that the rates of $\nu_e$ or $\bar{\nu}_e$ appearance in a wide-band
beam will be substantial and that high-statistics measurements will be
dominated by systematic uncertainties, especially uncertainties in the
modelling of neutrino-nucleus scattering.
It is therefore crucial that these systematic uncertainties are
reduced in order for the next generation experiments to achieve the
precision and sensitivity defined by the collaborations in the various
proposals.

The current generation of neutrino-oscillation experiments employ
neutrino-interaction models developed in the 1970's and 1980's
\cite{LlewellynSmith:1971zm,Smith:1972xh,Rein:1980wg}.
In the energy region of interest to the LBL programme 
(0.1\,GeV--10\,GeV) the dominant reaction types, in order of threshold,
are: quasi-elastic scattering; resonant and coherent pion-production;
and deep inelastic scattering.  
High statistics neutrino-scattering measurements made in the past
decade by K2K
\cite{Gran:2006jn,Rodriguez:2008aa,Mariani:2009zzb,Nakayama:2004dp},
MiniBooNE
\cite{AguilarArevalo:2007ab,AguilarArevalo:2010zc,AguilarArevalo:2009eb,AguilarArevalo:2010bm,AguilarArevalo:2010xt,AguilarArevalo:2009ww,AguilarArevalo:2010cx}
and SciBooNE
\cite{AlcarazAunion:2009ku,Nakajima:2010fp,Kurimoto:2009wq,Kurimoto:2010rc}
indicate that the quasi-elastic scattering and pion-production models
do not describe nature. 

Even with this degree of activity, the precision with which the basic
neutrino-nucleon cross sections are known is still not better
than 20\%--30\%. 
There are two main reasons for this: the poor
knowledge of neutrino fluxes and the fact that all the recent
cross-section measurements have been performed on nuclear targets.  It
is important to recall that current neutrino experiments measure
events that are a convolution of an energy-dependent neutrino flux
with an energy-dependent cross section with energy-dependent nuclear
effects.  Experiments have, for example, measured an effective
neutrino-carbon cross section. Extracting a neutrino-nucleon cross
section from these measurements requires separating nuclear-physics
effects that can be done only with limited precision.
For many experiments, using the same nuclear targets in their near
and far detectors is a good start.  However, even with the same
nuclear target near and far, the presence of oscillations leads to
different neutrino fluxes at the near and far detectors. 
This means that there is a different convolution of cross section with
nuclear effects near and far, so there is no automatic cancellation
between the near-and-far detectors at the precision needed for the LBL
programme.  
Furthermore, these effects are exacerbated in measurements of
anti-neutrino cross sections because the event rates are significantly
reduced.  Finally, the intrinsic differences between $\nu_{\mu}$- and
$\nu_e$-interaction cross sections must be measured with a precision
commensurate with the precision goals of the LBL programme (see
section \ref{SubSubSect:NuE}). 

In summary, to ensure a successful LBL programme, a thorough
comparison of measured neutrino-nucleon cross sections with
theoretical models is needed so that all these convoluted effects can
be understood.

\subsubsection{Charged-current quasi-elastic scattering}
\label{SubSubSect:CCQE}

Neutrino-nucleon charged-current quasi-elastic (CCQE) scattering,
$\nu_l\ n\rightarrow l^-\ p$, is the most abundant neutrino reaction in
the $1$~GeV energy region and also the most important in
investigations of the oscillation signal.  
Despite its importance and
apparent simplicity, the CCQE cross section is known with limited
accuracy.  
The main reasons for the poor understanding of this
reaction \cite{Gallagher:2011zza, Morfin:2012kn} are the large
neutrino-flux uncertainties (both the overall normalisation and the
energy spectrum) and the fact that all recent CCQE cross-section
measurements were made on bound nucleons with many complications
coming from nuclear effects.

In the standard theoretical approach to describe the CCQE cross
section, a weak-current transition matrix element is expressed in
terms of three independent form factors 
\cite{LlewellynSmith:1971zm}.
The two vector form factors are known from electron-scattering
experiments, thanks to the conserved vector-current hypothesis
\cite{Bradford:2006yz,Budd:2004bp,Bodek:2003ed,Budd:2003wb}.  
Assuming the partially conserved
axial-current hypothesis leaves one independent axial-vector
form-factor for which one usually assumes a dipole form and this, in
turn, leaves only one free parameter: the axial mass ($M_A$). Within
this simple theoretical framework, an investigation of CCQE scattering
is equivalent to an $M_A$ measurement.  Experience from electron
scattering tells us that dipole expressions provide a reasonable
approximation to electric and magnetic form factors, and {\it
extrapolation} of this argument to the axial form factors seems to be
a justified, though not completely obvious, assumption. $M_A$
determines both the overall CCQE cross section and also the shape of
the distribution of events in $Q^2$, the square of four-momentum
transfer. The preferred way to measure $M_A$ is to analyse the shape
of the $d\sigma/dQ^2$ spectrum because this mitigates the dependence
on the overall flux normalisation.

Another problem with measuring the CCQE cross section stems from the
fact that the energy spectrum of all neutrino beams are broad making
it difficult to separate the various dynamic mechanisms in
neutrino-nucleon (-nucleus) interactions. 
The situation is much more complex than for electron scattering where
good knowledge of the initial and final electron states allows a
model-independent measurement of $Q^2$.
For these reasons, neutrino cross-section measurements are always
inclusive and there is even reason to consider the limitations of the
commonly-assumed impulse approximation \cite{Ankowski:2008df} in
which it is assumed that the neutrino interacts with an individual
bound nucleon and thus one can neglect collective effects (all the
major Monte Carlo (MC) event generators do not include (continuous)
random-phase approximation corrections).

Nuclear effects include Fermi motion and nucleon binding.
Clearly, in investigations of CCQE, it is important
to use the best Fermi motion models, which means employing the
spectral-function formalism \cite{Benhar:2005vg} that has been
validated in electron scattering.  
Moreover, it is important to consider a two body current contribution
to the cross section \cite{Martini:2009uj,Nieves:2011pp}; these
currents give rise to events that can be easily confused with genuine
CCQE events unless one investigates final-state nucleons carefully.

\begin{figure}
  \centering
  \mbox{
    \subfigure{
      \includegraphics[width=0.6\textwidth]%
        {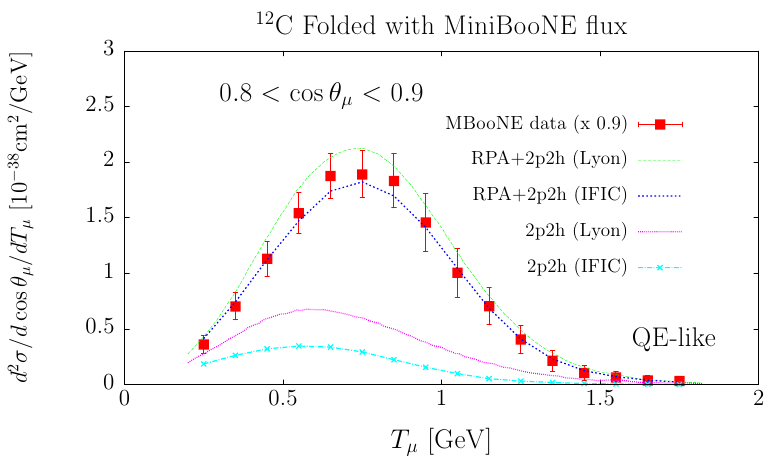}}\quad
    \subfigure{
      \includegraphics[width=0.35\textwidth]%
        {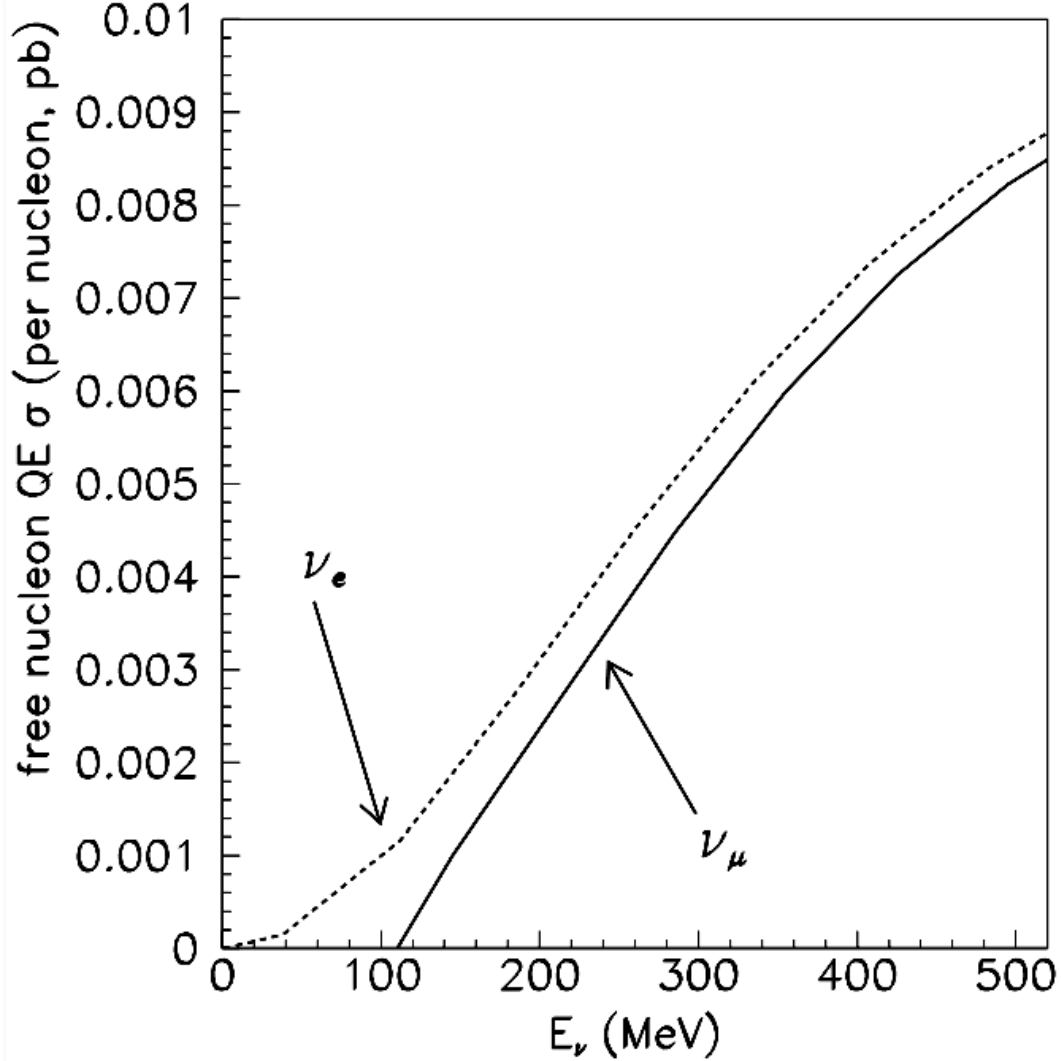}}
  }
  \caption{
    (left) MiniBooNE flux averaged CCQE-like cross section normalised
    per neutron with experimental points rescaled by
    0.9. $\cos\theta_\mu\in (0.8, 0.9)$. 
    Predictions from two theoretical models are compared and
    contributions from np-nh mechanism are also shown separately.
    (right) The total charged-current quasi-elastic cross-section for
    $\nu_{\mu}$ and $\nu_e$ neutrinos.
  }
  \label{fig:qe}
\end{figure}

Recent interest in CCQE scattering was triggered by several large
$M_A$ measurements, in particular the high-statistics muon-carbon
double-differential cross sections from the MiniBooNE collaboration
\cite{AguilarArevalo:2010zc}.
Here, ``large'' is relative to values obtained from older, mostly
light nuclear target, neutrino \cite{Bodek:2007ym} and pion
electroproduction data \cite{0954-3899-28-1-201}.  
The MiniBooNE detector is not sensitive to final-state nucleons, which
are produced below Cherenkov threshold.  
What MiniBooNE measures can be described as CCQE-like events---defined
as those with no pion in the final state---with data-driven
corrections for the contribution from pion production and absorption. 
Several theoretical groups have attempted to explain the MiniBooNE
CCQE double-differential cross-section data with models containing
significant contributions from $n$p-$n$h mechanisms, which allow $n$
particles and $n$ holes, with $n\geq 2$, in the final state ($n$p-$n$h
mechanisms are also called meson exchange currents (MEC),
multi-nucleon knock-out, or two-body currents).    
The Valencia/IFIC group performed a fit with its model to the
two-dimensional MiniBooNE CCQE data, obtaining 
$M_A=1.077\pm 0.027$\,GeV \cite{Nieves:2011yp}.  
Good qualitative agreement was obtained by the
Lyon group \cite{Martini:2011wp}.   
These two models are shown compared to MiniBooNE double-differential
muon data in figure \ref{fig:qe}. 
Qualitative agreement has also been obtained with an optical-potential
model \cite{Meucci:2012yq}, while slightly worse agreement was found
with the super-scaling approach~\cite{Amaro:2011qb} and transverse
enhancement (TE) model~\cite{Bodek:2011ps, Sobczyk:2012ah}.  
A general
observation is that theoretical models are usually able to explain the
normalisation effect of the large $M_A$ value from MiniBooNE but their
predictions do not agree with the full two-dimensional muon data set.

Theoretical models of the MEC contribution give quite different
estimates of the significance of the effect in the case of
anti-neutrino scattering. 
Recently, MiniBooNE showed the first
high-statistics anti-neutrino CCQE cross section and in particular a
ratio of neutrino and anti-neutrino CCQE-like cross sections
(defined as explained above) as a function of energy. 
These data may
allow some comparison between the models, but higher precision data on
multiple nuclear targets are needed.

For CCQE events one can calculate the energy of the incoming neutrino
using just the final charged-lepton three-momentum assuming the
target nucleon was at rest.  
Clearly, the effects of Fermi motion and binding energy limit the
accuracy of the neutrino-energy reconstruction and introduce some
model-dependent bias.
The neutrino
energy is used for oscillation studies since that is the only experimental
parameter which affects the oscillation probability.  
Additional
complications come from events which mimic CCQE interactions,
e.g., from real pion production and absorption.  
The MiniBooNE data for the muon double-differential cross section can
be described using the standard CCQE model with a large value of $M_A$
(although it is better to call this an effective parameter $M_A^{eff}$
as proposed in \cite{AguilarArevalo:2007ab}).  
However, use of the CCQE model with $M_A^{eff}$ in the oscillation
signal analysis introduces some bias since the presence of two-body
current contributions changes the mapping from neutrino energy to
charged-lepton momentum, as noted in several recent studies
\cite{Nieves:2012yz,Morfin:2012kn,Lalakulich:2012hs,Lalakulich:2012gm,Martini:2012fa}.

Separation of two-body-current contributions should be possible by
looking at final-state nucleons
\cite{Lalakulich:2012gm,Sobczyk:2012me}.
This is, however, a very challenging goal because of nucleon
final-state interactions and contamination from real-pion production
and absorption events.  
One needs very good resolution of final-state nucleons with a low
threshold for the momentum of reconstructed tracks.
Liquid argon TPCs have been suggested as candidate instruments to
improve MC cascade models \cite{P}.

\subsubsection{Resonance Region}
\label{SubSubSect:Res}

The neutrino-interaction landscape in the few-GeV region is a complex
mix of resonance production, shallow-inelastic-scattering physics,
where resonance production merges into deep-inelastic scattering, and
coherent processes. 
The dominant production mechanism in this region is the production of
a $\Delta(1232)$ baryon followed by its decay to a single pion final
state.
A challenging process to study experimentally, most experiments being
complicated by the fact that the neutrinos interact in an extended
nuclear target; the final state particles must leave the nucleus to be
observed and along the way can be scattered, absorbed or undergo
charge-exchange reactions.
These final-state interactions must somehow be decoupled from the
underlying neutrino-nucleon cross-sections---a process which is
model-dependent---making interpretation of the data challenging. 
The resonance-production channel presents the largest background to
current neutrino-oscillation experiments and it is therefore important 
to understand its contribution. 
Moreover, future experiments such as
LBNE \cite{Akiri:2011dv} and LBNO \cite{Rubbia:LBNO} are designed to
operate at neutrino energies of 3\,GeV--7\,GeV where this transition
region between quasi-elastic scattering and deep inelastic scattering
is most important.  
For these experiments, a much better understanding of this region is
required if they are to have maximum sensitivity to CP-invariance
violation in the neutrino sector.
\begin{figure}
  \begin{center}
    \mbox{
      \subfigure{
        \includegraphics[width=0.45\textwidth]%
          {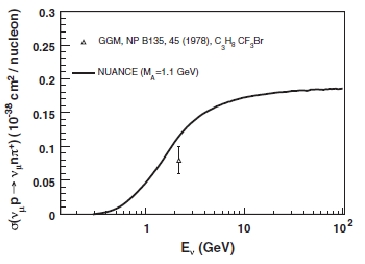}
      }\quad
      \subfigure{
        \includegraphics[width=0.50\textwidth]%
          {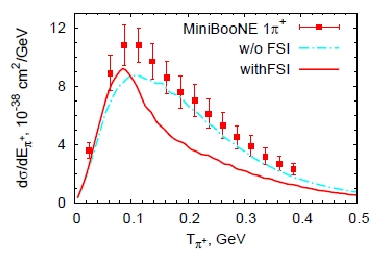}
      }
    }
  \end{center}
  \caption{
    (left) Existing measurements of the $\nu_{\mu} p \rightarrow
    \nu_{\mu} n \pi^{+}$ cross section as a function of neutrino
    energy \cite{RevModPhys.84.1307}. 
    Data points come from the Gargamelle bubble chamber
    data \cite{Krenz:1977sw}.  
    The line is the prediction from the NUANCE Monte Carlo event
    generator.
    (right) Distribution of $\nu_{\mu} C \rightarrow \mu^{-} \pi^{+} X$ 
    cross section  as a function of the pion momentum from the GIBBU
    simulation \cite{Lalakulich:2011ne}, compared with MiniBooNE
    data \cite{AguilarArevalo:2010xt}.
  } 
  \label{Fig:res}
\end{figure}

The quality of experimental data in the resonance region is varied. 
Whilst there has been recent work on neutrino-induced single-pion
production mechanisms in experiments such as MiniBooNE, data on
multi-pion and other final-state production mechanisms are sparse or
non-existent.
Figure \ref{Fig:res} (left) shows, for example, the only data
on the $\nu_{\mu} p \rightarrow \nu_{\mu} n \pi^{+}$ channel.
In recent years experiments such as K2K
\cite{Rodriguez:2008aa,Mariani:2010ez,Tanaka:2006zm}, MiniBooNE
\cite{AguilarArevalo:2010zc,AguilarArevalo:2009eb,AguilarArevalo:2010bm} 
and SciBooNE \cite{Kurimoto:2009wq,Kurimoto:2010rc} have presented
data on neutral-current $\pi^{0}$ (NC\,$\pi^{0}$) production, charged
current $\pi^{+}$ (CC~$\pi^{+}$) production and the charged-current
$\pi^{0}$ (CC~$\pi^{0}$) channel.  
Improved knowledge of the NC~$\pi^{0}$ production cross section is
vital as it is a dominant systematic error in
$\parenbar{\nu}_{e}$-appearance oscillation experiments.  
The CC~$\pi^{+}$ and CC~$\pi^{0}$ channels have been studied by
MiniBooNE \cite{AguilarArevalo:2010bm} which has produced differential
cross sections in the final-state particle momenta and angles. 
The cross-section results differ from the current Monte Carlo models
by up to 20\% in the case of the charged-pion mode and by up to a
factor of two for the neutral-pion mode, suggesting a discrepancy in
both the understanding of the neutrino-nucleon cross section and the
final state effects. 
Figure \ref{Fig:res} (right) shows the differential cross section for
CC~$\pi^{+}$ production on $^{12}\mbox{C}$ as a function of pion
kinetic energy from MiniBooNE compared to the sophisticated GIBUU
simulation \cite {Lalakulich:2011ne}.
The model appears to favour no, or at least a very small, component of
final-state interactions even though it is known that final-state 
interactions have a large effect.  
The solution to this puzzle lies in
understanding both the neutrino-nucleon cross section and final-state
effects independently. 
Such a program of study would involve the
comparison of the final-state topologies of the CC~$\pi$ reaction on
different nuclei. 
A critical element, however, is knowledge of the
neutrino-nucleon cross section on an $H_{2}$ or $D_{2}$ target. 
This is an anchor point, allowing the analysers to tune their models
to the nucleon cross section before comparison with nuclear data.
Light nucleon data was last taken by the bubble-chamber experiments in
the 1970's and 1980's. 
More complete, and better understood, data on light nuclei is
now necessary to understand the resonance-production models. 
A dedicated light-target detector in the $\nu$STORM facility is
therefore of interest.  
It should be emphasised that this is the state of data
from neutrino-induced interactions.  
Data on anti-neutrino resonance
production are even more sparse and there is no data on resonance
production in an electron-neutrino beam.  
One of the primary means of studying CP-invariance violation is to
investigate differences between measurements of oscillations of
neutrinos and anti-neutrinos. 
Poor knowledge of the cross-sections present one of the largest
systematic errors limiting these analyses and so a precise
determination of these cross sections is vital.

Another pion-production process is the coherent neutrino-nucleus
interaction. 
In this process the neutrino interacts with the entire nucleus at very
low momentum-transfer, resulting in a forward-going pion and leaving
the nucleus in the ground state. 
This process can proceed via the charged and neutral currents for both
neutrinos and anti-neutrinos.
Neutral-current interactions which result in a $\pi^{0}$ in the final
state are of particular interest for oscillation experiments
investigating $\nu_{e}$ appearance as they form a large part of the
background.
The process has been observed at high (greater than 5\,GeV) neutrino
energy \cite{Kullenberg:2009pu} and agrees with the standard
Rein-Seghal model \cite{Rein:1981ys} predictions, which are based on
PCAC with pion dominance.
However, in the 1\,GeV--3\,GeV range, the landscape becomes unclear as
the available data are limited.  
Both MiniBooNE \cite{AguilarArevalo:2008xs} and
SciBooNE \cite{Kurimoto:2010rc} have measured the neutral-current mode
at an average neutrino energy of 0.8\,GeV. 
The charged-current mode is mode is more puzzling. 
Isospin symmetry implies that the charged-current process should occur
with twice the rate of the neutral-current process.
However K2K \cite{Hasegawa:2005td} and
SciBooNE \cite{Hiraide:2009zz} have reported no evidence for the
charged-current coherent process. 
It is now becoming clear that it is not appropriate to continue the
high-energy theory down to lower energies and that other models
involving microscopic $\Delta$ dominance are more reliable
\cite{Amaro:2008hd,Nakamura:2009iq}.
Testing these models requires data on a number of different types of
target nucleus and over a range of neutrino energies.  
This is crucial since the contribution of this process to the $\nu_e$
backgrounds in the first oscillation maximum must be predicted
accurately for the LBL experiments. 

\subsubsection{Differences in the energy-dependent cross sections of $\nu_{\mu}$- and $\nu_e$-nucleus interactions}
\label{SubSubSect:NuE}

To determine the mass hierarchy of neutrinos and to search for
CP-invariance violation in the neutrino sector, current and upcoming
accelerator-based neutrino-oscillation experiments such as T2K
\cite{Abe:2011ks} and NOvA \cite{Ayres:2004js} as well as future
proposed experiments such as LBNE \cite{Akiri:2011dv} and the Neutrino
Factory \cite{Akiri:2011dv} plan to make precision measurements of the
neutrino flavour oscillations $\parenbar\nu_\mu \to \parenbar\nu_e$ or 
$\parenbar\nu_e \to \parenbar\nu_\mu$.  
An important factor in the
ability to fit the difference in observed event rates between the near
and far detectors will be an accurate understanding of the cross
section of $\nu_{\mu}$- and $\nu_e$-nucleus interactions.
Uncertainties on differences in expected event rates due to
differences between these cross-sections will contribute to
experimental uncertainties in these flavour-oscillation measurements.

There are obvious differences in the cross sections due to the
difference in mass of the outgoing lepton.  
These can be calculated by including the lepton-mass term in the
cross-section expression. 
Figure \ref{fig:qe} (right) \cite{Day:2012gb}, shows these expected
differences in the cross sections as a function of neutrino energy.  
Another such calculable difference occurs because of radiative
corrections.
Radiative corrections from a particle of mass $m$ are proportional to
$\log(1/m)$, which implies a significant difference since the muon is
$\sim 200$ times heavier than the electron \cite{De_Rujula:1979jj}.  
This turns into a difference of $\sim 10$\% in the cross sections.
In addition to these differences, there are other more subtle
differences due to the coupling of poorly-known or unknown form
factors to the lepton tensor that reflect the differences in the
outgoing lepton mass.
These effects have been investigated in some detail~\cite{Day:2012gb}
but must be probed experimentally.

Regarding nuclear effects, while there are no differences expected in
the final-state interactions, there are expected differences in the
initial reaction cross-sections between $\nu_{\mu}$- and
$\nu_e$-nucleus interactions.
Since the lepton tensor, reflecting the mass of the outgoing lepton,
couples to the hadron-response functions, there is a difference in
nuclear effects at the interaction vertex due to the $\mu$-to-$e$ mass
difference.
The expected difference in the $\nu_{\mu}$- and $\nu_e$-nucleus
cross-section ratio is around 5\% when using a spectral-function model
\cite{Benhar:1994hw} for the initial nucleon momentum compared to the
relativistic Fermi gas model \cite{Smith:1972xh,Bodek:1980ar}.  
There is another 5\% difference expected for multi-nucleon ($n$p-$n$h)
contribution \cite{Martini:2012:private}.
These differences in cross sections extend up into the resonance
region with the low-$Q^2$ behaviour of $\Delta$ production exhibiting
10\% differences at values of $Q^2$ where the cross section has
levelled off.

While each of the individual effects outlined above may not be large
compared to current neutrino-interaction uncertainties, they are large
compared to the assumed precision of oscillation measurements in the
future LBL programme.
Moreover, the sum of these effects could be quite significant and the
uncertainty in our knowledge of the size of these effects will
contribute directly to uncertainties in the neutrino-oscillation
parameters determined from these experiments---and these uncertainties
can only be reduced with good quality $\parenbar{\nu}_e$ scattering
data.
$\nu$STORM is the only source of a well-understood and well-controlled
$\parenbar{\nu}_e$ neutrino beam with which these cross-section
differences can be studied systematically.

\subsubsection{Effects of neutrino-nucleus interaction systematics on
               oscillation measurements}
\label{SubSubSect:xsec-systematics}

A neutrino-oscillation experiment must compare neutrino-scattering
event rates with a prediction in order to extract oscillation
parameters.
Many systematic errors in such analyses can be mitigated by using of a
near detector with similar target nuclei, but, importantly, several
systematic uncertainties still remain.   
Neutrino oscillation is a function of the true energy of the neutrino,
but experiments must infer the energy of neutrino interactions from
measurements of the outgoing charged-lepton partner (which also
identifies the neutrino flavour).

As discussed in section~\ref{SubSubSect:CCQE}, the problem we face is
that the micro-physics of the nuclear environment can change the
mapping between the charged-lepton momentum and the neutrino energy.
This mapping is model-dependent because the form factors for
axial-currents have not yet been measured precisely since the
uncertainty in the reconstructed neutrino energy is inherently larger
than the widths generated by nuclear effects.  
The model-dependence of these predictions adds a systematic
uncertainty that cannot be mitigated without data sets that are fine
enough in final-state-particle resolution while covering enough of the
kinematic phase-space and target nuclei.  
The systematic uncertainty due to this model dependence cannot be
mitigated by a near detector unless and until the model calculations
are sufficiently detailed to allow falsification with final-state
particle data.
Another issue that contributes to the systematic errors is the
migration of events between near (and far) detector data samples.
In the main, these arise because final-state particles can scatter
hadronically within the target nucleus before escaping into the
detector medium.  As discussed in section~\ref{SubSubSect:Res}, the
exact kinematics of the final-state particles in the resonance region
must be predicted, and then measured, in order to reduce these
uncertainties. 
Finally there is the very real effect of differences in the
$\parenbar{\nu}_{\mu}$ and $\parenbar{\nu}_e$ interaction cross
sections, which must be measured with high precision.

The stated goals for the precision of the proposed next generation
long-baseline neutrino oscillation experiments such as LBNE, LBNO and
T2HK cannot be reached without mitigating these systematic
uncertainties, even with high precision near detectors.
$\nu$STORM is the only experimental facility with the precision and
flexibility needed to tackle all of these neutrino-interaction
cross-section uncertainties.

%% file: 02-Motivation/02-01-Steriles/02-01-Steriles.tex
\subsection{Sterile neutrino search}
\label{SubSect:SterileSearch}

\subsubsection{Sterile neutrinos in extensions of the Standard Model}
\label{sec:sterile-bsm}

Sterile neutrinos---fermions that are uncharged under the 
$SU(3) \times SU(2) \times U(1)$ gauge group---arise naturally in many
extensions of the Standard Model and even where they are not an
integral part of a model, they can usually be accommodated easily. A
detailed overview of sterile neutrino phenomenology and related model
building considerations is given in \cite{Abazajian:2012ys}.

In Grand Unified Theories (GUTs), fermions are grouped into multiplets
of a large gauge group, of which $SU(3) \times SU(2) \times U(1)$ is a
subgroup.  If these multiplets contain not only the known quarks and
leptons, but also additional fermions, these new fermions will, after
the breaking of the GUT symmetry, often behave like gauge singlets
(see for instance
\cite{Bando:1998ww,Ma:1995xk,Shafi:1999rm,Babu:2004mj} for GUT models
with sterile neutrinos). 

Models by which the smallness of neutrino masses are explained using a
``see-saw'' mechanism generically contain sterile neutrinos.  
While in the most generic see-saw scenarios, these sterile neutrinos
are extremely heavy ($\sim 10^{14}$~GeV) and have very small mixing
angles ($\sim 10^{-12}$) with the active neutrinos, slightly
non-minimal see-saw models can easily feature sterile neutrinos with
eV-scale masses and with per-cent level mixing with the active
neutrinos.
Examples for non-minimal see-saw models with relatively light sterile
neutrinos include the split see-saw scenario~\cite{Kusenko:2010ik},
see-saw models with additional flavour symmetries (see
e.g. \cite{Mohapatra:2001ns}), models with a Froggatt-Nielsen
mechanism~\cite{Froggatt:1978nt, Barry:2011fp}, and extended see-saw
models that augment the mechanism by introducing more than three
singlet fermions as well as additional symmetries
\cite{Mohapatra:2005wk, Fong:2011xh, Zhang:2011vh}.  

Finally, sterile neutrinos arise naturally in ``mirror models'', in which
the existence of an extended ``dark sector'', with non-trivial
dynamics of its own, is postulated. If the dark sector is similar to
the visible sector---as is the case, for instance in string-inspired
$E_8 \times E_8$ models---it is natural to assume that it also contains
neutrinos \cite{Berezhiani:1995yi, Foot:1995pa, Berezinsky:2002fa}.

\subsubsection{Experimental hints for light sterile neutrinos}
\label{sec:sterile-hints}

While the theoretical motivation for the existence of sterile
neutrinos is certainly strong, what has mostly prompted the interest
of the scientific community in this topic is the fact that there are
several experimental results that show deviations from the Standard
Neutrino  Model predictions which can be interpreted as hints for
oscillations involving light sterile neutrinos with masses on the 
order of an eV.

The first of these hints was obtained by the LSND collaboration, which
carried out a search for $\bar\nu_\mu \to \bar\nu_e$ oscillations over
a baseline of $\sim 30$~m~\cite{Aguilar:2001ty}. 
Neutrinos were produced in a stopped-pion source in the decay of pions
at rest ($\pi^+ \to \mu^+ + \nu_\mu$) and the subsequent decay 
$\mu^+ \to e^+ \bar\nu_\mu \nu_e$.  
Electron anti-neutrinos were detected through the inverse-beta-decay
reaction $\bar\nu_e p \to e^+ n$ in a liquid-scintillator detector.  
Backgrounds to this search arise from the decay chain 
$\pi^- \to \bar\nu_\mu + (\mu^- \to \nu_\mu \bar\nu_e e^-)$ if
negative pions produced in the target decay before they are captured
by a nucleus and from the reaction $\bar\nu_\mu p \to \mu^+ n$ which
is only allowed for the small fraction of muon anti-neutrinos produced
by pion decay in flight rather than stopped-pion decay. 
The LSND collaboration found an excess of $\bar\nu_e$-candidate events
above this background with a significance of more than $3\sigma$. 
When interpreted as $\bar\nu_\mu \to \bar\nu_e$ oscillations through
an intermediate sterile state $\bar\nu_s$, this result is best
explained by sterile neutrinos with an effective mass-squared
splitting $\Delta m^2 \gtrsim 0.1$~eV$^2$ relative to the active
neutrinos, and with an effective sterile-sector induced
$\bar\nu_\mu$--$\bar\nu_e$ mixing angle 
$\sin^2 2\theta_{e\mu, \rm eff} \gtrsim 2 \times 10^{-3}$, depending
on $\Delta m^2$. 

The MiniBooNE experiment \cite{AguilarArevalo:2012va} was designed to
test the neutrino-oscillation interpretation of the LSND result using
a different technique, namely neutrinos from a horn-focused pion beam.
By focusing either positive or negative pions, MiniBooNE could run
either with a beam consisting mostly of neutrinos or in a beam
consisting mostly of anti-neutrinos. 
In both modes, the experiments observed an excess of electron-like
events at sub-GeV energies.  
The excess has a significance above $3\sigma$ and can be interpreted
in terms of $\parenbar\nu_\mu \to \parenbar\nu_e$ oscillations
consistent with the LSND observation 
\cite{AguilarArevalo:2012va,Aguilar-Arevalo:2013pmq}.

A third hint for the possible existence of sterile neutrinos is
provided by the reactor anti-neutrino anomaly. 
In 2011, Mueller et al.\ published a new ab-initio computation of the
expected neutrino fluxes from nuclear reactors~\cite{Mueller:2011nm}.
Their results improve upon a 1985 calculation
\cite{Schreckenbach:1985ep} by using up-to-date nuclear databases, a
careful treatment of systematic uncertainties and various other
corrections and improvements that were neglected in the earlier
calculation.
Mueller et al.\ find that the predicted anti-neutrino flux from a
nuclear reactor is about 3\% higher than previously thought.  
This result, which was later confirmed by Huber~\cite{Huber:2011wv},
implies that short-baseline reactor experiments have observed a
deficit of anti-neutrinos compared to the
prediction~\cite{Mention:2011rk,Abazajian:2012ys}. 
It needs to be emphasised that the significance of the deficit depends
crucially on the systematic uncertainties associated with the
theoretical prediction, some of which are difficult to estimate
reliably.
If the reactor anti-neutrino deficit is interpreted as 
$\bar\nu_e \to \bar\nu_s$ disappearance via oscillation, the required
2-flavour oscillation parameters are $\Delta m^2 \gtrsim 1$~eV$^2$ and
$\sin^2 2\theta_{ee,\rm eff} \sim 0.1$.

Short-baseline oscillations in this parameter range could also explain
another experimental result: the gallium anomaly. The GALLEX and SAGE
solar neutrino experiments used electron neutrinos from intense
artificial radioactive sources to demonstrate the feasibility of their
radio-chemical detection principle
\cite{Anselmann:1994ar,Hampel:1997fc,Abdurashitov:1996dp,Abdurashitov:1998ne,Abdurashitov:2005tb}.
Both experiments observed fewer $\nu_e$ from the source than expected.
The statistical significance of the deficit is above the 99\%
confidence level and can be interpreted in terms of short-baseline
$\bar\nu_e \to \bar\nu_s$ disappearance with 
$\Delta m^2 \gtrsim 1$~eV$^2$ and 
$\sin^2 2\theta_{ee,\rm eff} \sim 0.1$--$0.8$ 
\cite{Acero:2007su,Giunti:2010wz,Giunti:2010zu}. 

\subsubsection{Constraints and global fit}
\label{sec:sterile-global-fit}

\begin{figure}[ht]
  \begin{center}
    \includegraphics[width=0.45\textwidth]%
      {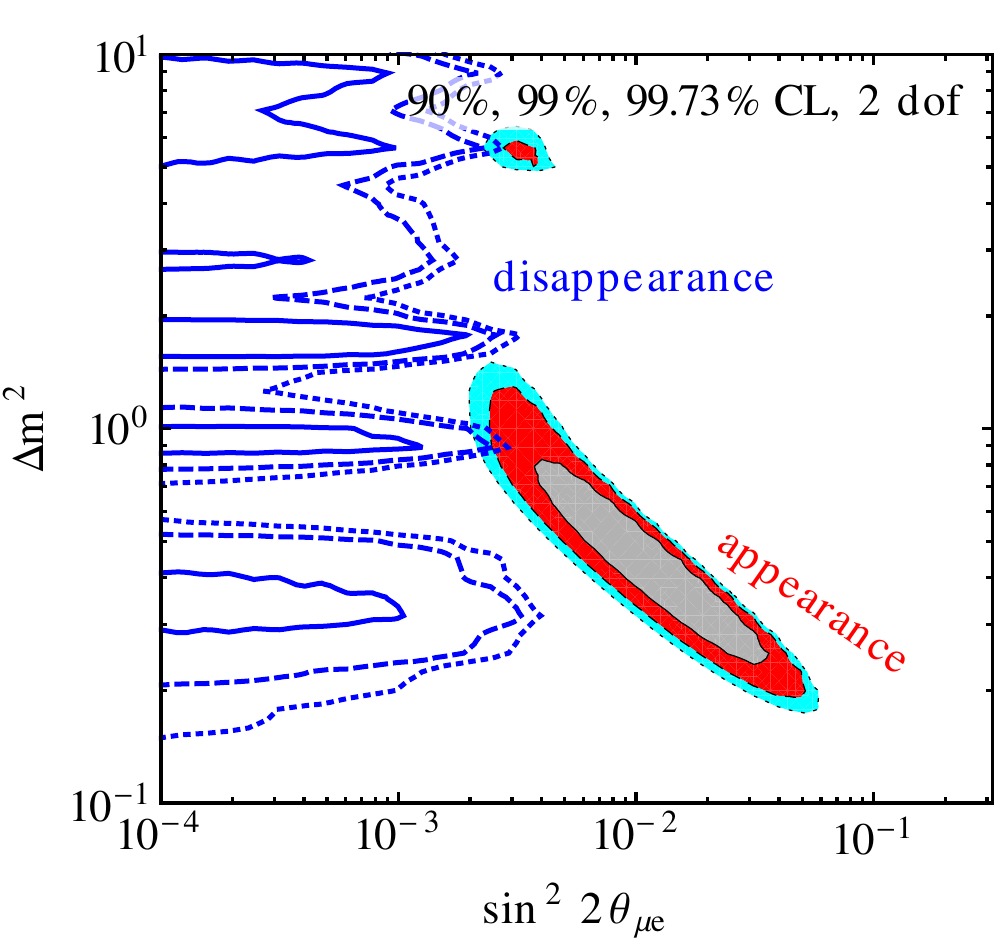} \quad\quad\quad\quad
    \includegraphics[width=0.45\textwidth]%
      {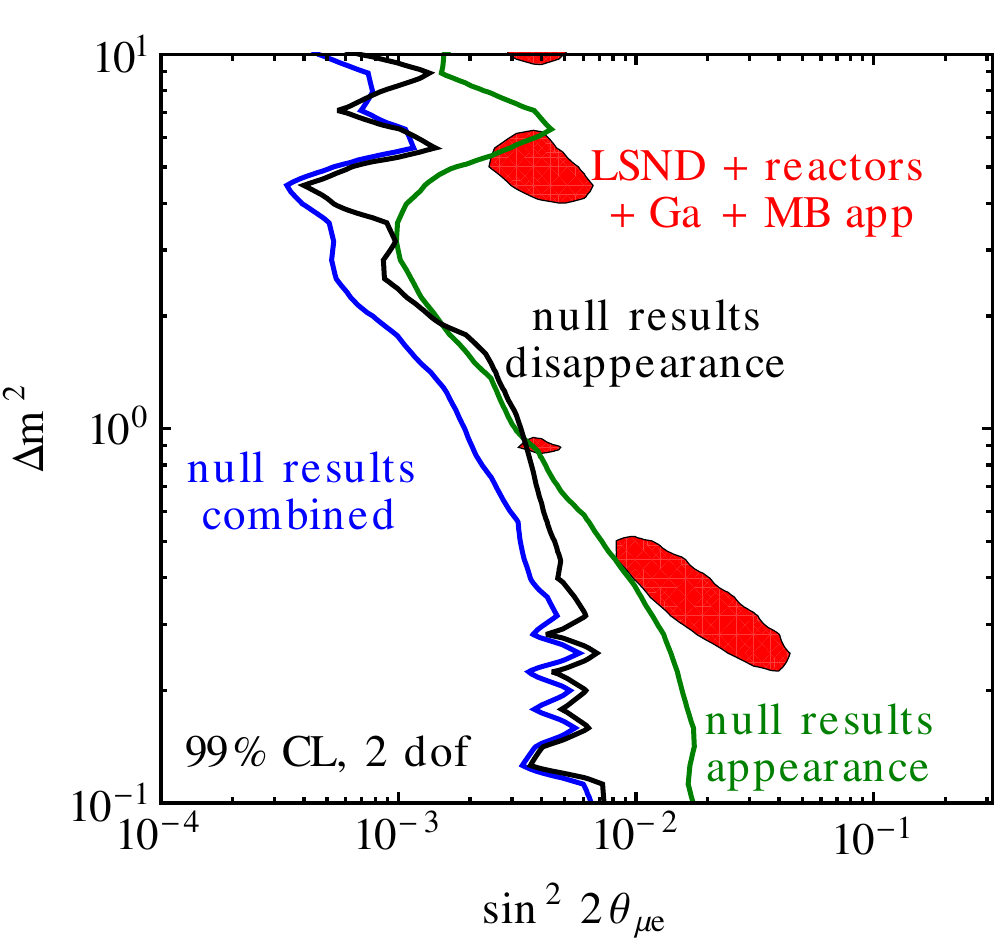}
  \end{center}
  \caption{Global constraints on sterile neutrinos in a 3+1 model. In the left panel,
    we show that $\protect\parenbar{\nu}_e$ appearance data (coloured region:
    LSND~\cite{Aguilar:2001ty}, MiniBooNE~\cite{AguilarArevalo:2012va},
    KARMEN~\cite{Armbruster:2002mp}, NOMAD~\cite{Astier:2001yj},
    E776~\cite{Borodovsky:1992pn}, ICARUS~\cite{Antonello:2012fu}) is only
    marginally consistent with disappearance data (blue contours: atmospheric
    neutrinos~\cite{Ashie:2005ik}, solar neutrinos~\cite{Cleveland:1998nv,
    Kaether:2010ag, Abdurashitov:2009tn, Hosaka:2005um, Aharmim:2007nv,
    Aharmim:2005gt, Aharmim:2008kc, Aharmim:2011vm, Bellini:2011rx,
    Bellini:2008mr}, MiniBooNE/SciBooNE~\cite{AguilarArevalo:2009yj,
    Cheng:2012yy} MINOS~\cite{Adamson:2010wi, Adamson:2011ku}, reactor
    experiments~\cite{Declais:1994su, Declais:1994ma, Kuvshinnikov:1990ry,
    Vidyakin:1987ue, Kwon:1981ua, Zacek:1986cu, Apollonio:2002gd, Boehm:2001ik,
    DBneutrino, Abe:2012tg, Ahn:2012nd, Gando:2010aa},
    CDHS~\cite{Dydak:1983zq}, KARMEN~\cite{Reichenbacher:2005nc} and
    LSND~\cite{Auerbach:2001hz} $\nu_e$--${}^{12} \text{C}$ scattering data and
    gallium experiments~\cite{Hampel:1997fc, Abdurashitov:1998ne,
    Kaether:2010ag, Abdurashitov:2005tb}). In the right panel, we have split
    the data into those experiments which see unexplained signals (LSND,
    MiniBooNE appearance measurements, reactor experiments, gallium
    experiments) and those which don't. For the analysis of reactor data, we
    have used the new reactor flux predictions from~\cite{Mueller:2011nm}, but
    we have checked that the results, especially regarding consistency with
    LSND and MiniBooNE $\bar\nu$ data, are qualitatively unchanged when the old
    reactor fluxes are used.  Fits have been carried out in the GLoBES
    framework~\cite{Huber:2004ka, Huber:2007ji} using external modules
    discussed in~\cite{GonzalezGarcia:2007ib, Maltoni:2007zf,
    Akhmedov:2010vy, Kopp:2013vaa}.}
  \label{fig:regions-3p1}
\end{figure}

While the previous section shows that there is an intriguing
accumulation of hints for the existence of new oscillation
effects---possibly related to sterile neutrinos---in short-baseline
experiments, these hints are not undisputed. 
Several short-baseline oscillation experiments (KARMEN
\cite{Armbruster:2002mp}, NOMAD \cite{Astier:2001yj}, E776
\cite{Borodovsky:1992pn}, ICARUS \cite{Antonello:2012fu}, atmospheric
neutrinos \cite{Ashie:2005ik}, solar neutrinos
\cite{Cleveland:1998nv,Kaether:2010ag,Abdurashitov:2009tn,Hosaka:2005um,Aharmim:2007nv,Aharmim:2005gt,Aharmim:2008kc,Aharmim:2011vm,Bellini:2011rx,Bellini:2008mr},
MINOS \cite{Adamson:2010wi, Adamson:2011ku}, and
CDHS \cite{Dydak:1983zq}) did not confirm the observations from LSND,
MiniBooNE, reactor experiments, and gallium experiments, and place
very strong limits on the relevant regions of parameter space in
sterile-neutrino models.  
To assess the viability of these models it is necessary to carry out a
global fit to all relevant experimental data sets
\cite{Abazajian:2012ys,Kopp:2011qd,Giunti:2011cp,Karagiorgi:2011ut,Giunti:2011hn,Giunti:2011gz}. 
In figure \ref{fig:regions-3p1}, which is based on the analysis
presented in \cite{Abazajian:2012ys,Kopp:2011qd,Kopp:2013vaa}, we
show the current constraints on the parameter space of a $3+1$ model
(a model with three active neutrinos and one sterile neutrino).
We have projected the parameter space onto a plane spanned by the
mass-squared difference, $\Delta m^2$, between the heavy, mostly
sterile, mass eigenstate and the light, mostly active, ones and by the
effective amplitude $\sin^2 2\theta_{e\mu, \rm eff}$ for
sterile-mediated $\nu_\mu \to \nu_e$ oscillations.

We see that there is severe tension in the global data set: the
parameter region flavoured by the hints from LSND, MiniBooNE, reactor
neutrinos and gallium experiments is incompatible, at the 99\%
confidence level, with constraints from other experiments.
Similarly, the parameter region flavoured by the global
$\parenbar\nu_e$ appearance data, has only very little overlap with
the region flavoured by $\parenbar\nu_\mu$ and $\parenbar\nu_e$
disappearance experiments. 
Using a parameter goodness-of-fit test \cite{Maltoni:2003cu} to
quantity this tension, p-values on the order of a
$\text{few} \times 10^{-5}$ are found for the compatibility of
appearance and disappearance data.
The global fit improves somewhat in models with more than one sterile
neutrino, but significant tension remains
\cite{Abazajian:2012ys,Kopp:2011qd}.

One can imagine several possible resolutions to this puzzle:
\begin{enumerate}
  \item One or several of the apparent deviations from the S$\nu$M
    oscillation framework discussed in section \ref{sec:sterile-hints}
    have explanations not related to sterile neutrinos; 
  \item One or several of the null results that favour the
    no-oscillation hypothesis is or are in error;
  \item There are more than two sterile-neutrino flavours.
    Note that scenarios with one sterile neutrino with an eV-scale
    mass are already in some tension with cosmology (see, however,
    \cite{Joudaki:2012uk}), but the existence of one sterile neutrino
    with a mass well below 1\,eV is actually preferred by cosmological
    fits
    \cite{GonzalezGarcia:2010un,Hamann:2010bk,Giusarma:2011ex,Mangano:2011ar}. 
    Cosmological bounds on sterile neutrinos can be avoided in
    non-standard cosmologies \cite{Hamann:2011ge} or by invoking
    mechanisms that suppress sterile-neutrino production in the early
    universe \cite{Bento:2001xi,Dolgov:2004jw}; and
  \item There are sterile neutrinos plus some other kind of new
    physics at the eV scale (see for instance
    \cite{Akhmedov:2010vy,Karagiorgi:2012kw} for an attempt in this
    direction).
\end{enumerate}

We conclude that our understanding of short-baseline neutrino
oscillations is currently in a rather unsatisfactory state.
Several experiments hint at deviations from the established
three-neutrino framework.
However, none of these hints can be considered conclusive; moreover,
when interpreted in the simplest sterile neutrino models, the
parameter sets favoured by the data are in severe tension with
existing constraints on the parameter-space of these models.
An experiment searching for short-baseline neutrino oscillations with
good sensitivity and well-controlled systematic uncertainties has
great potential to clarify the situation either by finding a new type
of neutrino oscillation or by deriving a strong and robust constraint
on any such oscillation.
While the former outcome would constitute a major discovery, the
latter would also receive a lot of attention since it would provide
the world's strongest constraints on a large variety of theoretical
models postulating ``new physics'' in the neutrino sector at the eV
scale. 

%% file: 02-Motivation/02-03-RnD/02-03-RnD.tex
\subsection{Technology test-bed}
\label{SubSect:RnD}

\subsubsection{Muon beam for ionisation cooing studies}
\label{SubSect:MuBm}

Muon ionisation cooling improves by a factor $\sim 2$ the stored-muon
flux at the Neutrino Factory and is absolutely crucial for a Muon
Collider of any centre-of-mass energy to achieve the required
luminosity.
The Muon Ionisation Cooling Experiment (MICE) \cite{MICE:2005zz} will
study four-dimensional ionisation cooling and work is underway to
specify the scope of a follow-on six-dimensional (6D) cooling
experiment.
MICE is a ``single-particle'' experiment; the four-momenta of single
muons are measured before and after the cooling cell and then input and
output beam emittances are reconstructed from an ensemble of
single-muon events.
A 6D cooling experiment could be done in the same fashion, but doing
the experiment with a high-intensity pulsed muon beam is preferred.  
One feature of $\nu$STORM is that an appropriate low-energy muon beam
with these characteristics can be provided in a straightforward
fashion. 

\begin{figure}[ht]
  \begin{center}
    \includegraphics[width=0.85\textwidth]%
      {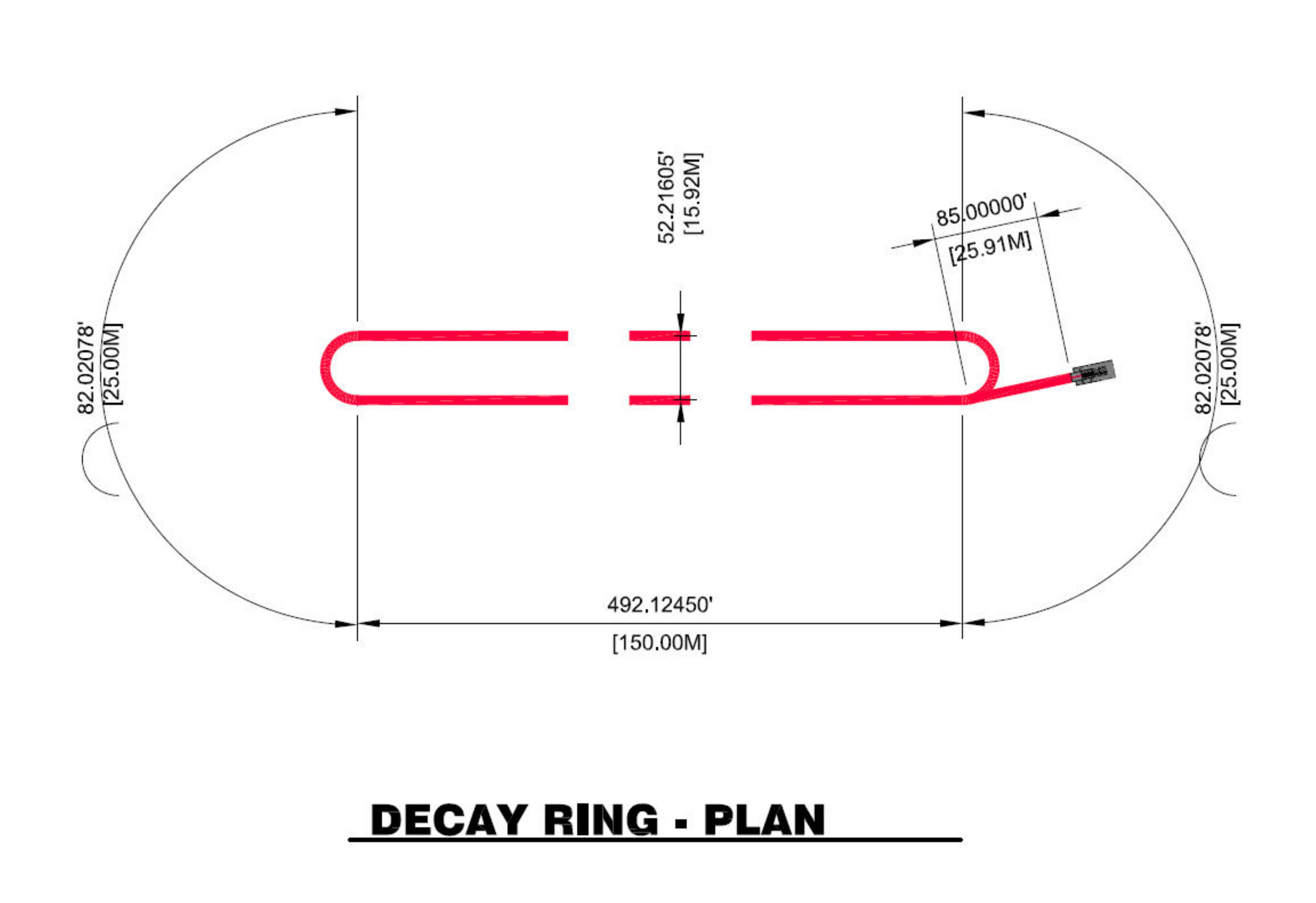}
  \end{center}
  \caption{Schematic of the $\nu$STORM decay ring.}
  \label{fig:Decay_ring}
\end{figure}

Figure \ref{fig:Decay_ring} shows a schematic of the decay ring for
$\nu$STORM.
As is described in section \ref{SubSect:AccelFacility} below, 5\,GeV/c
pions are injected at the end of the straight section of the ring.
Given the 150\,m length of the straight, only $\sim 40$\% of the
pions decay in the injection straight.
Since the arcs are set for the central muon momentum of 3.8\,GeV/c,
the pions remaining at the end of the straight will not be transported
by the arc.
The power contained within the pion beam that reaches the end of the
injection straight is 4\,kW--5\,kW making it necessary to dump the
undecayed pion beam into an appropriate absorber.

\begin{figure}
  \begin{center}
    \includegraphics[width=0.52\textwidth]%
      {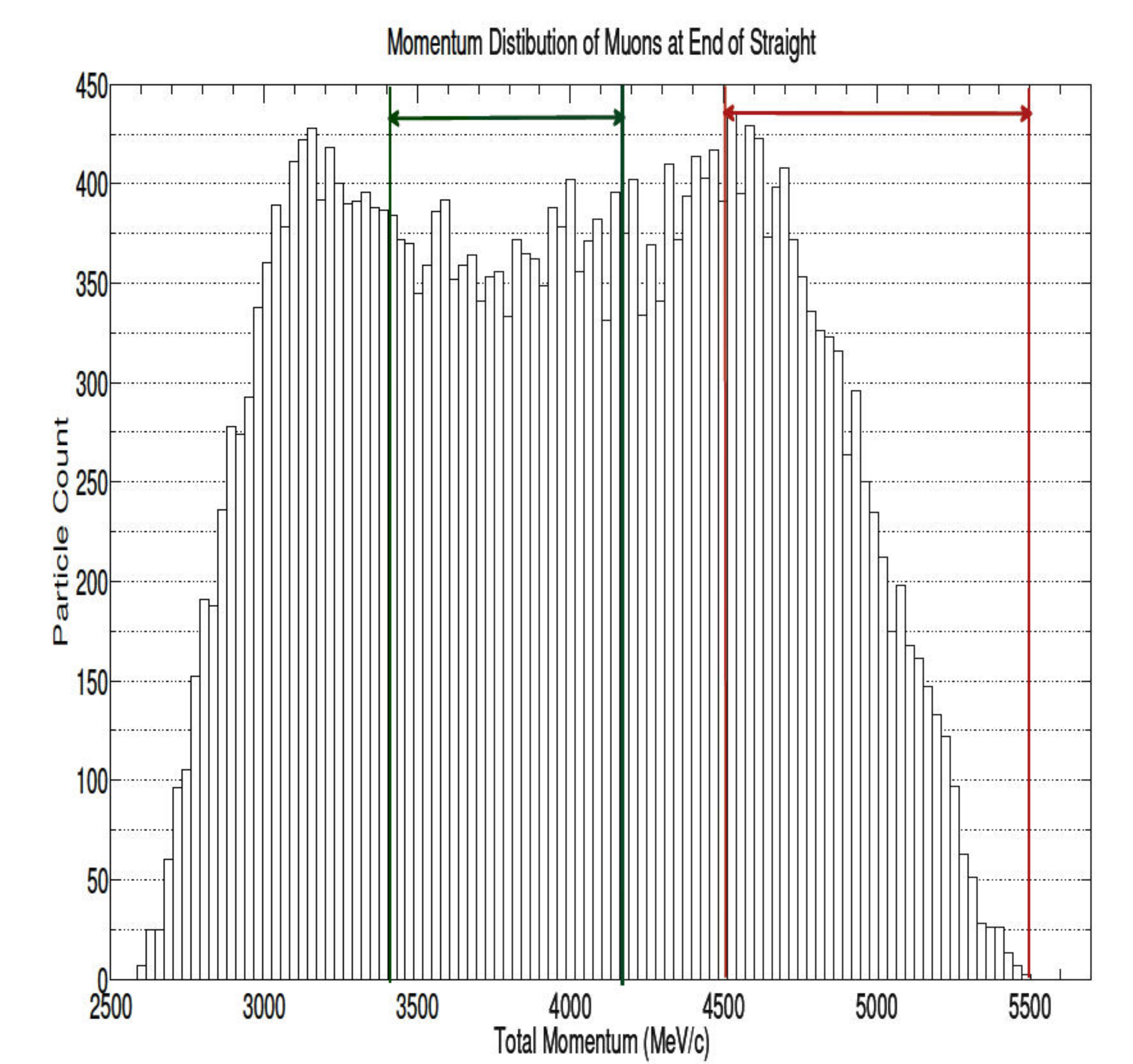} \quad\quad
    \includegraphics[width=0.36\textwidth]%
      {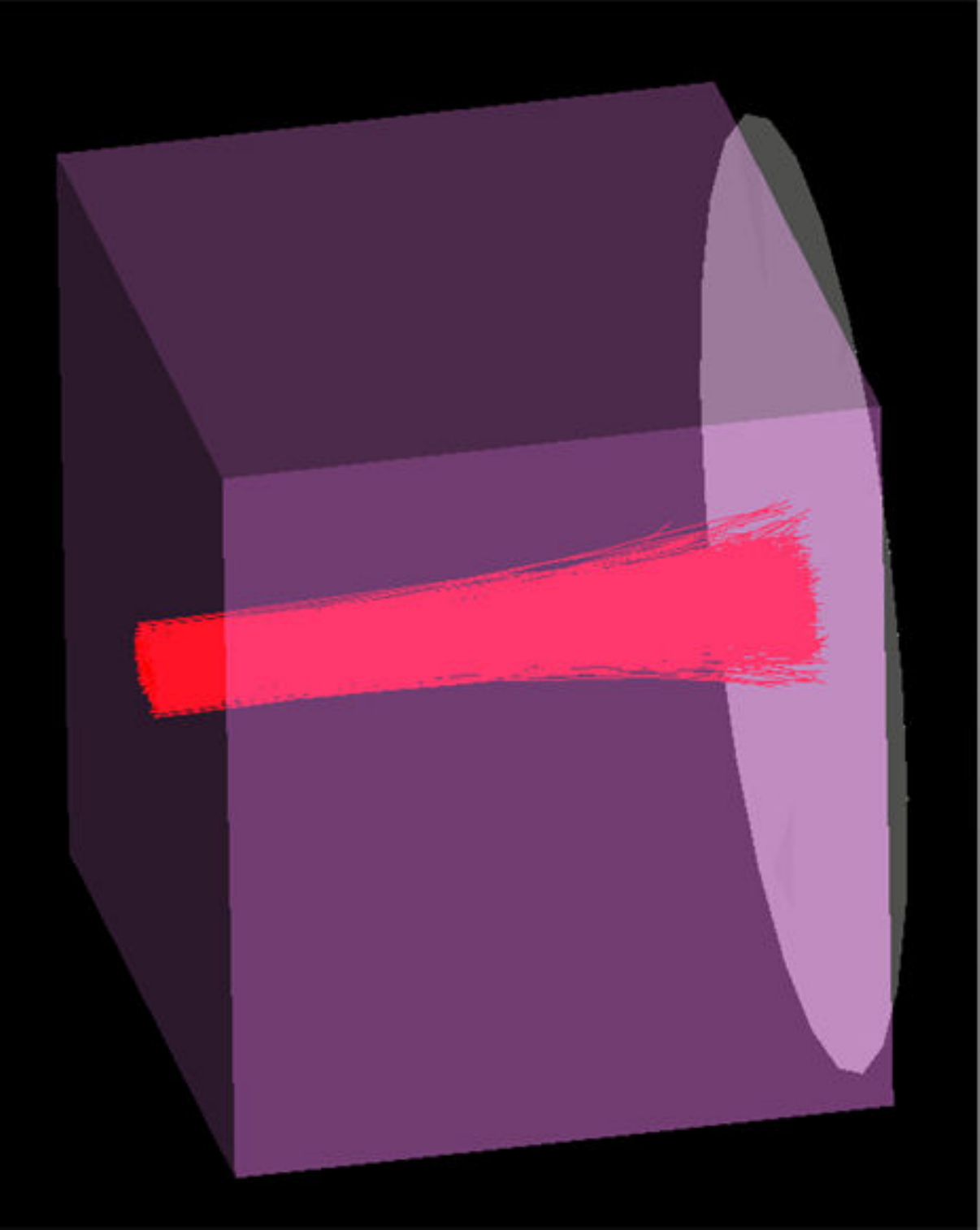}
  \end{center}
  \caption{
    Left panel: Momentum distribution of muons after the first
    straight.
    Right panel: Visualisation of muons in the degrader.
  } 
  \label{fig:degrader}
\end{figure}

\begin{figure}
  \begin{center}
    \includegraphics[width=0.48\textwidth]%
      {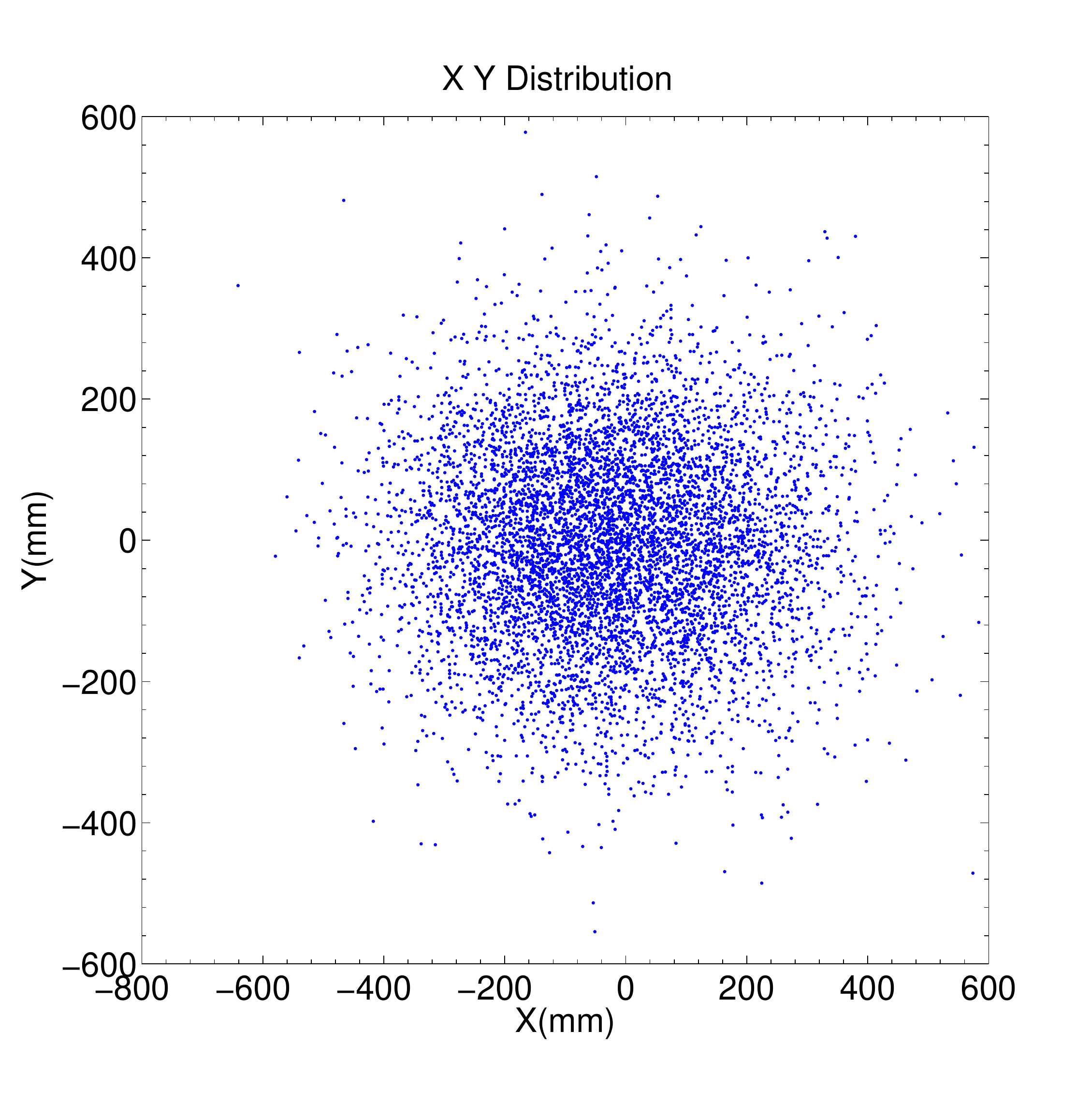} \quad\quad
    \includegraphics[width=0.46\textwidth]%
      {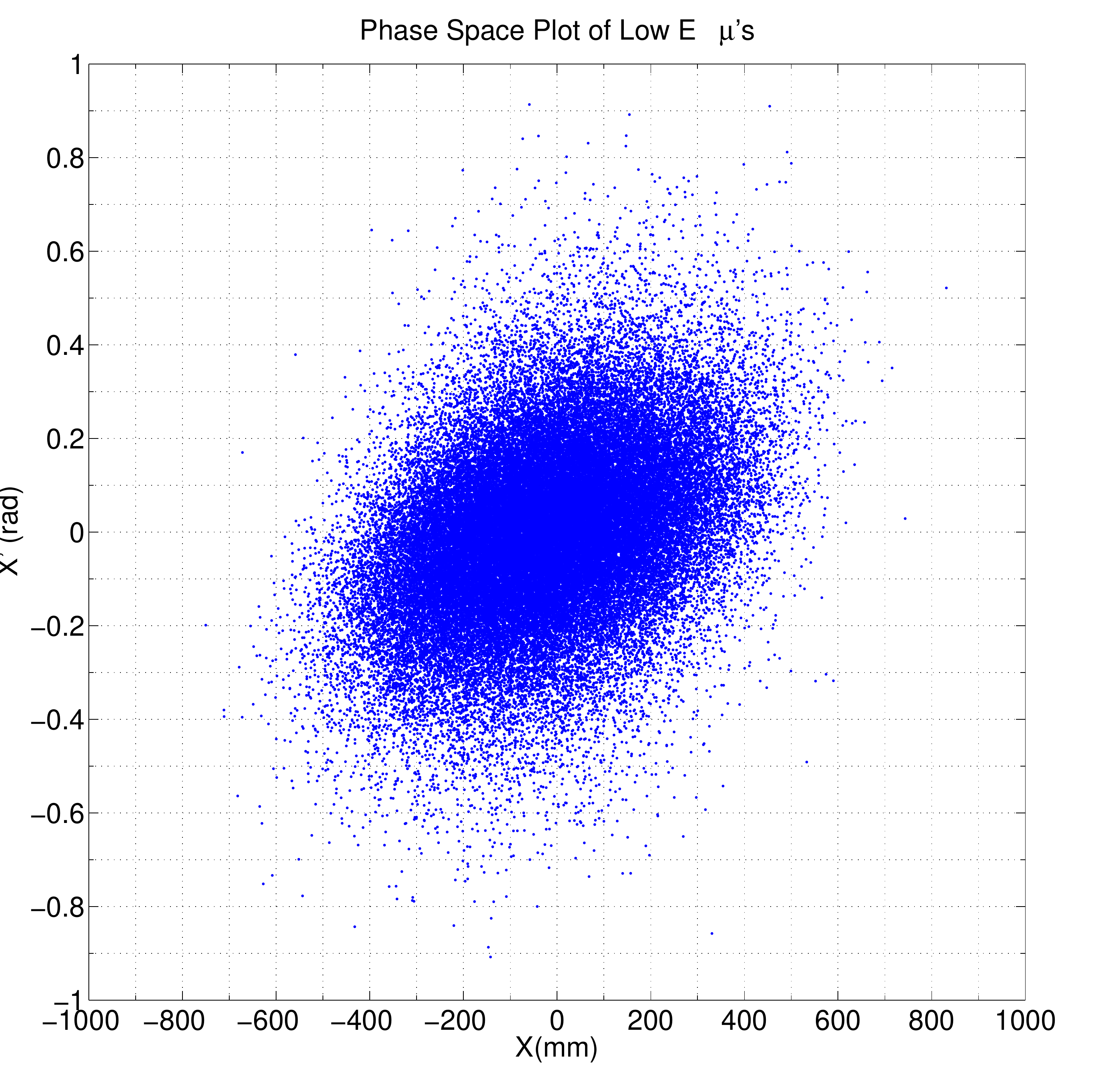}
  \end{center}
  \caption{
    Phase-space of the muon beam as it leaves the degrader.
    Left panel: $x-y$ distribution; Right panel: $x-x^\prime$ 
    distribution.
  }
  \label{fig:XY-dist}
\end{figure}

\begin{figure}
  \begin{center}
    \includegraphics[width=0.98\textwidth]%
      {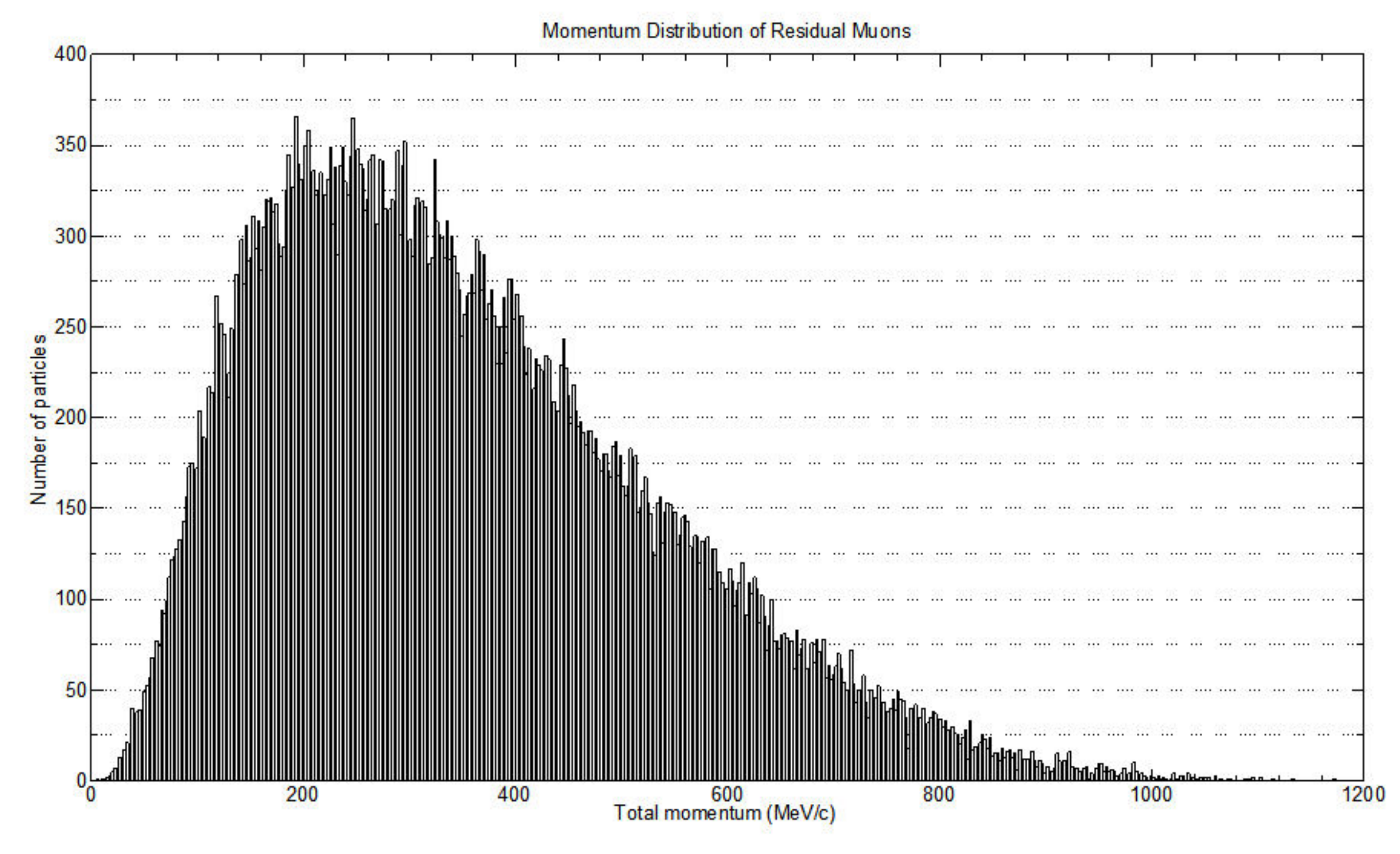}
  \end{center}
  \caption{Muon momentum distribution after degrader.}
  \label{fig:lowE}
\end{figure}

The same optics that are used for injection can be used to extract the
pions at the end of the straight and transport them to an absorber as
shown in figure \ref{fig:Decay_ring}.  
However, if the absorber is ``redefined'' to be a ``degrader" capable
of stopping the pions but allowing muons above a certain energy to
pass, then a low-energy muon beam appropriate for a 6D muon cooling
experiment can be produced.  
The left panel of figure \ref{fig:degrader} shows the momentum
distribution for the first pass of muons at the end of the decay-ring
straight.
The green band indicates the momentum acceptance of the decay ring.  
The red band covers the same momentum band as the input pions, these
muons will be extracted along with the remaining pions.
If the degrader is sized appropriately, a muon beam of the desired
momentum for a 6D cooling experiment will emerge downstream of the
degrader.
The right panel of figure \ref{fig:degrader} shows a visualisation of a
G4Beamline \cite{G4BeamLine:WWW,Roberts:2008zzc} simulation of the
muons in the pion momentum band ($5\pm10\%$\,GeV/c) propagating
through a 3.48\,m thick iron degrader. 
The left panel of figure \ref{fig:XY-dist} shows the $x-y$ distribution
of the muon beam exiting the degrader while the right panel shows the
$x-x^\prime$ distribution.
Figure \ref{fig:lowE} shows the muon momentum distribution of the
muons that exit the degrader.
Our initial estimate is that in the momentum band of interest for a 6D
cooling experiment (100--300\,MeV/c), we will have approximately
$10^{10}$ muons in the 1$\mu$sec spill.

\subsubsection{Neutrino cross-section measurements for Super Beams}
\label{SubSubSect:nuXSectfor SB}

\begin{figure}
  \begin{center}
    \includegraphics[width=0.95\textwidth]%
    {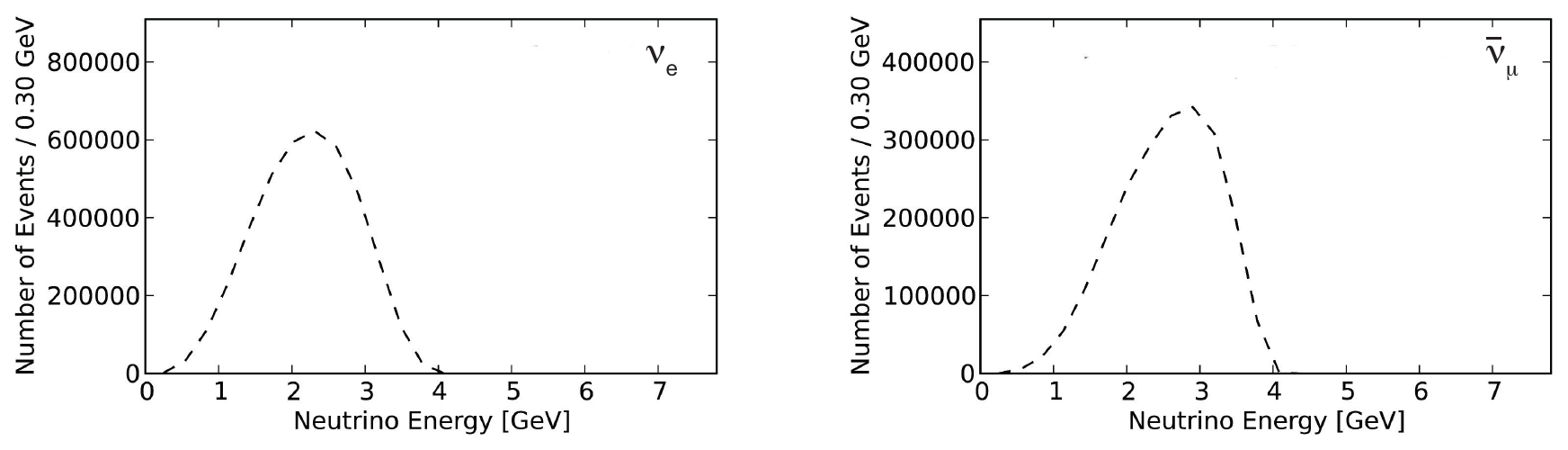}  \\
    \centering{
      \includegraphics[width=0.35\textwidth]%
      {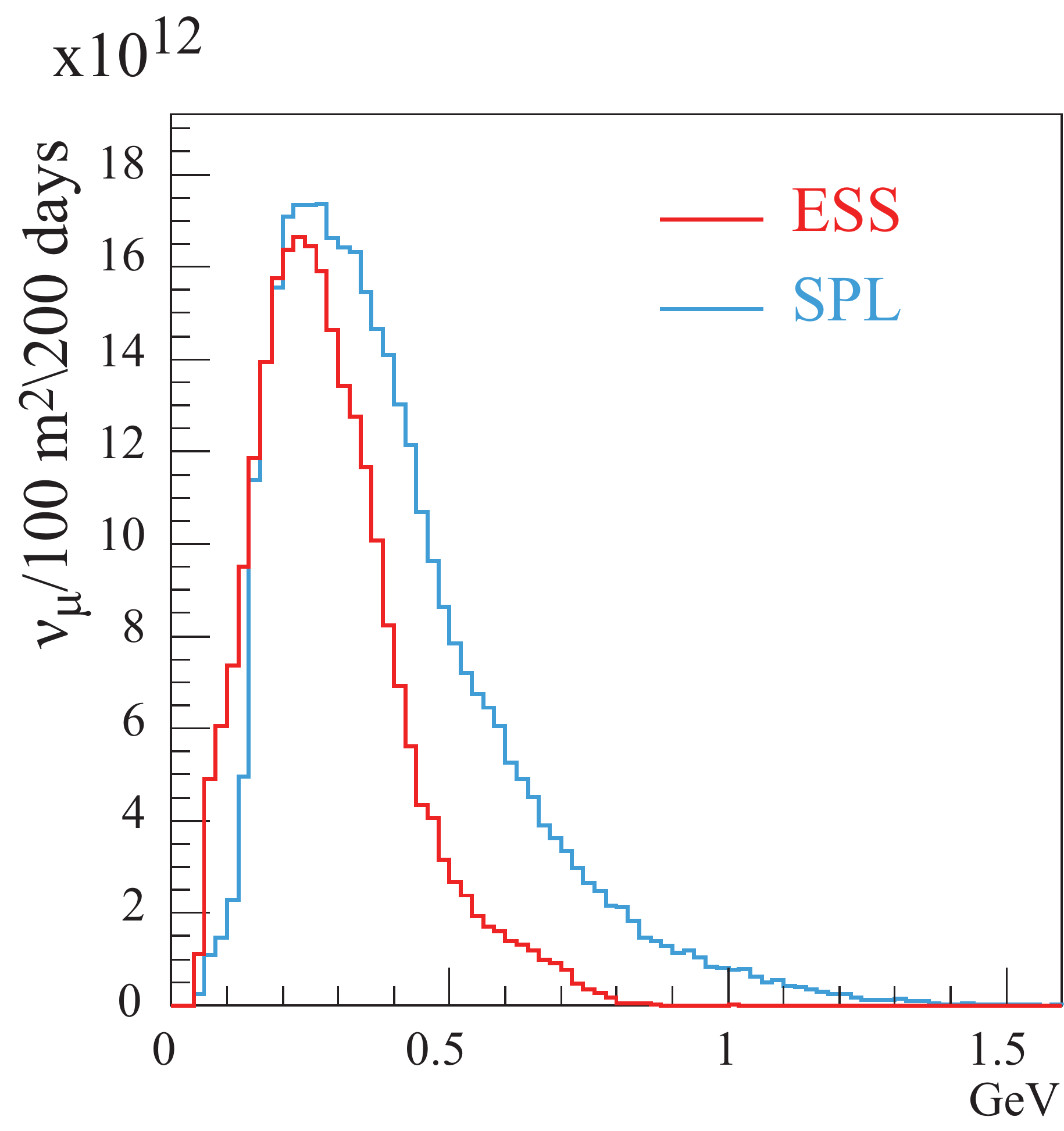}
    }
  \end{center}
  \caption{
    Top row: $\nu_e$ (left) and $\bar{\nu}$ (right) event rates per
    100\,T in a detector placed $\sim 50$\,m from the end of one of
    the straight sections (for a stored $\mu^+$ beam).
    Bottom row: neutrino energy spectrum for the SPL (with proton
    energy of 4.5\,GeV) and ESS (with proton energy of 2.5\,GeV) based 
    super beams.
  } 
  \label{Fig:Fluxes}
\end{figure}

The neutrino spectrum produced by the $\nu$STORM 3.8\,GeV/c stored muon
beam is shown in figure \ref{Fig:Fluxes}.
The $\nu$STORM flux at low neutrino energy ($< 0.5$\,GeV) is
relatively low.
The neutrino energy spectrum that would be produced at a low-energy
super-beam such as the SPL-based beam studied in
\cite{Baussan:2012wf} or the recent proposed super beam at the
European Spallation Source (ESS) \cite{Baussan:2012cw} is shown also
shown in figure \ref{Fig:Fluxes}.
Both the SPL and ESS based super beams propose to use the MEMPHYS
water Cherenkov detector \cite{deBellefon:2006vq,Agostino:2012fd}.
To enhance the event rate in the low neutrino-energy region of
importance to such facilities, the possibility of capturing muons with
a central momentum below 3.8\,GeV/c is being studied.

The ``degrader'' introduced above provides an alternative technique by
which the required low-energy neutrino beam could be produced.
Muons with momentum in the range 4.5\,GeV/c to 5.5\,GeV/c extracted
and directed towards the degrader will decay to produce neutrinos with
energies of around 300\,MeV. 
A detector placed a few tens of metres behind the degrader would make
it possible to measure the $\parenbar{\nu}_e N$ and
$\parenbar{\nu}_\mu N$ cross sections required for SPL or ESS based
super beams.

%% file: 03-nuSTORM-facility/03-nuSTORM-facility.tex
\section{The \boldmath{$\nu$}STORM facility; overview}
\label{Sect:nuSTORM_overview}

\input 03-nuSTORM-facility/03-01-Accel/03-01-Accel
\input 03-nuSTORM-facility/03-02-SterileDetectors/03-02-SterileDetectors
\input 03-nuSTORM-facility/03-03-nuScattDetectors/03-03-nuScattDetectors

%% file: 03-nuSTORM-facility/03-01-Accel/03-01-Accel.tex
\subsection{Accelerator facility}
\label{SubSect:AccelFacility}

\begin{figure}
  \begin{center}
    \includegraphics[width=0.90\textwidth]
      {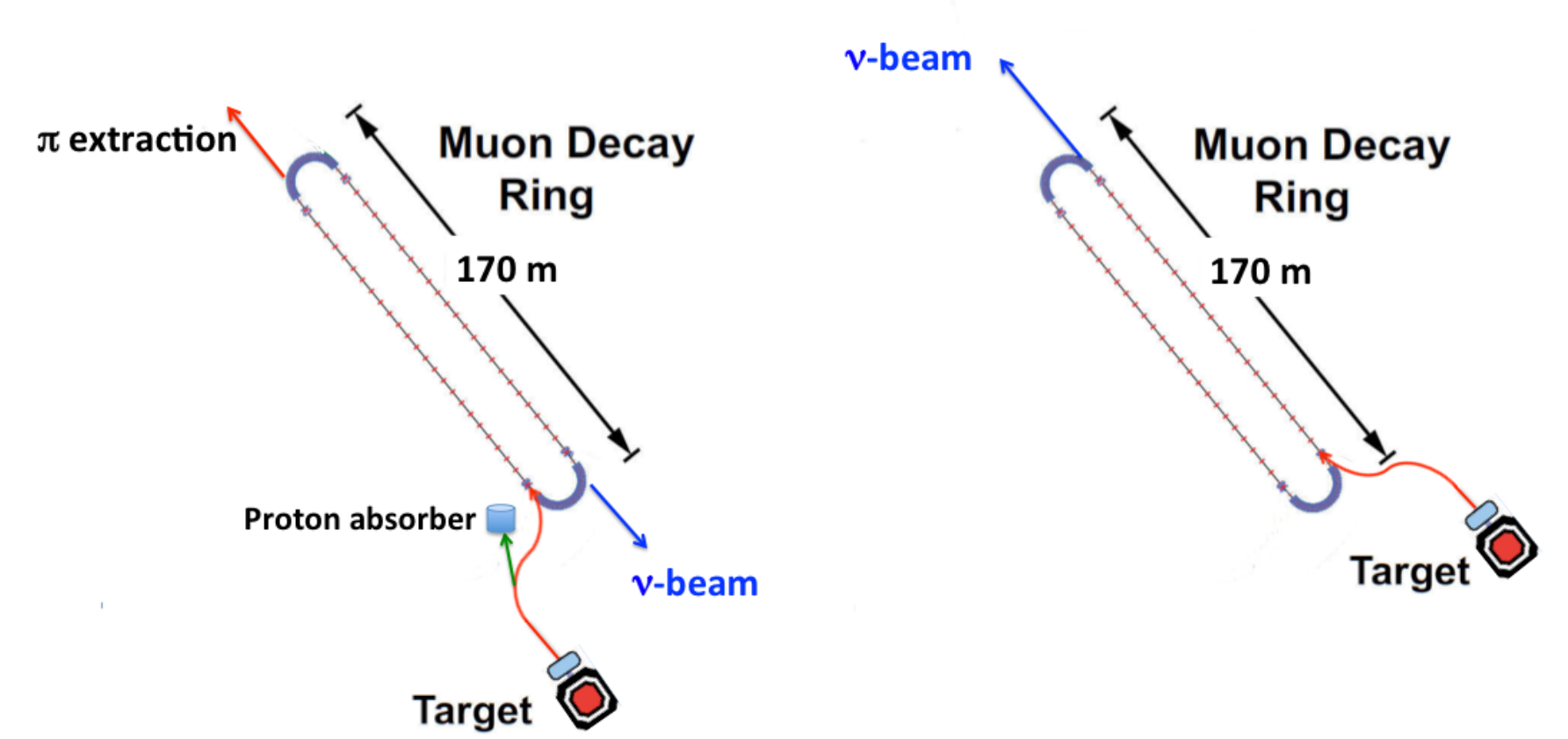}
  \end{center}
  \caption{
    The $\nu$STORM facility, with a Decay Ring having 150\,m straights
    and 25\,m, 180$^\circ$ arcs. 
    The left figure shows a possible CERN layout where injection of
    pions produced from the full 10\,$\mu s$ spill of 100\,GeV protons
    from the SPS would correspond to $\sim 7$ turns of the muon beam
    in the storage ring.
    The FNAL option, to the right, injects pions produced during only
    1 $\mu s$ 60\,GeV proton extraction from the Main Ring. 
  }
  \label{Fig:Accel:Schema}
\end{figure}

The concept for the facility proposed in \cite{Kyberd:2012iz} is shown
in figure \ref{Fig:Accel:Schema}.
The neutrino beam is generated from the decay of muons confined within
a race-tracked shaped storage ring. 
A high-intensity proton source places beam on a target, producing a
large spectrum of secondary pions.  
Forward pions are focused by a collection element (horn) into a
transport channel. 
Pions decay within the first straight of the decay ring and a fraction
of the resulting muons are stored in the ring. 
Muon decay within the straight sections will produce neutrino beams of
known flux and flavour via: 
$\mu^+ \rightarrow e^+\,\nu_e\,\bar{\nu}_\mu$ or
$\mu^- \rightarrow e^-\,\bar{\nu}_e\,\nu_\mu$.
A storage ring of 3.8\,GeV/c is proposed to obtain the desired spectrum
of $\sim 2$\,GeV neutrinos; pions have then to be captured at a
momentum of approximately 5\,GeV/c. 
In table \ref{Tab:Accel:Param} the  parameters for the Fermilab
baseline option (60\,GeV protons on target) and the proposed parameters
for a CERN option (100\,GeV protons on target) are shown.
We assume that  similar production (target and capture) and ring
layout is used for the FNAL and the CERN implementations. 
There may be constraints for the CERN option in the North Area that
can have impact on the design of the injection and of the ring itself
(see section \ref{SubSect:nuSTORMatCERN}).
\begin{figure}
  \begin{center}
    \includegraphics[width=0.90\textwidth]
      {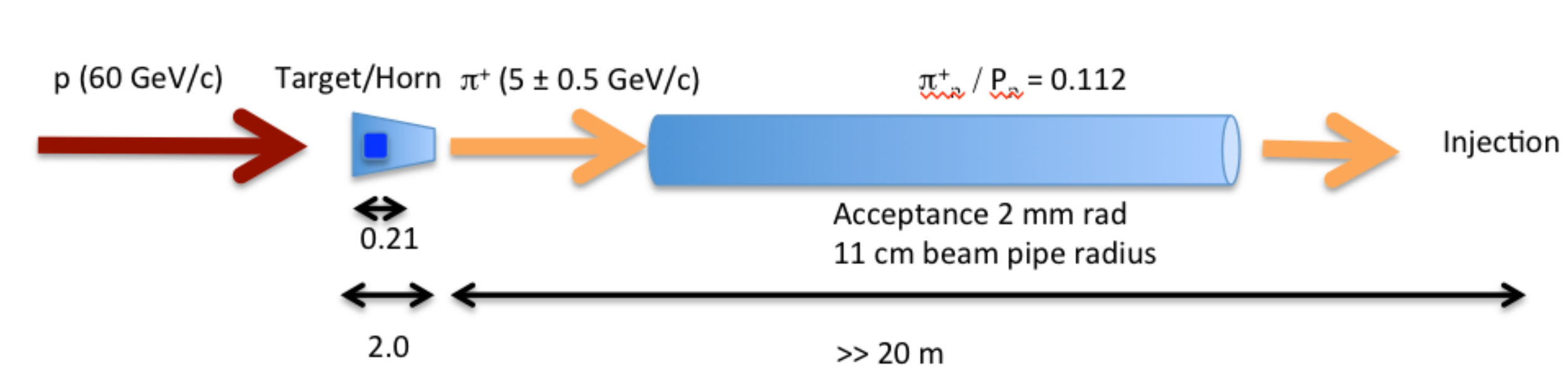}
  \end{center}
  \caption{
    Block diagram of the production section of the $\nu$STORM facility.
  }
  \label{Fig:Accel:ProdSect}
\end{figure}
\begin{table}
  \caption{
    Summary of parameters for $\nu$STORM at Fermilab and at CERN. 
    For Fermilab the performance is based on simulations with a tantalum
    target and a NuMI-like horn operating at 300\,kA.
  }
  \label{Tab:Accel:Param}
  \begin{center}
    \begin{tabular}{|l|l|l|}
      \hline
      {\bf Neutrino characteristics}             & {\bf Fermilab} & {\bf CERN}                          \\
      Aimed neutrino energy [GeV]                & 1.0 to 3.0     & 1.0 to 3.0                          \\
      Flux measurement precision [\%]            & 1.0            & 1.0                                 \\
      Protons on target (POT)                    & $10^{21}$           & 2.3$10^{20}$                            \\
      Useful $\mu$ decays $[10^{18}]$            & 1.00           & $100/60=1.67$                       \\
      \hline
      {\bf Production, horn and injection}       &                &                                     \\
      Target (Ta) diameter/length [m], material       & 0.01/0.21         & - / -                                \\
      Pulse length [$\mu$s]                      & 1.0              & 10.5                                \\
      Proton energy [GeV/c]                      & 60             & 100                                 \\
      Pion energy [GeV/c]                        & $5.0\pm10$\%   & $5.0\pm10$\%                        \\
      Horn diameter/length [m]                   & - / 2.0              & - / -                         \\
      Reflector diameter/length [m]              & -             & - / -                         \\
      Current Horn/Reflector [kA]                & 300            & - / -                             \\
      Estimated collection efficiency            & 0.8            & 0.8                                 \\
      Estimated transport efficiency             & 0.8            & 0.8                                 \\
      Estimated injection efficiency             & 0.9            & 0.9                                 \\
      Acceptance [mm rad]                        & 2.0            & 2.0                                 \\
      $\pi/{\rm pot}$ within momentum acceptance & 0.11      & $0.11 \times \frac{100}{60}=0.187$ \\
      Length of target [m]                       & 0.21           & 0.21                                \\
      Distance between target and horn [m]       & inside              & inside                                 \\
      Length of horn [m]                         & 2.0              &  -                                  \\ 
      Distance between horn and injection [m]    & 20              & 20                                    \\
      \hline
      {\bf The muon storage ring}                &                &                                     \\
      Momentum of circulating muon beam [GeV/c]  & 3.8            & 3.8                                 \\
      Momentum of circulating pion beam [GeV/c]  & $5.0\pm10$\%   & $5.0\pm10$\%                        \\
      Circumference [m]                          & 350            & 350                                 \\
      Length of straight [m]                     & 150            & 150                                 \\
      Ratio of Lstraight to ring circumference [$\Omega$] & 0.43  & 0.43                                \\
      Dynamic aperture, $\rm A_{dyn}$                     & 0.7            & 0.7                                 \\
      Acceptance [mm rad]                        & 2.0            & 2.0                                 \\
      Decay length [m]                           & 240            & 240                                 \\
      Fraction of $\pi$ decaying in straight ($\rm F_s$) & 0.41          & 0.41                                \\
      Relative $\mu$ yield ($N_\mu / {\rm POT}$)  & 0.002 & 0.002 \\
      \hline
      {\bf Detectors}                            &                &                                     \\
      Distance from target [m]                   &  20/1600              & 300/1800-2700                            \\
      \hline
    \end{tabular}
  \end{center}
\end{table}

\subsubsection{Production}

The production section of the facility is sketched in figure
\ref{Fig:Accel:ProdSect}.
A tantalum target is being studied at FNAL, including horn collection
and transport of pions up to the injection point. 
Different target materials, including low-$Z$ targets such as carbon,
will be considered.
Assessment of the feasibility of the target design and the choice of
target material will require the following studies to be made:
\begin{enumerate}
  \item {\it Heat removal:} \\
    A significant heat load is deposited by the beam
    on the target and has to be removed reliably by the cooling
    system;
  \item {\it Static and dynamic stresses:}  \\
    The target must withstand thermal-mechanical stresses arising from
    the beam-induced heating of the target;
  \item {\it Radiation damage:} \\
    Degradation of the material properties due to radiation damage
    must be accommodated;
  \item {\it Geometrical constraints:}
    The target has to fit inside the bore of the magnetic horn whilst
    having an appropriate geometry for effective pion production;
  \item {\it Remote replacement:} \\
    Once activated the target has to be remotely manipulated in the
    event of failure;
  \item {\it Minimum expected lifetime:} \\
    The target is expected operate without intervention between
    scheduled maintenance shutdowns; and
  \item {\it Safe operation:} \\
    The target design should minimise any hazard to the personnel or
    the environment.
\end{enumerate}
Beam structure on timescales below $\mu$s will not be ``seen'' by the
target.
The beam pulse has to be fast extracted to enhance background
rejection.

Simulations using a tantalum target show that, $N_\mu$, the number of
muons which decay in the production straight, would be comparable to
\cite{Kyberd:2012iz}:
\begin{equation}
  N_{\mu} = {\rm POT} \times (\pi{\rm~per~POT}) \times 
           \epsilon_{\rm col} \times \epsilon_{\rm trans} \times
           \epsilon_{\rm inj} \times (\mu{\rm~per~}\pi) \times
           {\rm A_{dyn}} \times \Omega ~ ;
\end{equation}
where POT is the number of protons on target, $\epsilon_{\rm col}$ is
the collection efficiency, $\epsilon_{\rm trans}$ is the transport
efficiency, $\epsilon_{\rm inj}$ is the injection efficiency, $\mu$
per $\pi$ is the chance that an injected pion results in a muon within
the ring acceptance, ${\rm A_{dyn}}$ is the probability that a muon
within the decay ring aperture is within the dynamic aperture, and
$\Omega$ is the fraction of the ring circumference that directs muons
at the far detector. 
$\nu$STORM assumes $10^{21}$ POT for a 4--5 year run using 60\,GeV
protons. 
From \cite{Kyberd:2012iz}, one obtains (with horn collection) 
$\sim 0.1 \pi/\rm{POT} \times \epsilon_{\rm col}$. 
The collection efficiency is 0.8.  
The transport efficiency (after collection to injection), and the
injection efficiency are assumed to be 0.8 and 0.9, respectively and
the probability that a $\pi$ decay results in a $\mu$ within the
acceptance times ${\rm A_{dyn}}$ is estimated to be 0.08. 
$\Omega$ is estimated to be 0.43. 
With these parameters the relative muon yield, $N_\mu / {\rm POT}$,
is $\sim 0.002$.

The number of pions produced off various targets by 60 GeV/c protons
has been simulated \cite{Kyberd:2012iz}. 
Target optimisation based on a conservative estimate for the
decay-ring acceptance of 2\,mm\,radian corresponds to a decay ring with
11\,cm internal radius and a $\beta$ function of 600\,cm. 
The optimal target length depends on the target material and the
secondary pion momentum. 
Results of the optimisation study are included in table
\ref{Tab:Accel:Param}.
Approximately 0.11 $\pi^+$/POT can be collected into a $\pm 10$\%
momentum acceptance off medium/heavy targets assuming ideal capture.

For the simulations, a NuMI design has been used; optimisation of the
horn inner shape could increase the number of collected pions. 
Simulations show that $\mu$/POT is an approximately linear function of
energy for the proton energies of interest.
These results are used to estimate the pion yield for the proposed SPS
proton beam energy. 
Ultimately, the CERN implementation (100 GeV proton case) remains to
be evaluated. 

To determine the available number of useful muons for the CERN case,
the values from the production studies in \cite{Kyberd:2012iz} have
been adjusted to take into account the linear dependence of $\mu$/POT
on proton energy.  

\subsubsection{Injection}

\begin{figure}[ht]
  \begin{center}
    \includegraphics[width=0.65\textwidth]
      {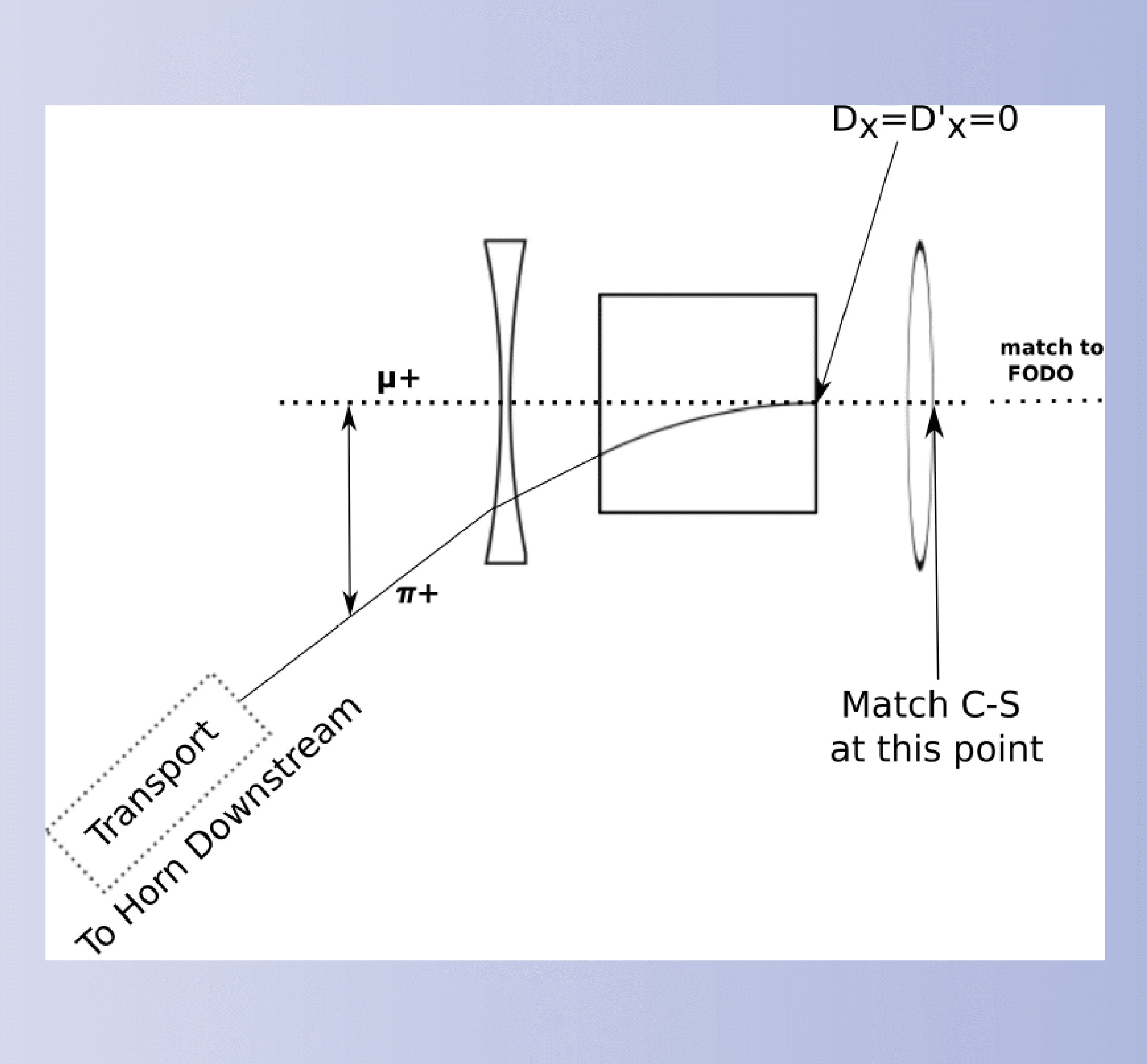}
  \end{center}
  \caption{
    Schematic diagram of stochastic injection into the $\nu$STORM
    ring.
  }
  \label{Fig:Accel:InjKick}
\end{figure}

Pion decay within the ring, and non-Liouvillean ``stochastic
injection'', are assumed to be optimised options. 
In stochastic injection, the $\simeq 5$\,GeV/c pion beam is
transported from the target into the storage ring and
dispersion-matched into a long straight section. 
Circulating and injection orbits are separated by momentum. 
Decays within the straight section provide muons that are within the
$\simeq 3.8$\,GeV/c ring momentum acceptance. 
With stochastic injection, muons from a beam pulse as long as the FNAL
Main Injector circumference (3\,000\,m) can be accumulated, and no
injection kickers are needed, see figure \ref{Fig:Accel:InjKick}.
For 5.0 GeV/c pions, the decay length is $\simeq 280$\,m; 
$\simeq 42$\% decay within the 150\,m decay ring straight. 

\subsubsection{Decay ring}

\begin{figure}
  \begin{center}
    \includegraphics[width=0.85\textwidth]
      {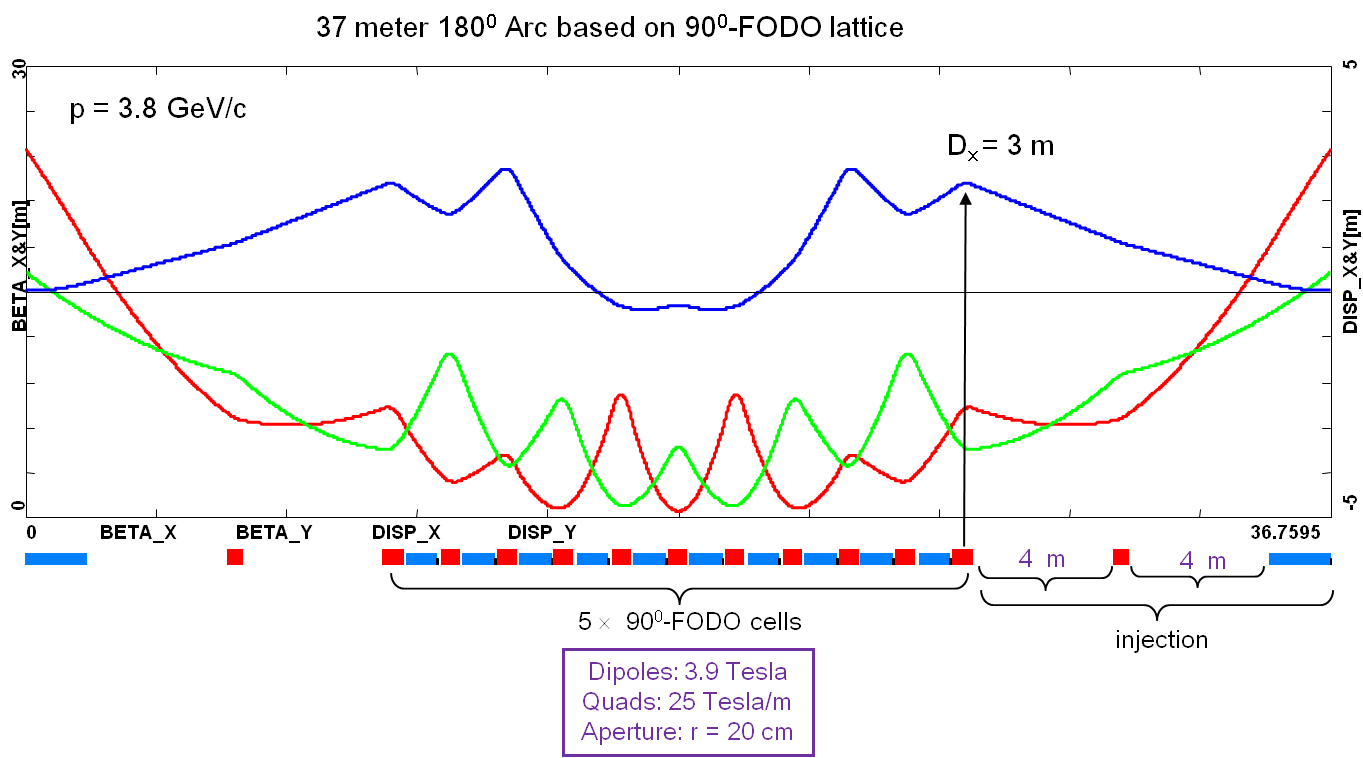}
  \end{center}
  \caption{
    Schematic of the magnets that make up the lattice in the storage-ring
    arcs together with the optical functions as indicated in the legend.
  }
  \label{Fig:Accel:ArcLatt}
\end{figure}

The decay ring is a compact racetrack design based on separate
function magnets.
The design goal is to maximise the momentum acceptance (around 
3.8\,GeV/c central momentum), while maintaining reasonable physical
apertures for the magnets in order to keep the cost down. 
This is accomplished by employing strongly focusing optics in the arcs
($90^\circ$ phase advance per FODO cell), featuring small $\beta$
functions ($\simeq 3$\,m average) and low dispersion ($\simeq 0.8$\,m
average).  
The linear optics for one of the $180^\circ$ arcs is illustrated in
figure \ref{Fig:Accel:ArcLatt}.

The current lattice design incorporates a missing-magnet dispersion
suppressor which will house the stochastic injection. 
With a dispersion of $\eta \simeq 1.2$\,m at the drift, the 5\,GeV/c
and 3.8\,GeV/c orbits are separated by $\simeq 30$\,cm; an aperture of
$\simeq \pm 15$\,cm is available for both the 5\,GeV/c $\pi$ and 
3.8\,GeV/c $\mu$ orbits.  
To maintain the high compactness of the arc, while accommodating
adequate drift space for the injection chicane to merge, two special
``half empty'' cells with only one dipole per cell were inserted at
both ends of the arcs to suppress the horizontal dispersion. 
This solution will limit the overall arc length to about 25\,m, while
keeping the dipole fields below 4\,T.
The arc magnets assume a relatively small physical aperture of radius
15\,cm, which limits the maximum field at the quadrupole magnet pole
tip to less than 4\,T.

\begin{figure}
  \begin{center}
    \includegraphics[width=0.85\textwidth]
      {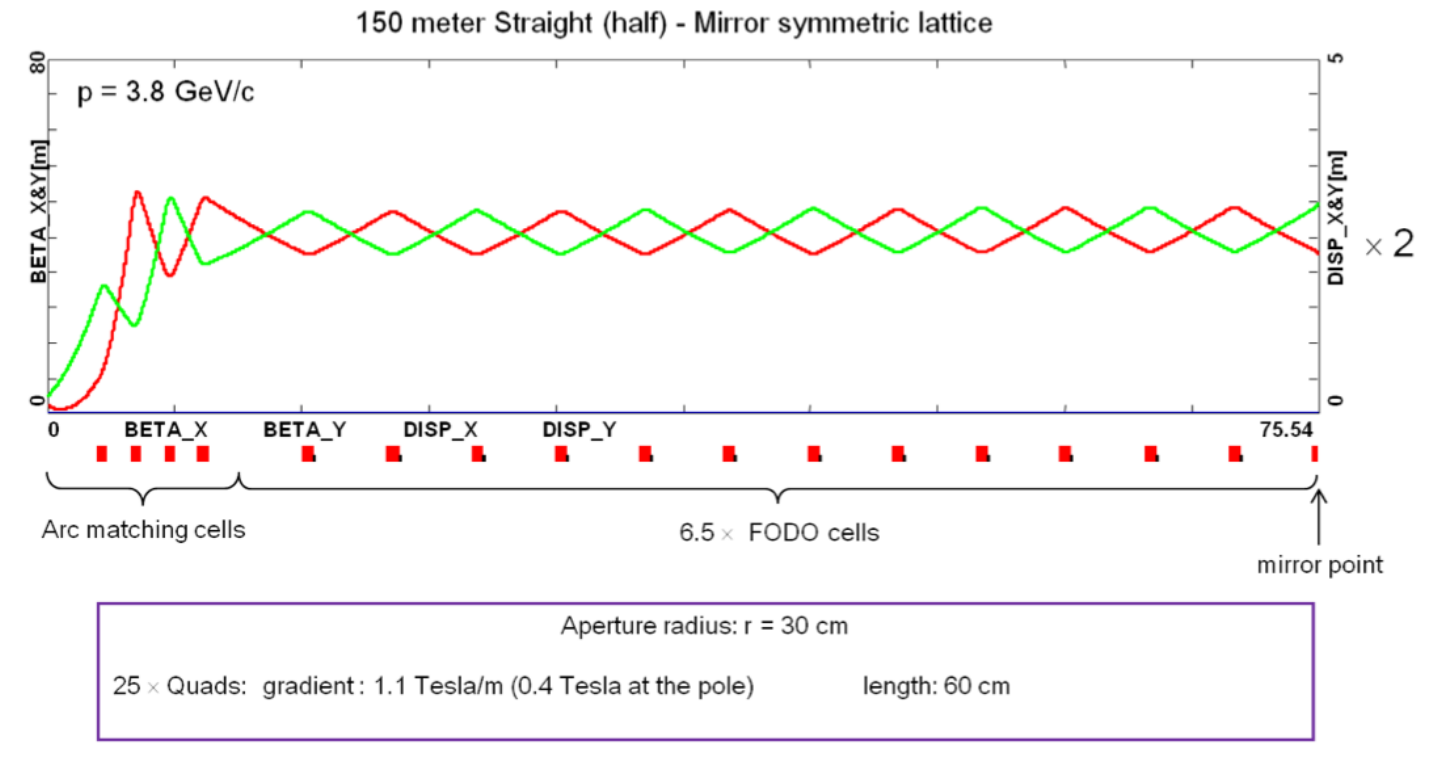}
  \end{center}
  \caption{
    Magnet lattice and optical functions of the decay straight.
  }
  \label{Fig:Accel:DecayStrght}
\end{figure}

On the other hand, the decay straight requires much larger values of
$\beta$-functions ($\simeq 40$\,m average) in order to maintain small
beam divergence ($\simeq 7$\,mrad).
The resulting muon beam divergence is a factor of 4 smaller than the
characteristic decay cone of $1/\gamma$ ($\simeq 0.028$ at 3.8\,GeV/c).
As illustrated in figure \ref{Fig:Accel:DecayStrght}, the decay
straight is configured with a much weaker focusing FODO lattice 
($30^\circ$ phase advance per cell). 
It uses normal conducting large aperture (r = 30 cm) quads with a
modest gradient of 1.1\,T/m (0.4\,T at the pole tip). 
Both the arc and the straight are smoothly matched via a compact
telescope insert, as illustrated in figure
\ref{Fig:Accel:DecayStrght}.

\begin{figure}
  \begin{center}
    \includegraphics[width=0.85\textwidth]
      {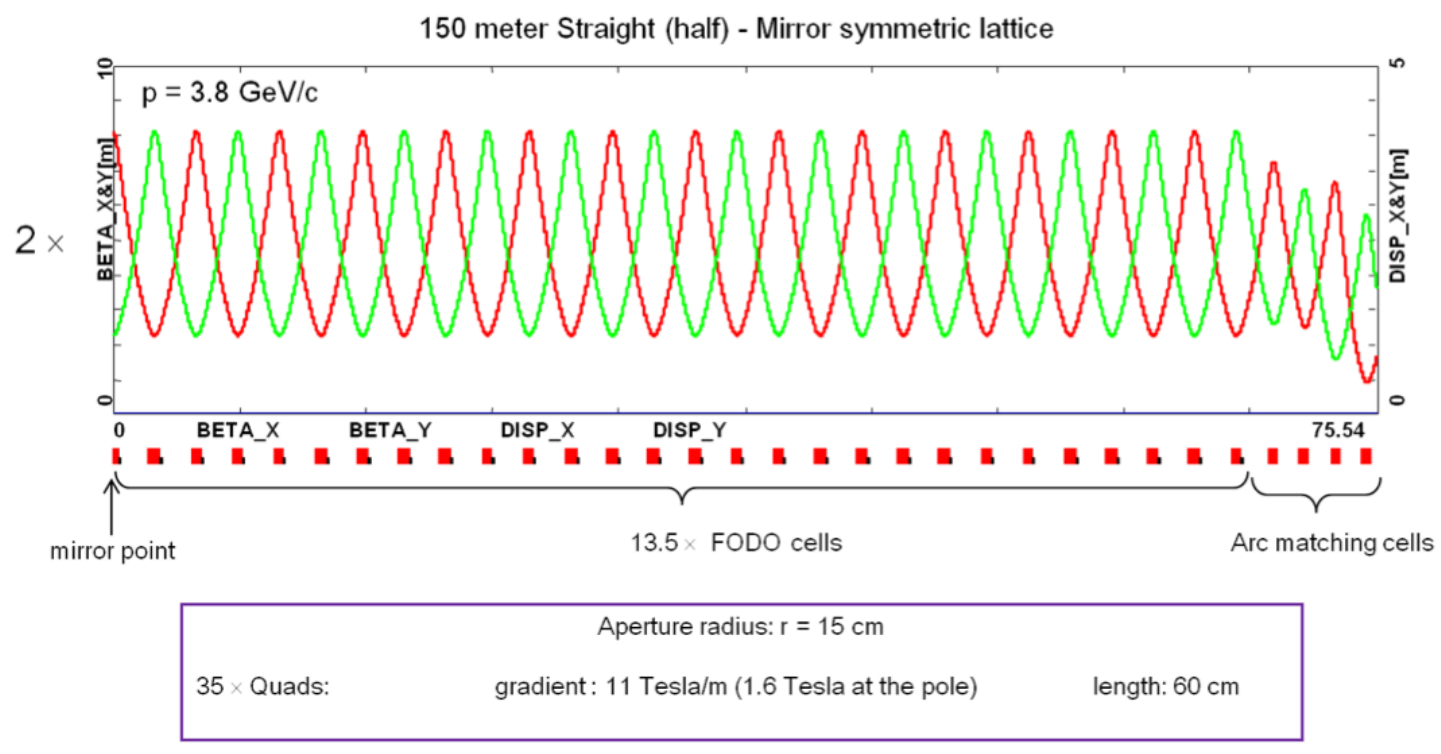}
  \end{center}
  \caption{
    Magnet positions and lattice functions for the ``return
    straight'', i.e. the straight that returns the muons to the decay
    straight.
  }
  \label{Fig:Accel:DecayStrght2}
\end{figure}

The ``other'' 150\,m straight, which is not used for neutrino
production, can be designed using a much tighter FODO lattice 
($60^\circ$ phase advance per cell), with rather small $\beta$
functions comparable to those in the arc ($\simeq 5$\,m average).
This way one can restrict the aperture of the straight to a radius of
15\,cm.
The second straight uses normal conducting quads with a gradient of
11\,T/m (1.6\,T at the pole tip).
Both the arc and the straight are smoothly matched, as illustrated in 
figure \ref{Fig:Accel:DecayStrght2}.

The complete racetrack ring architecture features the ``low-$\beta$''
straight matched to the $180^\circ$ arc and followed by the
``high-$\beta$'' decay straight connected to the arc with a compact
telescope insert.
To summarise the magnet requirements, both $180^\circ$ arcs were
configured with 3.9\,T dipoles and 25\,T/m quads (superconducting
magnets with an aperture of radius 15\,cm).
Both straights use normal-conducting magnets: the decay straight, 
1.1\,T/m quads with 30\,cm radius aperture and the other straight, 
11\,T/m quads with 15\,cm radius aperture. 
These magnets are challenging.
The transverse normalised acceptance of the ring is 78\,mm\,rad both
in $x$ and $y$ (or a geometric acceptance of 2.1\,mm\,rad) for the net
momentum acceptance of $\pm 10\%$. 

%% file: 03-nuSTORM-facility/03-02-SterileDetectors/03-02-SterileDetectors.tex
\subsection{Detectors for sterile neutrino search}
\label{SubSect:DetectSterileSearch}

The Super B Iron Neutrino Detector (SuperBIND), an iron and
scintillator sampling calorimeter similar in concept to the MINOS
detector, is the baseline detector for the sterile-neutrino search
focusing on muon-neutrino appearance and disappearance.
Two detectors of this type would be used for short-baseline
oscillation measurements; one 100\,Ton detector at 50\,m and a
1.6\,kTon detector $\sim 1.5$\,km from the storage ring.
The near detector is required to measure the characteristics of the
neutrino beam prior to oscillation for the reduction of systematic
uncertainties.
Simulations have been conducted for the far detector---a near detector
simulation is in preparation.

\begin{figure}
  \begin{center}
    \includegraphics[width=0.75\textwidth]%
      {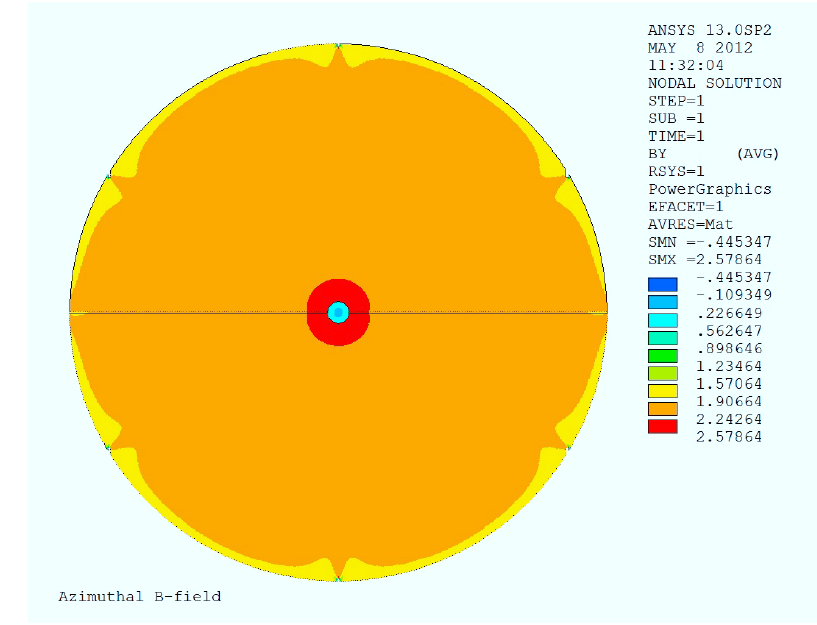}
  \end{center}
  \caption{
    A 2-D finite element magnetic field simulation of the SuperBIND
    iron plate.
  } 
  \label{fig:azB}
\end{figure}

The far detector has a circular cross-section 5\,m in diameter. 
The iron planes are to be 2\,cm thick and constructed from two
semi-circular pieces skip-welded at a central join.
The detector is magnetised using multiple turns of a superconducting
transmission line (STL) \cite{Ambrosio:2001ej} to carry a total of
250\,kA to induce a magnetic field between 1.5\,T and 2.5\,T within
the iron plate. 
To accommodate the STL, a 20\,cm bore runs through the centre of the
detector. A 2-D finite-element magnetic-field analysis of the iron
plate has been performed, with the results shown in figure
\ref{fig:azB}.

The scintillator detector planes are composed of two layers of
1$\times$1\,cm$^2$ scintillating bars providing vertical and
horizontal readout at each detection plane. 
A 1\,mm bore through the centre of each bar is provided for the
insertion of a wavelength shifting fibre. 
Each scintillator bar is read out from both ends using silicon
photo-multipliers.

\subsubsection{Far Detector Simulation}

A detailed detector simulation and reconstruction programme has been
developed for the determination of the detector response.
The simulation was based on software developed for the Neutrino
Factory Magnetised Iron Neutrino Detector (MIND) \cite{Bayes:2012ex}. 
GENIE \cite{Andreopoulos:2009rq} is used to generate neutrino events.
Events are passed to a GEANT4-based \cite{Apostolakis:2007zz}
simulation for the propagation of the final-state particles through
successive steel and scintillator layers.
This simulation includes hadron interactions simulated by the
QGSP\_BERT physics list \cite{Apostolakis:2007zz}.
Hits in the scintillator are grouped into clusters, smearing the
detector hit position, and energy deposition of the accumulated hits
is attenuated in a simple digitisation algorithm applied prior to
reconstruction.

\begin{figure}
  \begin{center}
    \includegraphics[width=0.75\textwidth]
      {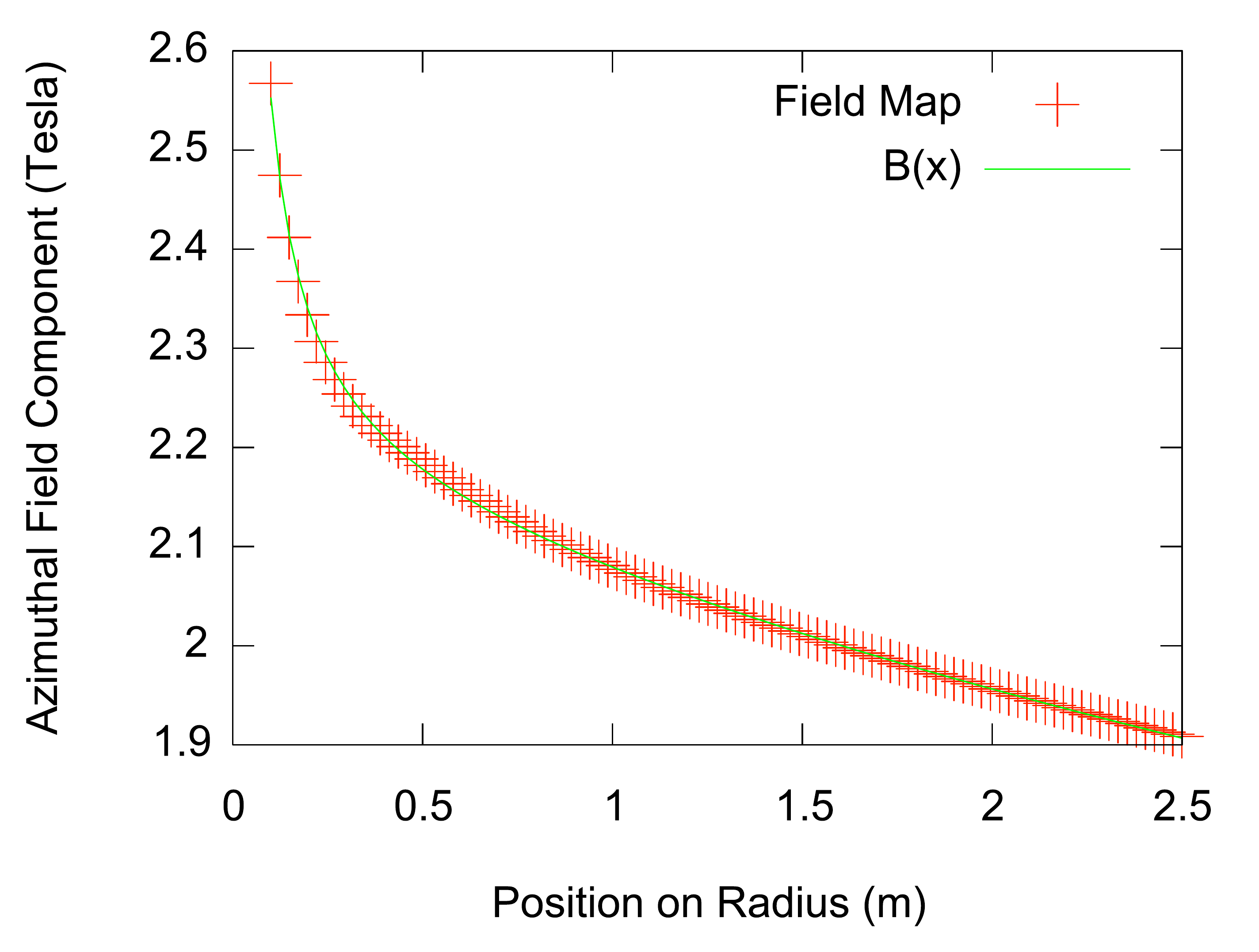}
  \end{center}
  \caption{
    The magnetic field magnitude as a function of radius along the
    45$^{\circ}$ azimuth with the parameterisation used in the detector
    simulation.
  } 
  \label{fig:Bmodel}
\end{figure}

Magnetisation within the iron is introduced by reducing the model of
figure \ref{fig:azB} to a toroidal magnetic field with a radial
dependence which follows the expression:
\begin{equation}
B_{\phi}(r) = B_{0} + \frac{B_{1}}{r} + B_{2}e^{-H r} ~ ;
\end{equation}
where $B_{0} = 1.53$\,T, $B_{1} = 0.032$\,T\,m, $B_{2} = 0.64$\,T, and 
$H = 0.28$\,m$^{-1}$. 
This parameterisation and the field along the 45$^\circ$ azimuthal
direction are shown in figure \ref{fig:Bmodel}. 

The reconstruction uses multiple passes of a Kalman-filter algorithm
for the purposes of identifying muon trajectories within events and to
determine the momentum and charge of an identified track. 
The algorithms are supplied by the RecPack software
package \cite{CerveraVillanueva:2004kt}. 
Geometrical information from the track including: the length of the
track; the direction of bending in the magnetic field; and the pitch
of the track are used at various points in this procedure to provide
information to the Kalman filter. 
The hadron reconstruction is not yet well developed so the
neutrino energy is reconstructed either by using the quasi-elastic
approximation, if no harmonisation is visible, or by smearing the true
hadron energy according to MINOS CalDet test beam
\cite{Michael:2008bc} results.

\subsubsection{Event Selection}

The reconstructed neutrino events are analysed to select events with
well reconstructed muons rather than those where muons are
mis-identified either in charge or particle identity. 
To achieve the target of 10$\sigma$ significance the background
efficiency must be reduced to less than parts in $10^{4}$. 
The selection of events is accomplished
with a multi-variate analysis facilitated by the ROOT based TMVA
package \cite{Hocker:2007ht}. 
This analysis outperforms the cuts based
analysis described previously \cite{Kyberd:2012iz} by offering a 
lower signal-energy threshold and increasing the sensitivity of the
experiment to oscillations.
\begin{table}[htdp]
\caption{Variables used in the analysis of events in the SuperBIND
  simulation. Variables in \subref{sub:mva} are used in the definition
  of the classifier, while the cuts in \subref{sub:fixed} are fixed. }
\subfigure[Variables used in the multivariate analysis.]{
	\begin{tabular}{rp{11cm}}
	\hline\hline
	Variable & Description \\
	\hline
	Track Quality & $\sigma_{q/p}/(q/p)$, the error in the trajectory curvature scaled by the curvature\\
	Hits in Trajectory & The number of hits in the trajectory \\
	Curvature Ratio & $(q_{init}/p_{range}) \times (p_{fit}/q_{fit})$: comparison of the initial guess of the curvature to the Kalman fit result. \\
	Mean Energy Deposition & Mean of energy deposition of hits in fit  of the trajectory \\
	Variation in Energy & $\sum^{N/2}_{i=0}\Delta E_{i} / \sum^{N}_{j=N/2}\Delta E_{j}$ where the energy deposited per hit $\Delta E_{i}~<~\Delta E_{i+1}$.\\
	\hline\hline 
	\end{tabular}
	\label{sub:mva}
          } \\
\subfigure[Preselection variables.]{
	\begin{tabular}{rp{11cm}}
	\hline\hline
	Variable & Description \\
	\hline
	Trajectory Identified & There must be at least one trajectory identified in event.\\
	Successful Fit & The longest identified trajectory must be successfully fit.\\
	Maximum Momentum & The momentum of the longest trajectory is less than 6 GeV/c. \\
	Fiducial & Longest trajectory must start prior to the last 1~m of the detector. \\
	Minimum Nodes & Fit to longest trajectory must include more than 60\% of hits assigned to trajectory by filter.\\
	Track Quality & $\sigma_{q/p}/(q/p) < 10.0$\\
	Curvature Ratio & $(q_{init}/p_{range}) \times (p_{fit}/q_{fit}) > 0$\\
	\hline\hline
	\end{tabular}
	\label{sub:fixed}
         	}
\label{tab:var}
\end{table}

\begin{figure}[htbp]
\begin{center}
\includegraphics[width=0.8\textwidth]{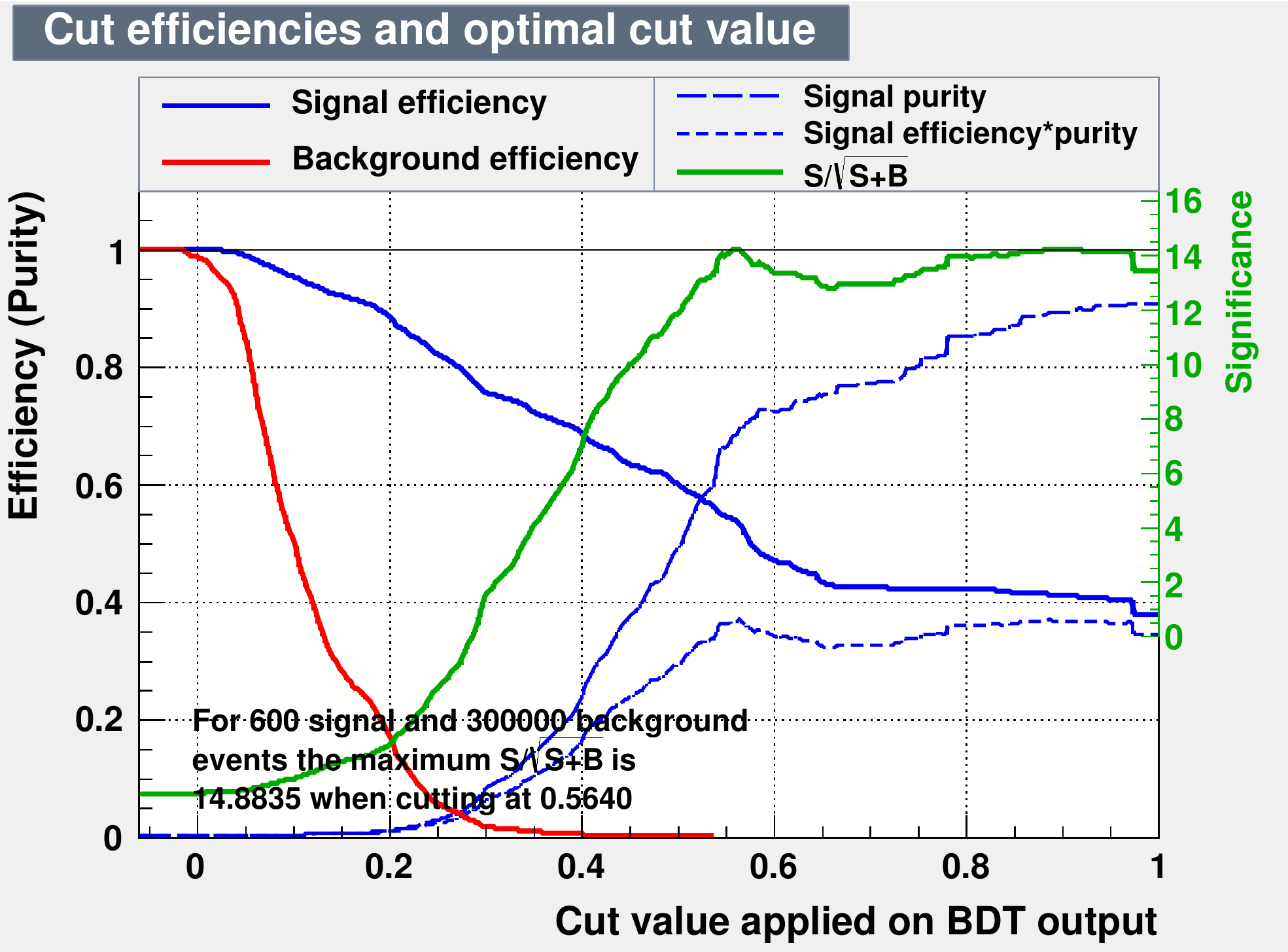}
\caption{Results from training the BDT method to simulations of
  $\nu_{\mu}$CC signal events and $\bar{\nu}_{\mu}$ background events,
  assuming a realistic number of events.}
\label{fig:eff}
\end{center}
\end{figure}

The analysis was trained to discriminate between the $\nu_{\mu}$
charged current (CC) interaction signal events and $\bar{\nu}_{\mu}$
neutral current (NC) interaction background events using a set of five
parameters to define a classifier variable. The majority of these
parameters were chosen based on the experience of the MINOS experiment
\cite{Adamson:2010uj}. Table \ref{tab:var}\subref{sub:mva} summarises
these parameters.  A set of preselection cuts, detailed in
Table~\ref{tab:var}\subref{sub:fixed} were applied to limit the
analysis to the subset of events containing useful data. These
preselection cuts were also used in the cuts based analysis with the
addition of cuts on the track quality and number of hits in a
trajectory. The multi-variate analysis was trained using a variety of
methods, but the best performance was achieved using Boosted Decision
Trees (BDT). Based on the performance of this method, shown in
figure \ref{fig:eff}, events are selected if the BDT classifier variable
is greater than 0.56.

\subsubsection{Sensitivities}

The appearance of $\nu_\mu$, via the channel $\nu_{e}\to\nu_{\mu}$,
gives $\nu$STORM broad sensitivity to sterile neutrinos and directly
tests the LSND/MiniBooNE anomaly. 
The oscillation probabilities for both appearance and disappearance
modes are:
\begin{eqnarray}
P_{\nu_e\to\nu_{\mu}} & = & 4|U_{e4}|^{2}|U_{\mu
  4}|^{2}\sin^{2}\left(\frac{\Delta m^2_{41}
  L}{4E}\right) \label{eq:app} ~ {\rm ;~and}\\
P_{\nu_{\alpha}\to\nu_{\alpha}} & = & 1 - [4|U_{\alpha 4}|^{2}(1 - |U_{\alpha
  4}|^{2})]\sin^{2}\left(\frac{\Delta m^2_{41} L}{4E}\right)
  ~. \label{eq:disapp} 
\end{eqnarray}
The detector is designed for the appearance signal
$\nu_{e}\to\nu_{\mu}$; the CPT conjugate of the channel with which
LSND observed an anomaly, $\bar{\nu}_{\mu}\to\bar{\nu}_{e}$. 
Although it is clear from equation \ref{eq:app} that the appearance
channel is doubly suppressed relative to the disappearance channel,
the experiment is much more sensitive to the appearance channel
because the backgrounds for wrong-sign muon searches can be suppressed
more readily.

\begin{figure}
  \begin{center}
    \includegraphics[height=0.75\textwidth,angle=270]%
    {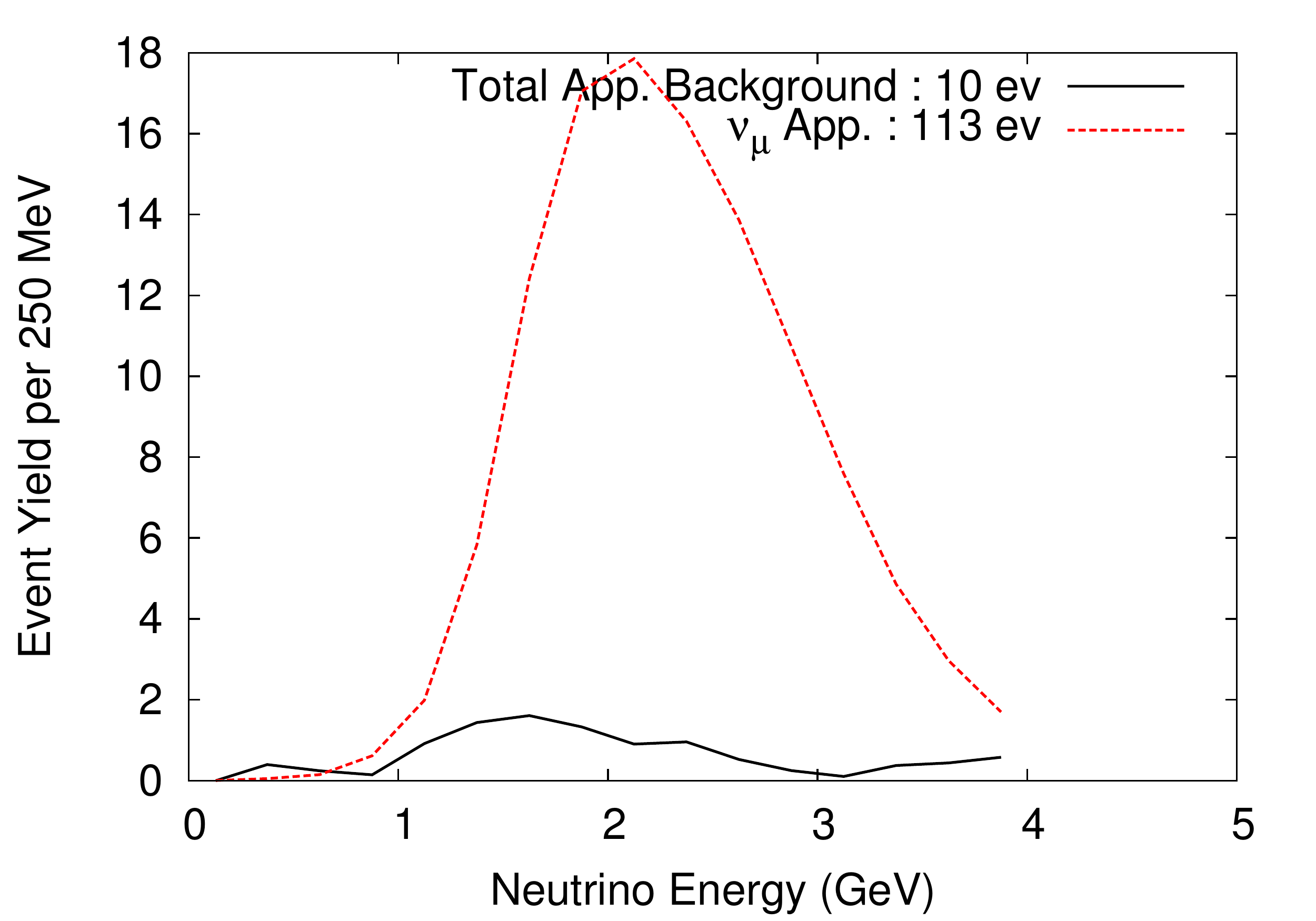}
  \end{center}
  \caption{
    The neutrino spectrum measured at the SuperBIND far detector using
    the simulated detector response.
  }
  \label{fig:spec}
\end{figure}

\begin{figure}[ht]
  \begin{center}
    \includegraphics[height=0.75\textwidth,angle=270]%
      {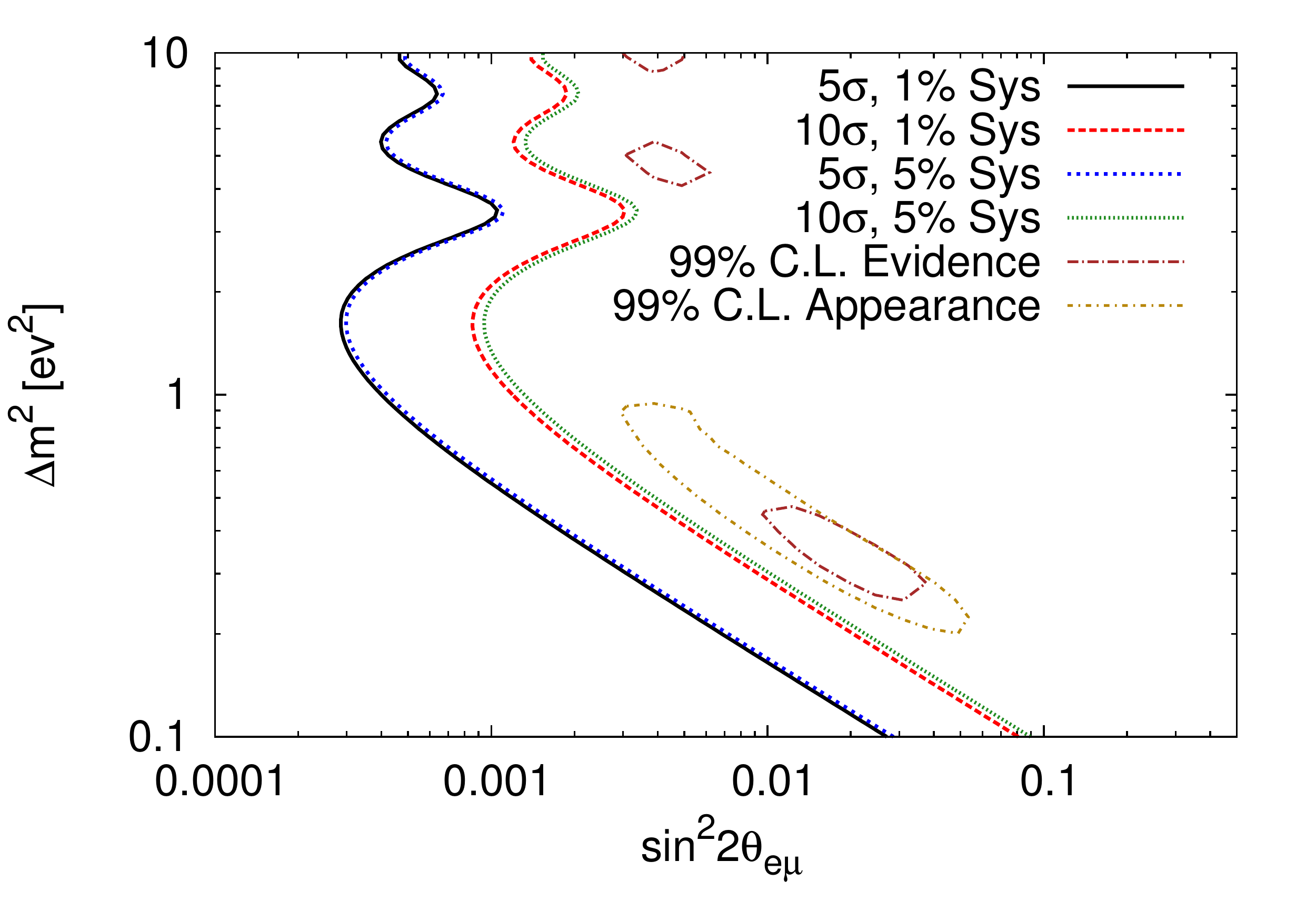}
  \end{center}
  \caption{ Contours of the $\chi^{2}$ deviation from the no-sterile
    neutrino hypothesis corresponding to 5$\sigma$ ($\chi^2=25$) and
    10$\sigma$($\chi^2=100$) variations. Two different sets of
    systematic uncertainties are represented; the default systematics
    with 1\% signal uncertainty and a 10\% background uncertainty and
    a conservative set that is five times larger. The 99\% confidence
    level contours from experiments showing evidence for unknown
    signals and contours derived from the accumulated data from all
    applicable neutrino appearance experiments, as described in
    figure \ref{fig:regions-3p1}.  }
  \label{fig:sens}
\end{figure}

The detector response derived from simulation is used to determine the
sensitivity of the experiment to the presence of sterile neutrinos.
The sensitivities and optimisation were computed using GLoBES
\cite{Huber:2004ka}. Modifications were made to simulate accelerator
effects such as the integration of the decay straight as outlined in
\cite{TunnellThesis,Tunnell:2012nu}.  
The detector response is summarised as
a ``migration'' matrix of the probability that a neutrino generated in
a particular energy bin $i$ is reconstructed in energy bin $j$.
Defined in this way, the migration matrix encapsulates both the
resolution of the detector and its efficiency.  Samples of all
neutrino interactions that could participate in the experiment are
generated to determine the response for each detection channel.  The
spectrum of expected signal and background for this simulation is
shown in figure \ref{fig:spec} assuming 1.8$\times 10^{18}$ $\mu^{+}$
decays collected over 5 years.  A contour plot showing the sensitivity
of the $\nu_{\mu}$ appearance experiment to sterile neutrinos is shown
in figure \ref{fig:sens}. These contours are shown with respect to the
derived variable 
$\sin^{2}2\theta_{e\mu} = |U_{e4}|^{2}|U_{\mu 4}|^{2}$. 
Systematic uncertainties are included in the contour as stated in the
legend.

The leading systematic uncertainty is expected to be related to the
neutrino flux appearing within the detector fiducial volume. This
systematic is anticipated to be on the order of 1\% for signal events
and 10\% for background events. These numbers assume that the neutrino
spectrum and rate will be well known from the storage ring. The
experiment is robust to a five fold increase in the systematic
uncertainties such that the appearance channel alone has the
sensitivity to probe the LSND anomaly with a confidence level
corresponding to more than 10$\sigma$.
\begin{figure}[h]
\begin{center}
\subfigure[Systematic effects]{
\includegraphics[width=0.48\textwidth]{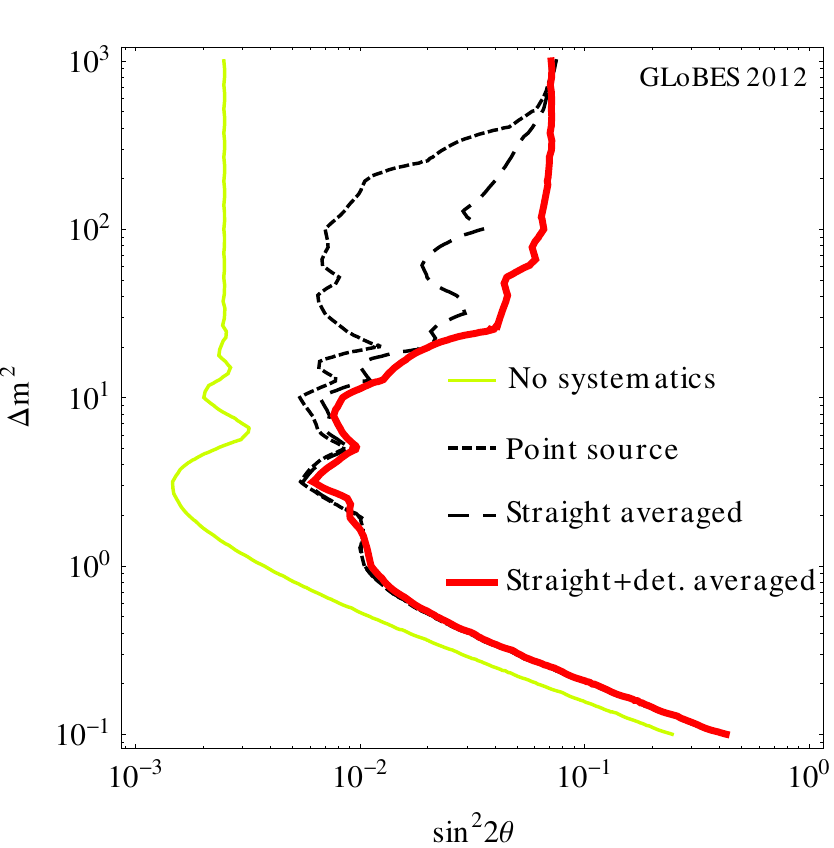} 
\label{figa:disapp}
}
\subfigure[Baseline optimisation]{
  \includegraphics[width=0.48\textwidth]{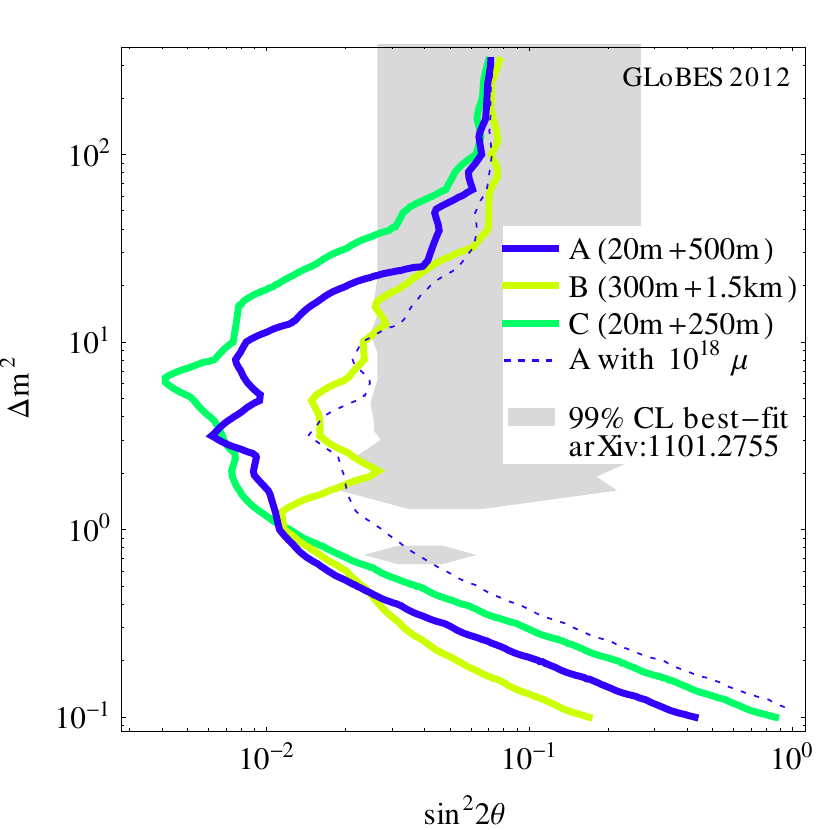}
  \label{figb:disapp}
}
\end{center}
\caption{\label{fig:disapp} Exclusion region in $\sin^2 2 \theta$-$\Delta m^2$ (right hand sides of curves) for $\nu_e$ disappearance for different geometry assumptions (left panel) and optimisation points (right panel); 90\% CL, 2 d.o.f. \subref{figa:disapp}: The curve ``no systematics'' represents a single detector at $d=500 \, \mathrm{m}$ using statistics only, whereas the other curves correspond to near-far detector setups, where the red thick curves include (conservative) full systematics, including a 10\% shape error, and geometry effects. \subref{figb:disapp}: Systematics are fully included, different two-distance optimisation points shown (distances to the end of the decay straight). Both panels: $E_\mu=2 \, \mathrm{GeV}$, $10^{19}$ useful muon decays per polarity, $d_1=20 \, \mathrm{m}$ ($200 \, \mathrm{t}$) and $d_2 = 500 \, \mathrm{m}$ ($1 \, \mathrm{kt}$), unless noted otherwise. Figure taken from reference \cite{Winter:2012sk}.}
\end{figure}

A disappearance experiment is more sensitive to the signal
normalisation than an appearance experiment. The neutrino flux is
extremely well understood for $\nu$STORM, but further understanding of
the measured spectrum resulting from the combination of efficiencies
and cross sections is required.  To control these effects a near
detector identical to the far detector is required. A study of
$\bar{\nu}_{\mu}$ disappearance in a SuperBIND detector is in
progress, but a study of $\nu_{e}$ disappearance using a generic
detector has been completed \cite{Winter:2012sk}. The results are
applicable to a $\bar{\nu}_{\mu}$ disappearance experiment. In absence
of a well developed near detector simulation, a conservative approach
was assumed. A 10\% systematic, uncorrelated between energy bins but
correlated between the near and far detector (shape error) becomes the
leading systematic in this case with further details and systematics
given in Ref.~\cite{Winter:2012sk}. Geometry effects are especially
important for the near detector as the beam divergence, both from the
muon decay kinematics and transverse beam components will lead to a
different beam spectrum for the near and far
detectors~\cite{Tang:2009na,Winter:2012sk}. The oscillations will also
average over the length of the decay
straight~\cite{Giunti:2009en,Winter:2012sk}. These effects are
illustrated in figure \ref{figa:disapp}. An optimisation of the baseline
distances are shown in figure \ref{figb:disapp}. The optimisation shows
that all options preform equally well for $\Delta m^2 \simeq 1$
eV$^2$, while larger values of $\Delta m^2 \simeq 1 \, \mathrm{eV}^2$
prefer shorter distances (from the end of the decay straight) for the
far detector.

%% file: 03-nuSTORM-facility/03-03-nuScattDetectors/03-03-nuScattDetectors.tex
\subsection{Detectors for neutrino scattering studies}
\label{SubSect:DetectNuScatt}

To explore fully the broad programme of $\nu N$ scattering studies
described in section \ref{SubSect:nuNScat} will require a number of
detectors optimised to address different aspects of the programme.
The two detectors described below are intended to indicate possible
options for further development.
The development of a detailed specification for the $\nu N$-scattering
detector suite is part of the programme of work we propose to carry
out (see section \ref{SubSect:PropProg}).
Physics topics offered by a high resolution detector such as the
options described below in $\nu$STORM are summarised in Appendix
\ref{App:PhysOfHiRes}.

\subsubsection{HIRESMNU: A High Resolution Near Detector \`a la LBNE}
\label{Sect:HiResMuNu}

Precision measurements of neutrino-interactions at the near-detector
(ND) are necessary to ensure the highest possible sensitivity to the
neutrino-oscillation studies in this proposal.
Regardless of the process under study---\nmne\ (\anmne) appearance  or
\nm\ (\anm) disappearance---the systematic error should be less than
the corresponding statistical error.  
The ND design must achieve the four principal goals: 
\begin{itemize}
  \item Measurement of the absolute and the relative abundance of the
    four species of neutrinos, \nm, \anm, \nue, and \ane, as a
    function of energy (\enu).
    Accurate determination of  the angle and the momentum of the
    electron in neutrino-electron neutral current scattering which
    will provide the absolute flux;
  \item Determination of the absolute \enu-scale, a factor which
    determines the value of the oscillation-parameter $\Delta m^2$;
  \item Measurement  of $\pi^0$s and of \pip\ and \pim\ produced in
    the NC and CC interactions.
    The pions are the predominant source of background for any
    oscillation study; and
  \item Measurement of $\nu$-nucleus cross-sections.
    The cross-section measurements of exclusive and inclusive CC and
    NC processes will furnish a rich panoply of physics relevant for
    most neutrino research.
    Knowing the cross sections at the \enu\ typical of the $\nu$STORM
    beam is essential for predicting both the signal and the
    background.
\end{itemize}

\begin{figure}
  \begin{center}
    \includegraphics[width=0.8\textwidth]%
      {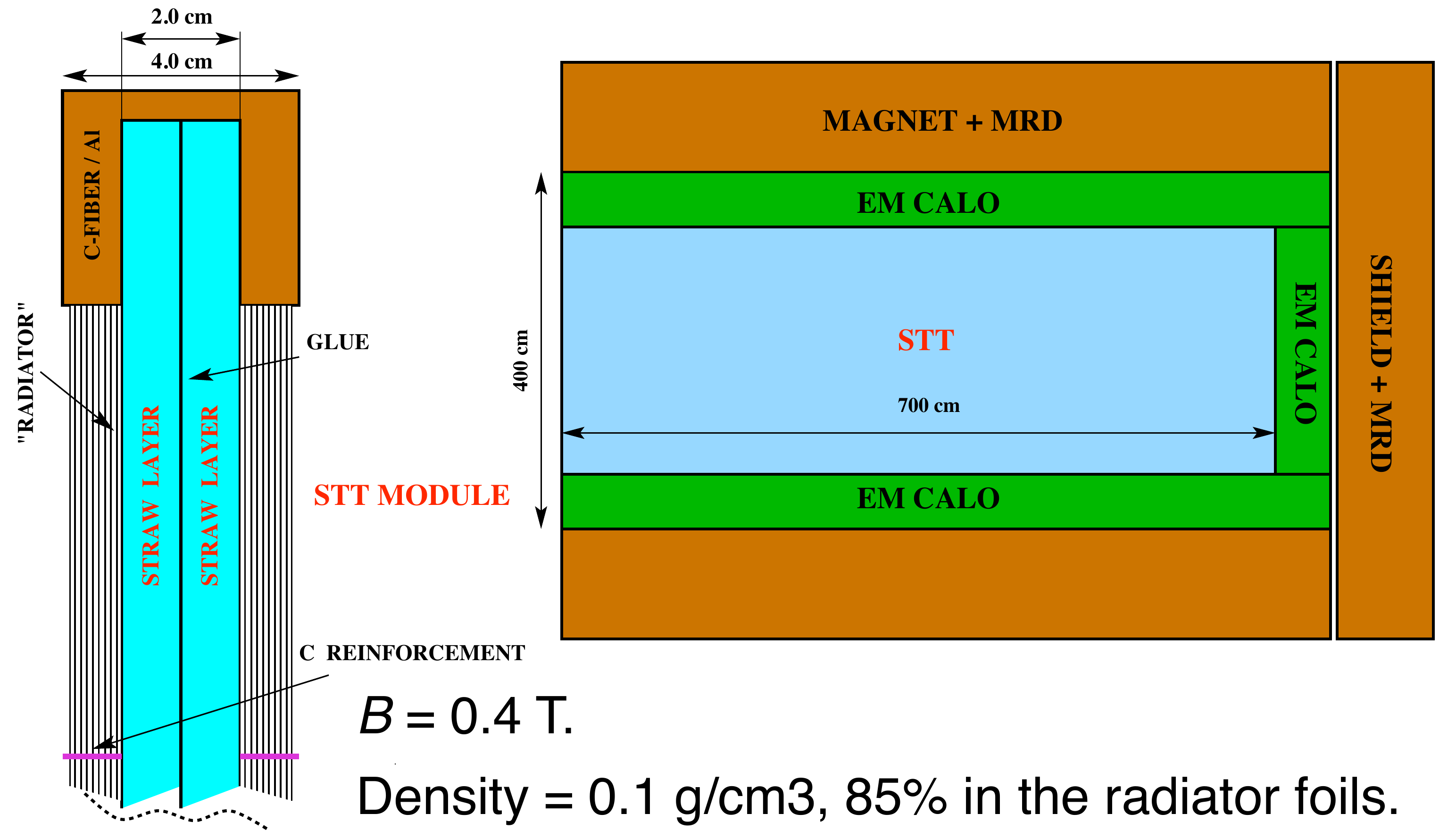}
  \end{center}
  \caption{
    Schematic of the ND showing the straw tube tracker (STT), the
    electromagnetic calorimeter (ECAL) and the magnet with the muon
    range detector (MRD). 
    The STT is based upon ATLAS \cite{ATLAS-TRT1,ATLAS-TRT2,ATLAS-TRT3}
    and COMPASS \cite{COMPASS1,COMPASS2} trackers.
    Also shown is one module of the proposed straw tube tracker
    (STT).
    Interleaved with the straw tube layers are plastic foil radiators,
    which provide 85\% of the mass of the STT.
    At the upstream end of the STT are layers of nuclear-target for
    the measurement of cross sections and the \piz's on these
    materials.
  }
  \label{fig-det-schematic}
\end{figure}

A high-resolution detector, the HIRESMNU, has been proposed as the
near detector for the LBNE project \cite{HIRESMNU1,HIRESMNU2}. 
Figure \ref{fig-det-schematic} shows a schematic of this the HIRESMNU
design.
The architecture of the detector \cite{HIRESMNU1,HIRESMNU2} builds
upon the experience of NOMAD \cite{NOMADnim}.
It embeds a $4 \times 4 \times 7$\,m$^3$ Straw-tube tracker (STT),
surrounded by a 4$\pi$ electromagnetic calorimeter (ECAL) in a  dipole
magnet with  $B \simeq 0.4$\,T.
Downstream of the magnet, and within the magnet yoke, are detectors
for muon identification.
The STT will have a low average density similar to liquid hydrogen,
about 0.1\,gm/cm$^3$, which is essential for momentum determination 
and the identification of electrons, protons, and pions.
The foil layers, up- and down-stream of the straw tubes, provide 
the transition-radiation and constitute most of the 7\,ton fiducial
mass.
The foil layers serve both as the mass on which the neutrinos will
interact and as generators of transition radiation (TR), which
provides electron identification. 

Along the beam, the total depth of the detector, in radiation lengths,
is sufficient for 50\% of the photons, largely  from \piz\ decay, 
to be observed as $e^+e^-$ pairs, which delivers superior resolution
compared with conversions in the ECAL.
Layers of nuclear-targets will be deployed at the upstream end of the
STT for the determination of cross sections on these materials. 

The HIRESMNU allows the cross-sections of exclusive and inclusive
processes to be measured, detailed studies of the multiplicity of
secondary particles to be carried out and the detailed
characterisation of the neutrino source.
It can identify all four neutrino species in $\nu$STORM.
Systematic studies of $\nu$-electron scattering, quasi-elastic
interactions, \nue/\ane-CC, neutral-current identification,
\piz\ detection, etc. have been carried out in the context of LBNE.   
The quoted dimensions, mass, and segmentation of HIRESMNU will be 
further optimised  for $\nu$STORM as the proposal evolves. 

\subsubsection{A pressurised gas TPC option for cross section measurements}
\label{Sect:Gas-argon-tpc}

\begin{figure}
  \begin{center}
    \includegraphics[width=0.7\textwidth]%
      {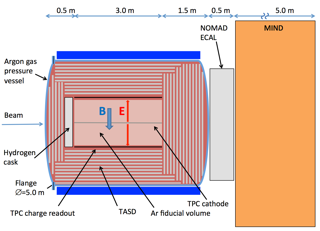}
  \end{center}
  \caption{
    Schematic of the pressurised argon gas-based TPC detector. 
    Both the TPC and scintillator calorimeter layers surrounding it
    are enclosed in a pressure vessel. 
    A 0.5\,T magnetic field is applied to the pressure vessel volume. 
    Downstream of the TPC are also an electromagnetic calorimeter
    (ECAL) and a magnetised iron neutrino detector (MIND). 
    The latter acts as a muon spectrometer for neutrino interactions
    occurring in the TPC and as an independent near detector for the
    sterile neutrino programme.
  }
  \label{fig-det-argon-tpc}
\end{figure}

A versatile detector has been proposed as a near detector for LBNO,
the Gas Argon Modular Apparatus for Neutrino Detection
($\gamma$$\nu$det) \cite{LBNO:EoI2CERN}. 
Based on a pressurised argon time projection chamber (TPC) located in
a large 5 m diameter pressure vessel, figure \ref{fig-det-argon-tpc},
this proposal is well suited for the precision measurement of cross
sections and detailed study of electron- and muon-neutrino
interactions at a $\nu$STORM near detector facility. 
A magnetic field is applied to the full volume of the pressure
vessel. 
A magnet design similar to that of the UA1/NOMAD spectrometer
dipole magnet provides a field with characteristics close to those
required for this detector. 
The required peak field magnitude is under study.
The pressure vessel can accommodate several layers of
scintillator-based calorimeter, such as the SuperBIND plastic
scintillator material, between the TPC and the pressure vessel inner
surface. 
One advantage of this design compared with a more compact pressure
vessel enclosing only the TPC is that there is less redundant/passive
material between the TPC and first layers of the scintillator detector
and fewer blind spots. 
The outer layers of the embedded scintillating material can be
interleaved with radiators to improve the containment within the
pressure vessel of more energetic secondaries from events of interest
occurring in the TPC. 
High energy muons of both signs $>$ 1 GeV from neutrino events
originating in the TPC would be measured downstream in a muon
spectrometer similar in design to the SuperBIND but 1/10th the size. 

\begin{figure}
  \begin{center}
    \includegraphics[width=0.7\textwidth]%
      {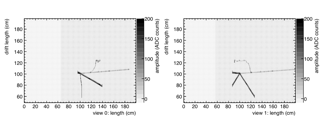}
    \hfill
    \includegraphics[width=0.7\textwidth]%
      {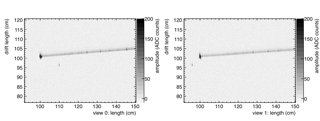}  
  \end{center}
  \caption{
    GENIE Monte Carlo simulations of a charged current quasi-elastic
    (CCQE) neutrino interaction in 20 bar pressurised argon gas (top)
    and liquid argon (bottom). 
    Three proton tracks are clearly resolved in the pressurised argon
    and completely undetectable in the liquid argon. 
    Of the two options, argon gas is better suited to study nuclear
    effects in neutrino interactions.
  }
  \label{fig-det-ccqe-argon-tpc}
\end{figure}

A gas TPC provides excellent vertexing capabilities, especially
relevant for the understanding of nuclear effects in neutrino
interactions. 
Figure \ref{fig-det-ccqe-argon-tpc} compares a CCQE neutrino
interaction in liquid argon ($\rho = 1.4$\,g.cm$^{-3}$) with
the same interaction in argon gas at 20\,bar 
($\rho = 0.034$\,g.cm$^{-3}$)
\cite{Curioni:LAGUNA-LBNO,CurioniLussi:Private}.
No magnetic field is applied in this GENIE Monte Carlo simulation
using the LBNO neutrino flux. 
With a 3\,mm pitch, the three proton tracks can clearly be resolved in
the argon gas but are completely undetectable in liquid argon. 
The liquid argon TPC (LAr TPC) would benefit from an even finer
granularity and lower kinetic-energy threshold for proton detection
(set to 40\,MeV). 
Here, the pressurised argon TPC provides a compelling case for
neutrino-nucleon interaction studies. 
By using argon as the target nucleus in the near detector, valuable
data are compiled for liquid argon-based neutrino detectors such as
those planned for long baseline projects. 
The TPC gas can be readily changed to other gases, for instance CH4 or
CO2, allowing studies of interactions on different materials similar
to for example the liquid scintillator of fully-active-scintillator
proposals. 

The $\nu_\mu$ flux at the LBNO near detector location peaks around
3\,GeV extending beyond 20\,GeV with a mean around 5\,GeV.  
The $\nu$STORM near detector flux peaks at 2.5\,GeV and has a cut off
at 4\,GeV which is set by the 3.8\,GeV/c $\mu$ beam and its 10\%
momentum spread. 
Event containment is therefore less of a challenge at $\nu$STORM
i.e. the present proposal for LBNO could be applied to the $\nu$STORM
facility with little re-optimisation. 
Simulation work is underway to optimise the detector configuration and
evaluate its physics performance. 

Engineering details will be addressed by the LBNO near detector task
which is due to report a conceptual design by end 2014. 
It will cover the dimensioning of the magnet and pressure vessel, the
feasibility and cost of a large flange on the pressure vessel, the
integration of TPC and scintillator calorimeter within the pressure
vessel, in particular feedthroughs for the supplies and readout of
those detector elements. 
As is mentioned in Section \ref{SubSubSect:Res} there is a strong
motivation for accurate measurements of the neutrino-nucleon
interaction cross sections on an $H_2$ or $D_2$ target. 
The argon-filled pressure vessel would be functionally equivalent to a
safety barrier, providing a second safety layer beyond the $H_2$
container vessel and enabling the safe operation of such an $H_2$ cask
in this near detector proposal. 
The $H_2$ cask design could be based on the cold neutron moderators in
routine operation at several neutron scattering facilities worldwide. 

%% file: 04-Implementing-nuSTORM/04-Implementing-nuSTORM.tex
\section{Implementing the \boldmath{$\nu$}STORM facility}
\label{Sect:Implement_nuSTORM}

\input 04-Implementing-nuSTORM/04-01-nuSTORMatCERN/04-01-nuSTORMatCERN
\input 04-Implementing-nuSTORM/04-02-nuSTORMatFNAL/04-02-nuSTORMatFNAL

%% file: 04-Implementing-nuSTORM/04-01-nuSTORMatCERN/04-01-nuSTORMatCERN.tex
\subsection{Implementing \boldmath{$\nu$}STORM at CERN}
\label{SubSect:nuSTORMatCERN}

\begin{figure}
  \begin{center}
    \includegraphics[width=0.85\textwidth]
      {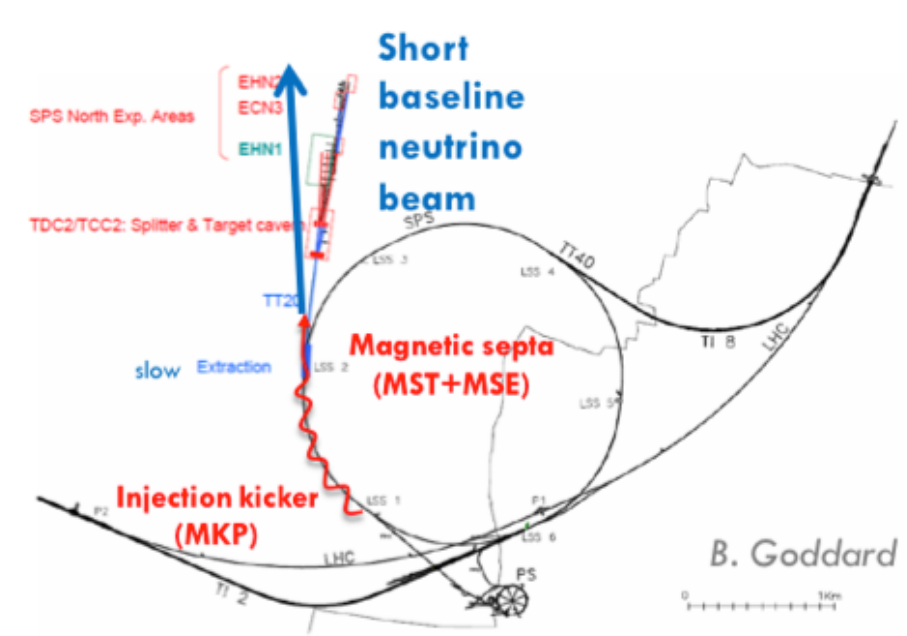}
  \end{center}
  \caption{
    Extraction of the SPS beam in the North Area.
  }
  \label{Fig:SPSExtract}
\end{figure}

The fast extraction of protons from the SPS is initiated by a kicker
in LSS1.
A septum in the TT20 beam line then extracts the beam from the SPS so
that it can be transported to the $\nu$STORM target.
This fast-extraction scheme (see figure \ref{Fig:SPSExtract}) has been
demonstrated for low intensities.
Several neutrino experiments are proposed for the North Area.
In developing the concept for implementing $\nu$STORM at CERN it will
be important to consider exploiting the present and planned
infrastructure in the North Area to the fullest extent.
 
\begin{figure}
  \begin{center}
    \includegraphics[width=0.75\textwidth]
      {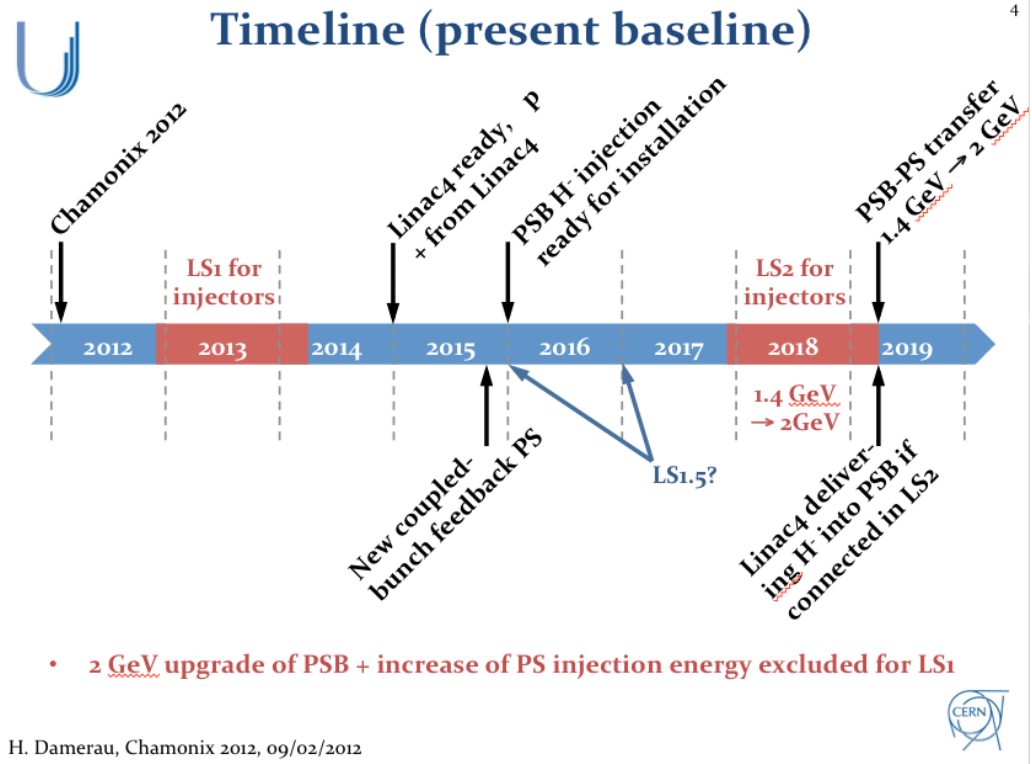}
  \end{center}
  \caption{
    Timeline for the CERN injector upgrades (presented by H.~Damerau at 
    the 2012 Chamonix workshop.
  }
  \label{Fig:Linac4TimeLine}
\end{figure}

To deliver the proton-beam phase space required by $\nu$STORM requires
that the LS2 upgrades to the injector systems, including the new
Linac4, are complete.
Figure \ref{Fig:Linac4TimeLine} shows the timeline for these
upgrades.
If the short-baseline programme proposed in \cite{CERN:EDMS1233951} is
executed on the timetable outlined by the proponents, $\nu$STORM will
be a highly-effective development of the short-baseline programme.

The 100\,GeV beam will deliver 0.56\,MJ 
($100\,{\rm GeV} \times 3.5\,10^{13} \times 1.6\,10^{-19}$).
Repeating this every 3.6\,s, the ultimate repetition rate, gives
156\,kW on target.  
The 6\,s interval between pulses would reduce the beam power by
roughly a factor of two compared to the FNAL option.
Using fast extraction, the proton-pulse duration will be 10.5\,$\mu$s
which is 10 times longer than for the beam pulse from the Main
Injector at FNAL.
Two such pulses of 2100 bunches, spaced by 50\,ms, are extracted
every SPS cycle.
The beam characteristics before and after the LS2 upgrades are shown
in table \ref{Tab:SPSbeamChar} \cite{CERN:EDMS1233951}.
\begin{table}
  \caption{
    Summary of the SPS beam characteristics at present and after the
    LS2 upgrade.
  }
  \label{Tab:SPSbeamChar}
  \begin{center}
    \begin{tabular}{|l|l|l|l|l|l|l|}
      \hline
      {\bf Parameter} & \multicolumn{2}{c|}{\bf SPS operation}  & 
                        \multicolumn{2}{c|}{\bf SPS record}     &
                        \multicolumn{2}{c|}{\bf After LIU 2020}  \\
          & LHC & CNGS & LHC & CNGS &  LHC & $\nu$STORM     \\
      \hline
Energy [GeV] & 450 & 400 & 450 & 400 & 450 & 100 \\
Bunch spacing [ns] & 50 & 5 & 25 & 5 & 25 & 5 \\
Bunch intensity [$10^{11}$]  & 1.6 & 0.105 & 1.3 & 0.13 & 2.2 & 0.17 \\
Number of bunches & 144 & 4200 & 288 & 4200 & 288 & 4200 \\
SPS intensity [$10^{13}$] & 2.3 & 4.4 & 3.75 & 5.3 & 6.35 & 7.0 \\
PS intensity [$10^{13}$] & 0.6 & 2.3 & 1.0 & 3.0 & 1.75 & 4.0 \\
SPS Cycle length [s] & 21.6 & 6.0 & 21.6 & 6.0 & 21.6 & 3.6\\
PS Cycle length [s] & 3.6 & 1.2 & 3.6 & 1.2 & 3.6 & $2 \times 1.2$ \\
PS beam mom. [GeV/c] & 26 & 14 & 26 & 14 & 26 & 14 \\
Beam Power [kW] & 77 & 470 & 125 & 565 & 211 & 156 \\
      \hline
    \end{tabular}
  \end{center}
\end{table}

The estimations made in \cite{CERN:EDMS1233951} indicate that 
$4.5 \times 10^{19}$\,POT/year may reasonably be expected.
If $\nu$STORM ran for five years with 100\,GeV protons, 
$5 \times 4.5\times10^{19} = 2.3 \times 10^{20}$\,POT would be
delivered.
With the assumption that the $\pi$/POT is proportional to the energy,
a further optimisation to gain a factor of two in POT would have
to be made.
 
\begin{figure}
  \begin{center}
    \includegraphics[width=0.85\textwidth]
      {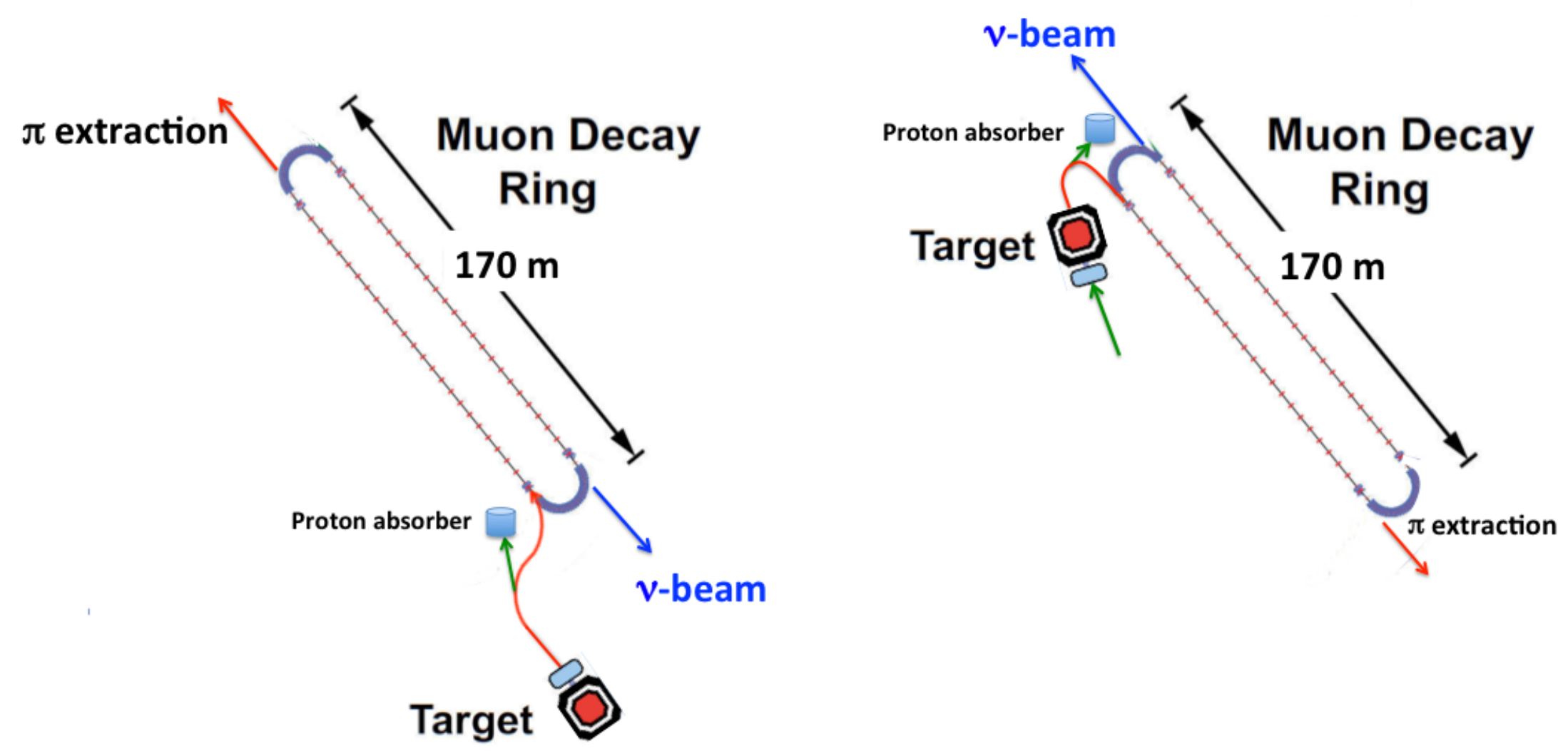}
  \end{center}
  \caption{
    The proton and pion transport have to be optimised taking into 
    account the placement of the $\nu$STORM ring and the direction 
    in which the beam is required.
    The space for the the proton absorber is constrained. 
    The figure sketches two possible configurations: to the left
    where the protons and the neutrinos have similar directions 
    and to the right when they have opposite directions.
  }
  \label{Fig:CERN-Config}
\end{figure}

The design of the target developed at FNAL has to be adapted.
The differing geometry of the CERN and FNAL options is shown in figure
\ref{Fig:CERN-Config}.
The 10.5\,$\mu$s pulse of protons from the SPS means that muons will
make a number of turns in the storage ring during pion injection.
To ensure the neutrino beam arises solely from the decay of muons, 
the injection of pions and the neutrino-beam extraction are at
different ends of the same arc.
The pion-injection channel and the proton absorber have to be designed
taking into account the proton and neutrino beam directions. 
The pion injection system is constrained by the limited space
available for the proton absorber and by the requirement that the
total length of the transport channel be minimised to limit pion decay
outside the storage ring. 

\begin{figure}
  \begin{center}
    \includegraphics[width=0.85\textwidth]
      {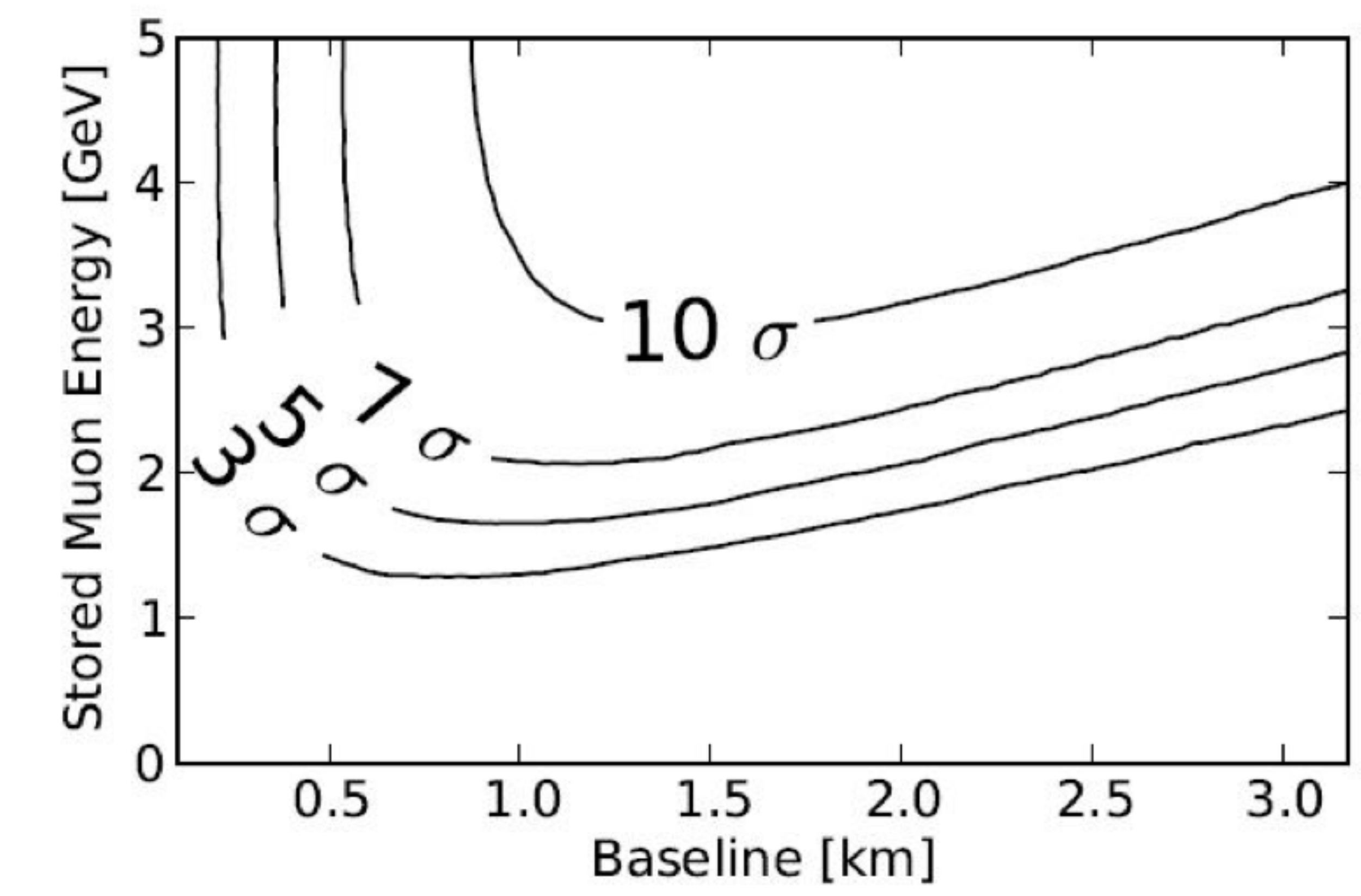}
  \end{center}
  \caption{
      Physics potential for different baselines and stored muon
      energies ($\chi^2$ contours were derived using the total event
      rate, without systematic errors, a signal efficiency of 0.5 and 
      background rejection of charge misidentification and NCs at 
      $10^{-3}$ and $10^{-2}$ respectively).
  }
  \label{Fig:Detector-Pos-Opt}
\end{figure}

It is important to investigate whether existing or planned beam lines,
target stations and detector caverns, or parts of them, can be
re-used.
The source-detector distance has to match the neutrino energy.
The physics potential of the facility for varying detector positions
for 3.8\,GeV/c stored muons is shown in figure
\ref{Fig:Detector-Pos-Opt}.
At FNAL, the position of the far detector is around 1.6\,km.
$\nu$STORM also needs space for a near detector at 20\,m to 50\,m from
the neutrino extraction point. 
While it is in general difficult to re-use target facilities due to
the high irradiation levels, if it were possible to design a target
station that could support both the CENF programme and $\nu$STORM, it
might be possible to reduce the implementation cost.

A study to lay out, in a cost effective and feasible manner, the
$\nu$STORM facility at CERN has to take into account the civil
engineering constraints, re-use of existing beam lines and detector
caverns etc. 
The facility could be placed in such a way as to exploit the CENF
\cite{CERN:EDMS1233951} or constructed underground at the SPS
level using existing caverns, BA 1, 4 or 5. 
Consideration should also be given to possible use of both straight
sections for physics.
The pion dump could be used to produce a muon beam suitable for the
implementation of a 6D ionisation cooling programme (see section 
\ref{SubSect:MuBm}).

\begin{figure}
  \begin{center}
    \includegraphics[width=0.90\textwidth]
      {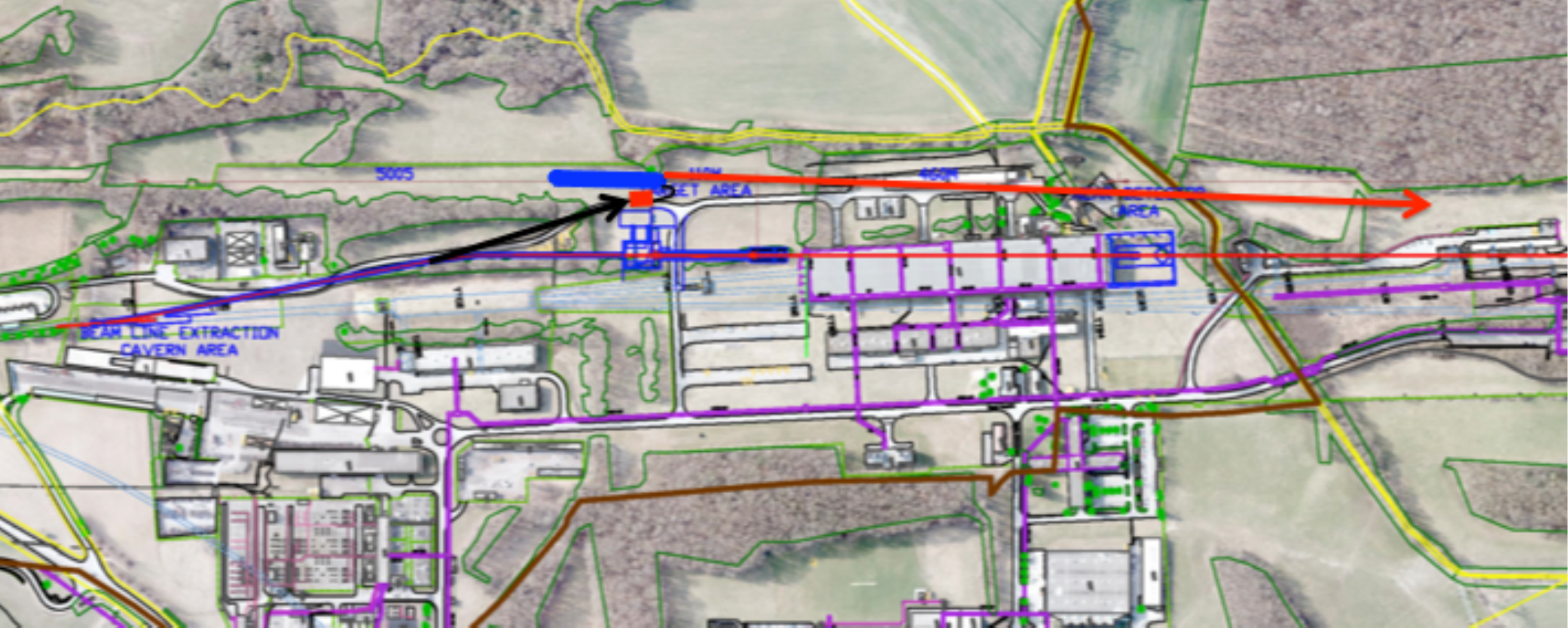}
  \end{center}
  \caption{
    An option using the North Area target station for CENF (preferably 
    prepared in advance for $\nu$STORM) and the far detector hall.
  }
  \label{Fig:NA-Case}
\end{figure}

\begin{figure}
  \begin{center}
    \includegraphics[width=0.85\textwidth]
      {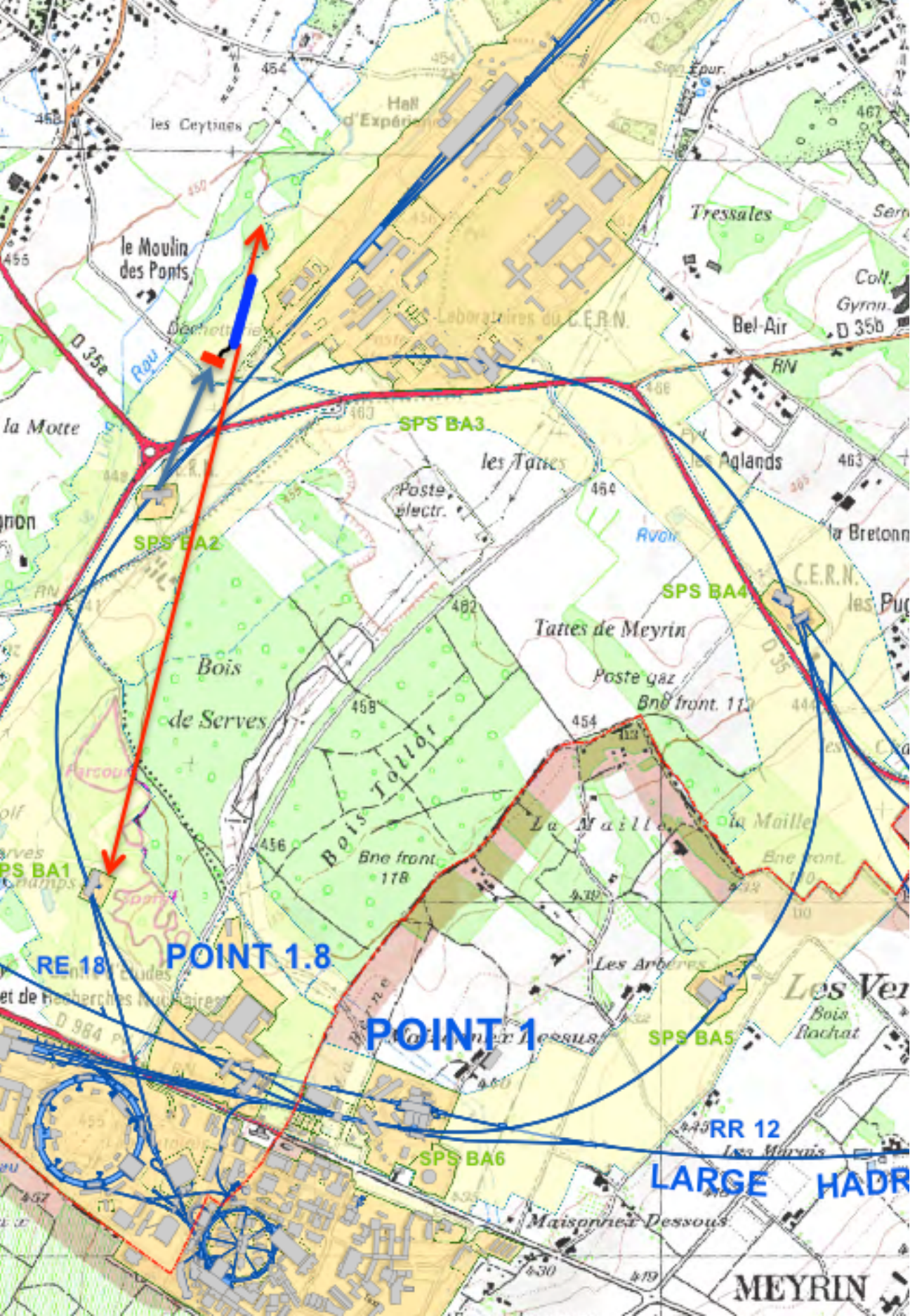}
  \end{center}
  \caption{
    Extraction of the proton beam at the SPS level with a detector
    hall in one of the SPS caverns. 
    The pion extraction is also shown.
  }
  \label{Fig:Underground}
\end{figure}

Figure \ref{Fig:NA-Case} shows $\nu$STORM in the North Area. 
The design of the pion-injection section for this case may be
difficult (space limitations and the requirement for high-field
magnets).
$\nu$STORM could also be situated 60\,m underground, at the SPS level,
directed to one of the existing SPS caverns in which the detector hall
would have to be built. 
A muon-cooling experiment could also be placed close to the decay ring
after the pion extraction channel (see figure\ref{Fig:Underground}).

\begin{figure}
  \begin{center}
    \includegraphics[width=0.85\textwidth]
      {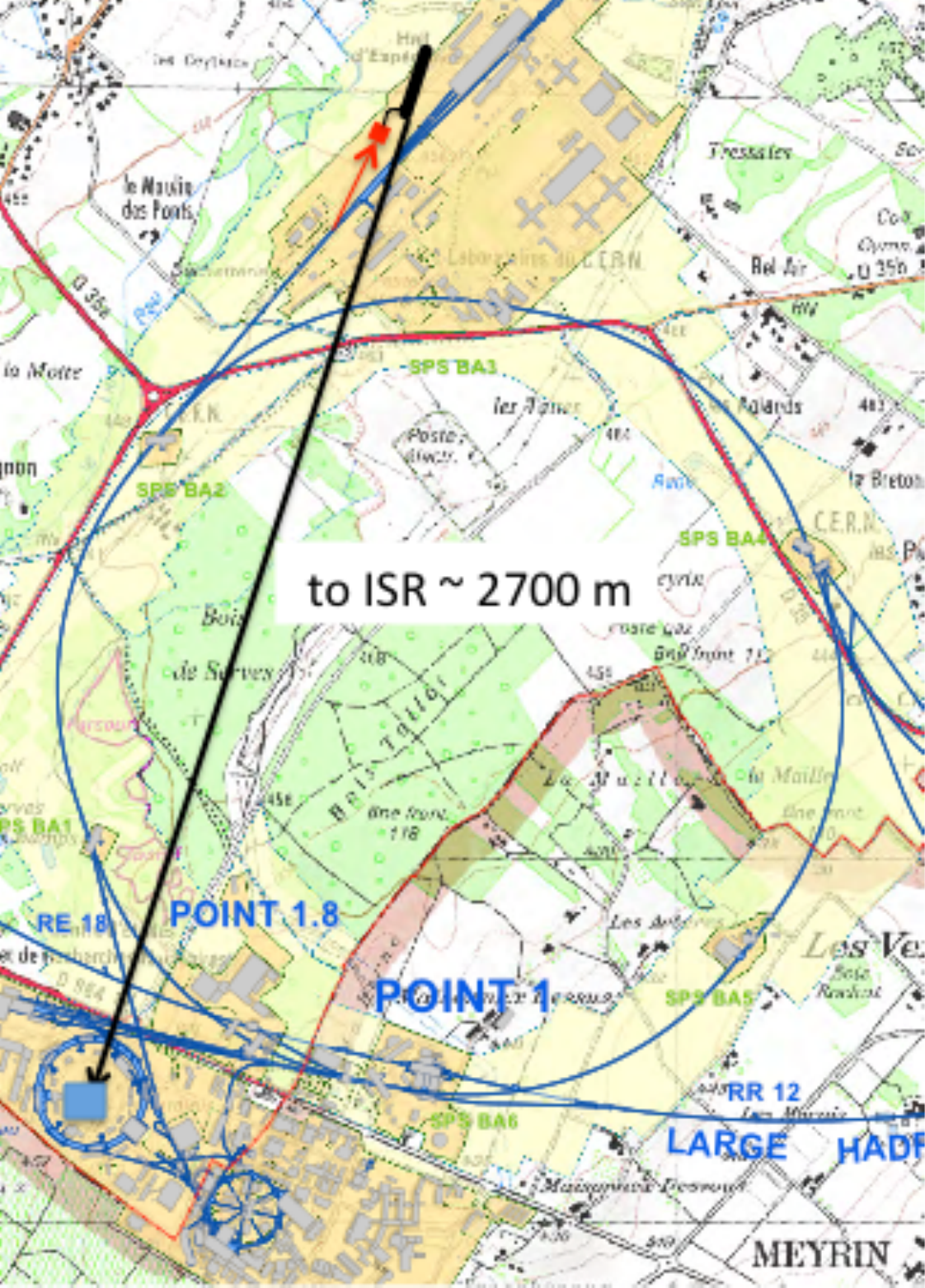}
  \end{center}
  \caption{
    This layout, with a far detector at the Meyrin site, aims to use
    part of the proton beam line, a new target station and new
    detector caverns.
  }
  \label{Fig:Baseline-ISR}
\end{figure}

Figure \ref{Fig:Baseline-ISR} shows an option in which the neutrino
beam is sent to the Meyrin site, the far detector being placed inside
the ISR ring.  
In this case, the design of the pion channel is similar to the FNAL
option.
However, the baseline is rather long (see figure
\ref{Fig:Detector-Pos-Opt}) and, for such a long baseline, the energy
of the pions and the stored muons would have to be selected
higher than the present baseline.

All of the options outlined above need to be considered in more
detail, in particular the proton beam lines and the pion transfer
channels, including the proton absorber, must be shown to be
feasible.
The target station should be similar to those already developed for
100\,GeV protons, however the capture system for a specific target
would need to be optimised.

%% file: 04-Implementing-nuSTORM/04-02-nuSTORMatFNAL/04-02-nuSTORMatFNAL.tex
\subsection{Implementing \boldmath{$\nu$}STORM at FNAL}
\label{SubSect:nuSTORMatFNAL}

\begin{figure}
  \centering{
    \includegraphics[width=0.9\textwidth]%
      {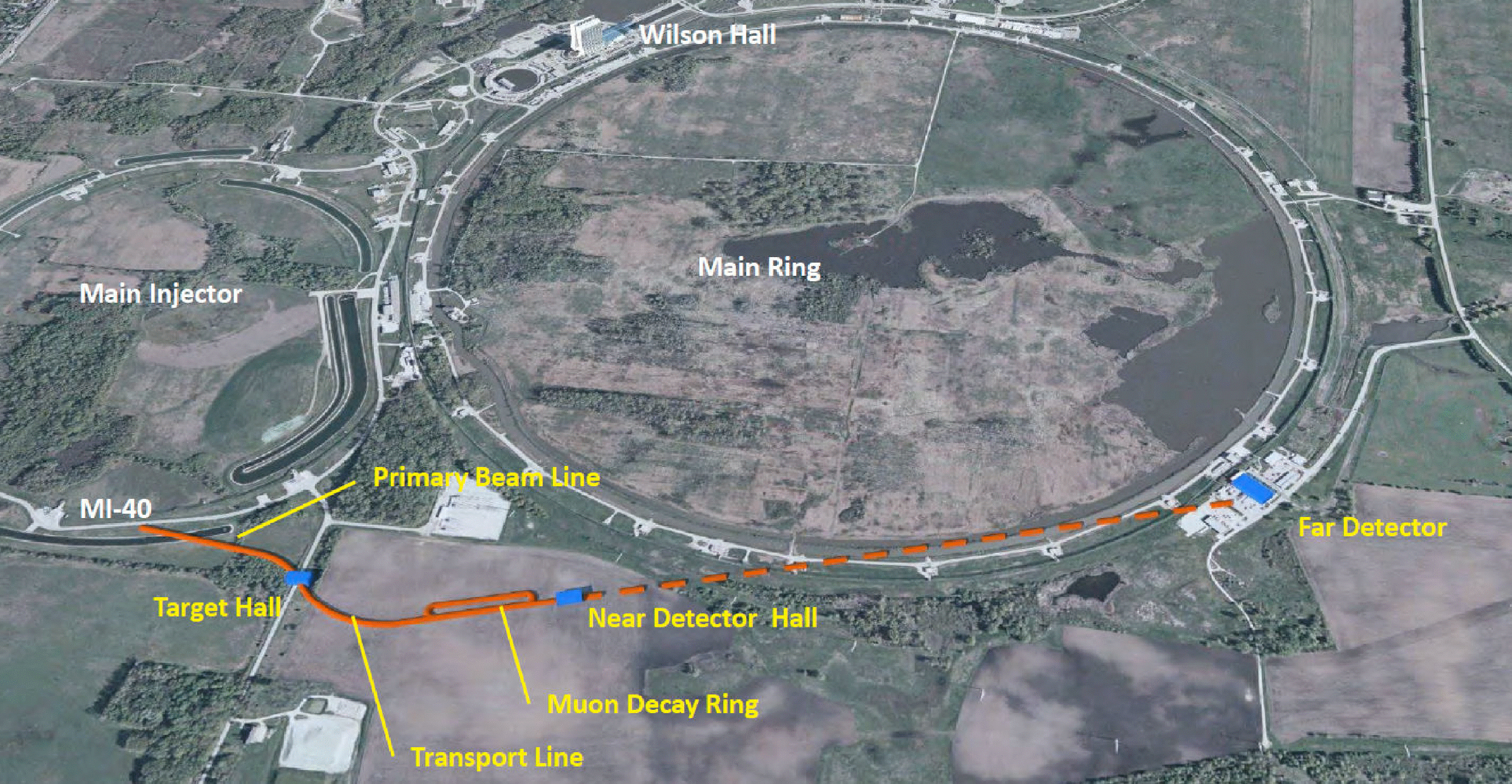}}
  \caption{Schematic of the $\nu$STORM facility on the FNAL site.}
  \label{fig:nuSTORM}
\end{figure}

The concept for siting $\nu$STORM at Fermilab follows ideas that were
developed nearly two decades ago for a short baseline 
$\nu_\mu \rightarrow \nu_\tau$ oscillation experiment
\cite{Kodama:1993ds,PDR:NumiSBL} that was to use protons extracted
from the Fermilab Main Injector using the proton abort line of
that machine.
Although this experiment was never carried out, the Main Injector
abort-beam absorber was assembled with the by-pass beam pipe that
would have been needed for this experiment.  $\nu$STORM will use
this by-pass. 
The basic siting concept for $\nu$STORM at Fermilab is shown in
figure \ref{fig:nuSTORM}.

\begin{figure}
  \centering{
    \includegraphics[width=0.85\textwidth]%
      {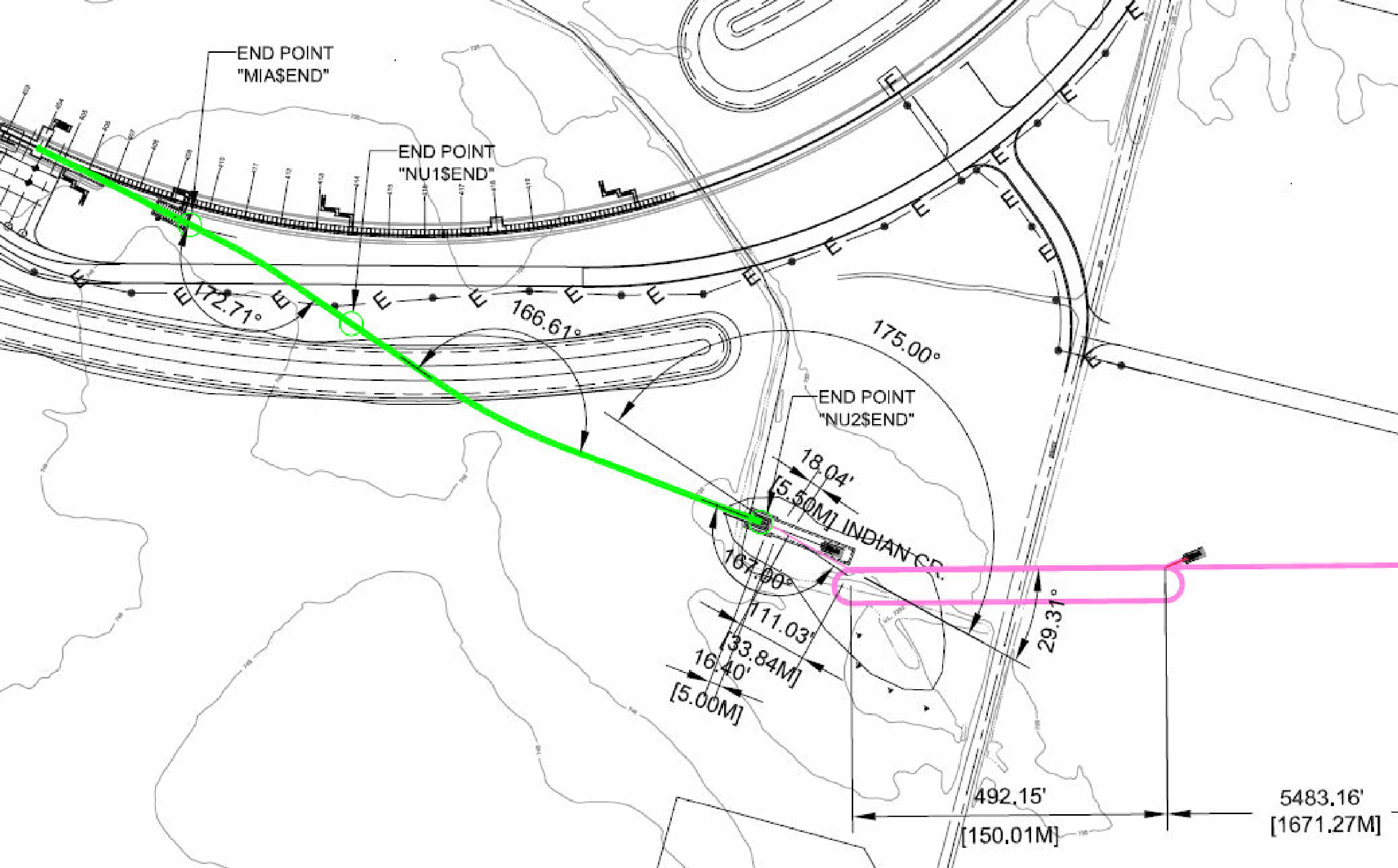}}
  \caption{
    Site detail of extraction of the beam from the FNAL Main
    injector, the target hall and the decay ring.
  }
  \label{fig:siteD}
\end{figure}

Protons from the Fermilab Main Injector will be brought to a new
target station located near the southern edge of the Fermilab site.  
The beam line will be designed for 120\,GeV protons, but the beam line
will be able to accommodate protons from 60\,GeV to 120\,GeV.  
Although the pion yield per proton on target increases linearly in the
60\,GeV to 120\,GeV range, the run conditions for $\nu$STORM
will have to take into consideration the other experiments running at
the time.  
A detail of the currently favoured siting option for beam line, target
hall, transport line and decay ring is shown in figure
\ref{fig:siteD}. 

\begin{figure}
  \centering{
    \includegraphics[width=0.85\textwidth]%
      {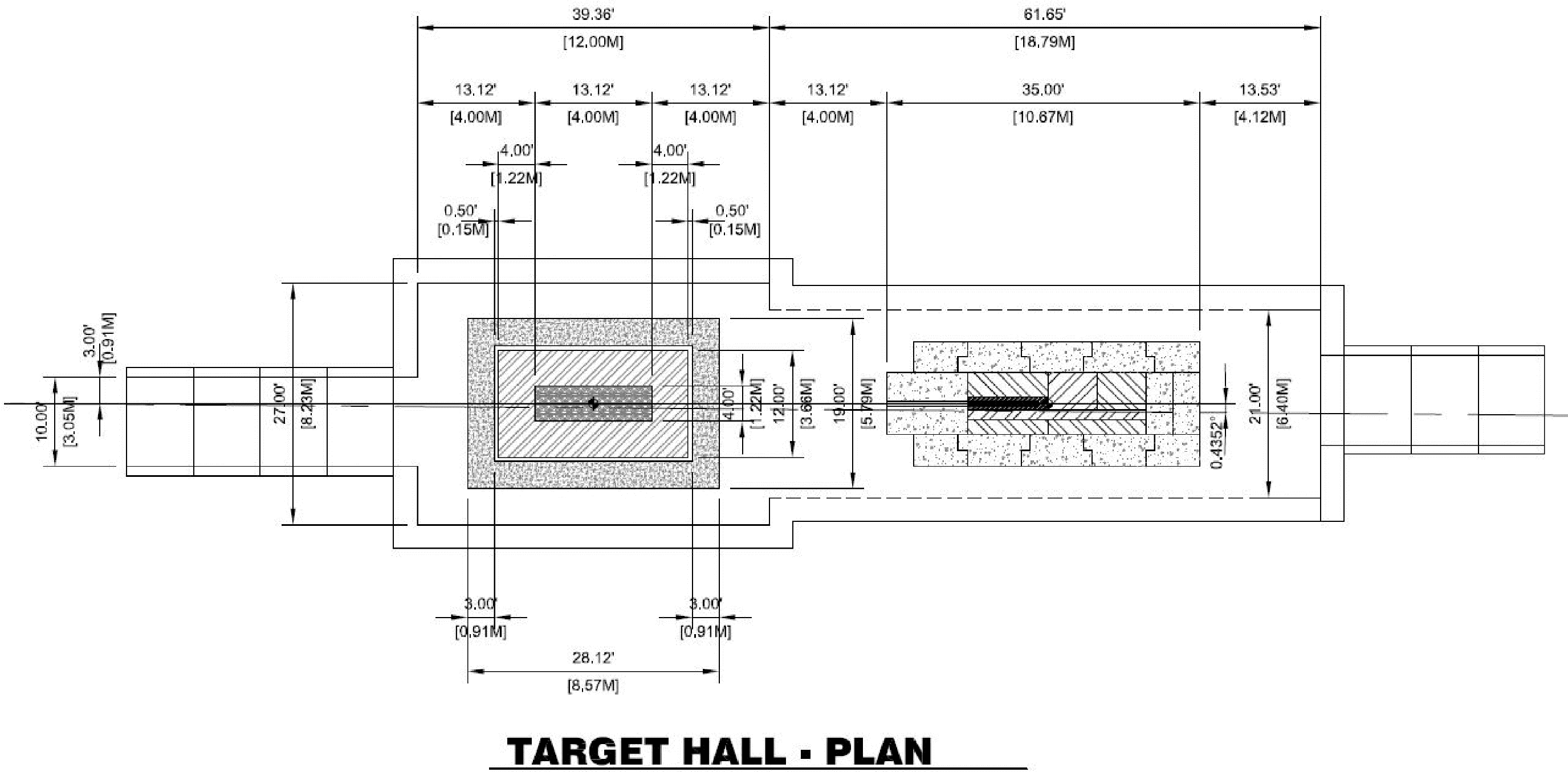}}
  \caption{Schematic of the target hall.}
  \label{fig:target}
\end{figure}

For $\nu$STORM at Fermilab, the baseline is 100\,kW on target which
represents approximately 1/7 of the 700\,kW proton power available
after completion of the Fermilab Proton Improvement Plan
\cite{PIP:2012dh}.
Current simulations for $\nu$STORM at Fermilab have assumed a tantalum
target and a NuMI-like horn operating at 300\,kA.  
A schematic of the current target station concept is given in figure
\ref{fig:target}.  
The pion capture and transport line starts 30 cm downstream of the
horn and transports pions to the decay ring.  
It is tuned to collect pions in the momentum acceptance of 5 $\pm$
0.5\,GeV/c. 
Pions are injected into the ring on an orbit separated from the
circulating muons, a process known as ``stochastic injection''. 

The current design for the injection section is shown in figure
\ref{Fig:Accel:InjKick}. 
The decay ring is approximately 350\,m in circumference and uses
compact arcs.
The ratio of the length of a single straight to the ring circumference
is 0.43.

There will be a near detector hall located approximately 50\,m from the
end of the straight (as shown in figure \ref{fig:nuSTORM}) and
$\nu$STORM will use the existing D0 assembly building (DAB) as the far 
(1.5\,km) detector hall.  
The pit area of DAB can accommodate a SuperBMIND of 1\,kT to 1.5\,kT
plus a LAr detector with a mass in the range of 500\,T to 1000\,T.   

\begin{figure}
  \centering{
    \includegraphics[width=0.65\textwidth]%
    {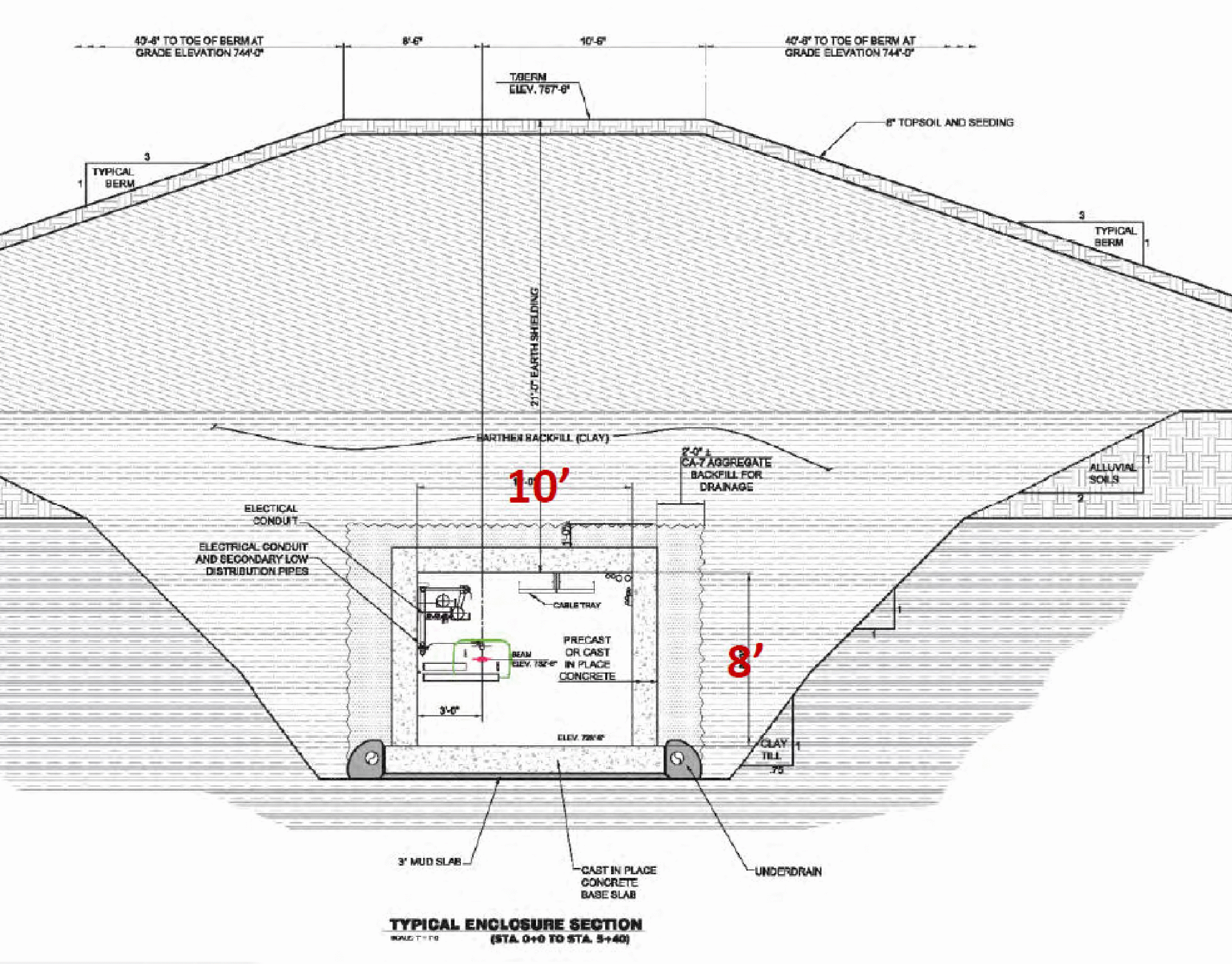}}
  \caption{Schematic of tunnelling.}
  \label{fig:tunnel}
\end{figure}

It is expected that all civil construction at the Fermilab site will
be at the Main Injector depth of 21\,ft below grade, although some
additional over burden may be required for the target hall.  
An engineering concept for the underground tunnelling is shown in figure
\ref{fig:tunnel}.  
The site location described above is ideal for $\nu$STORM.  
The services (water and power) are nearby, but the area is essentially
open and undeveloped so that $\nu$STORM construction will not
interfere (or have to accommodate) existing infrastructure. 
In addition, being able to use the D0 Assembly Building as the far
detector hall represents a significant cost saving. 

%% file: 05-Proposal/05-Proposal.tex
\section{Proposed programme}
\label{Sect:Proposal}

\subsection{Timeline}
\label{SubSect:TimeLine}

Formal consideration of $\nu$STORM began when the collaboration
submitted a Letter of Intent (LOI) to the Fermilab Physics Advisory
Committee (PAC) in June 2012 \cite{Kyberd:2012iz}.
The collaboration has been encouraged to submit a proposal in May
2013 \cite{Oddone:2012nuSTORM}.
Proton beams capable of serving the $\nu$STORM facility can be
provided at CERN and at FNAL.
With the encouragement of the CERN management, we have made an initial
investigation of the feasibility of implementing $\nu$STORM at CERN
(see section \ref{Sect:Implement_nuSTORM}).
In view of the fact that no siting decision has yet been taken, the
purpose of this Expression of Interest (EoI) is to request the
resources required to:
\begin{itemize}
  \item Investigate in detail how $\nu$STORM could be implemented at
    CERN; and 
  \item Develop options for decisive European contributions to the
    $\nu$STORM facility and experimental programme wherever the 
    facility is sited.
\end{itemize}

The timeline presented in figure \ref{Tab:TimeLine} identifies the
principal steps along the way to the preparation of the full Technical
Design Report (TDR) that is required before the project can be
considered for approval.
Should the collaboration's proposal to FNAL be accepted, project
approval would be by the DOE ``Critical Decision'' process.
In Europe, the usual CERN approval steps, followed by proposals to
national funding agencies, would be required.
In either case, the culmination of the next two years of effort will
be the TDR (in the US referred to as the Conceptual Design Report) for
the facility. 
With the exception of the site-specific elements noted below, the work
required to complete the design of the major systems is the same no
matter whether $\nu$STORM is implemented at CERN or at FNAL.
\begin{table}
  \caption{
    Indicative timeline for the preparation of the $\nu$STORM
    Technical Design Report. 
  }
  \label{Tab:TimeLine}
  \begin{center}
    \includegraphics[width=0.75\textwidth]
      {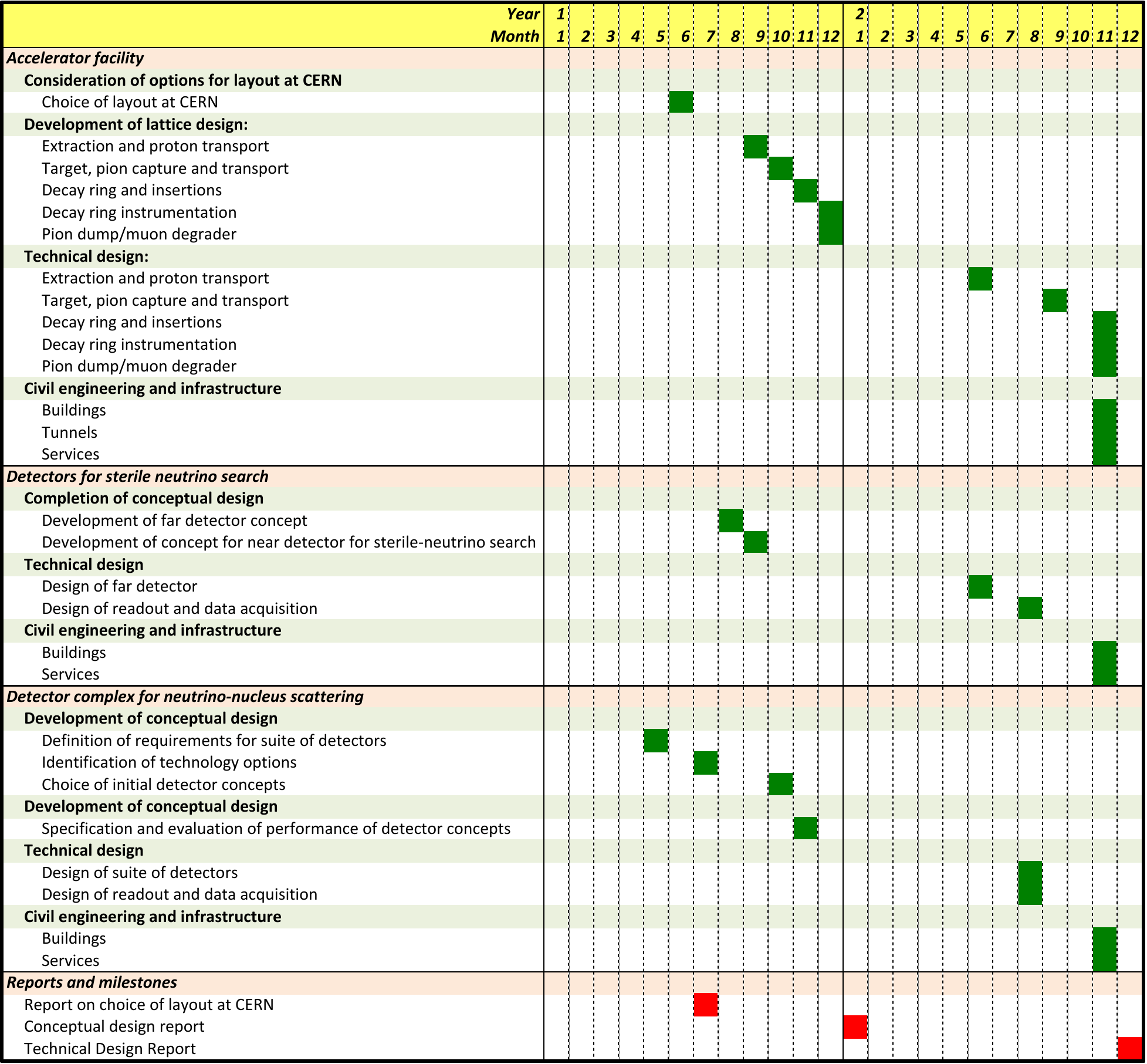}
  \end{center}
\end{table}

\subsection{Elements of the Project Breakdown Structure}
\label{SubSect:PBSElements}

While the civil construction, the provision of the necessary services
and the system integration will necessarily be the responsibility of
the host laboratory, the components and systems that make up the
accelerator complex, such as magnets or beam instrumentation, and the
neutrino detectors could be provided as in-kind contributions by the
international collaboration.
The list of tasks presented in table \ref{Tab:PBSEle} forms a
rudimentary Project Breakdown Structure (PBS) for the completion of
the TDR.
Those tasks which must be carried out by the host laboratory, supported
by the collaboration, are identified.
The design of large sections of the accelerator facility, beam-line
instrumentation and neutrino-detector systems are site-independent. 
\begin{table}
  \caption{
    Elements of the project breakdown structure that must be developed
    to determine the work required to deliver the Technical Design
    Report.
  }
  \label{Tab:PBSEle}
  \begin{center}
    \includegraphics[width=0.45\textwidth]
      {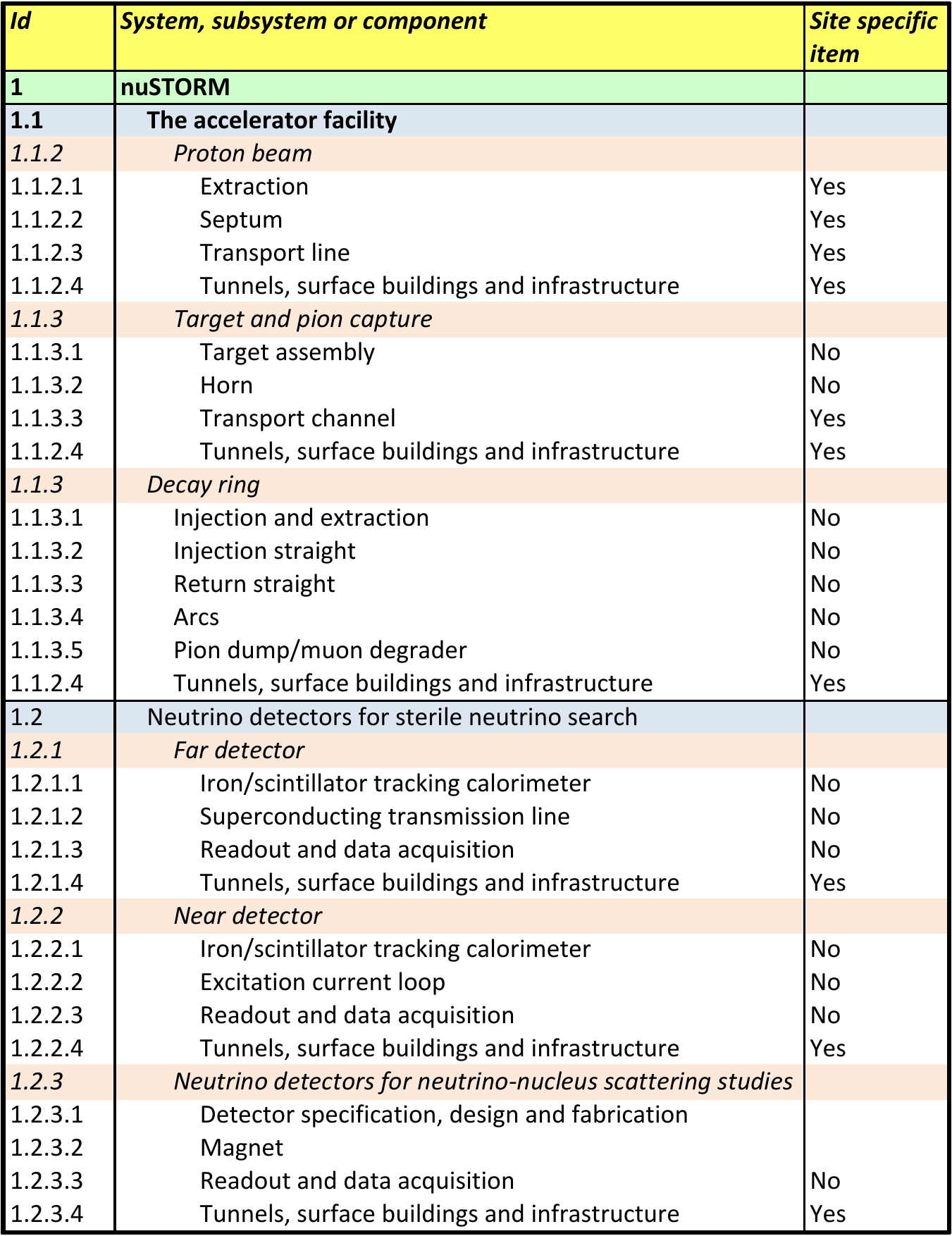}
  \end{center}
\end{table}

The optimisation and detailed design of the detectors required for the
sterile-neutrino search will be the responsibility of the $\nu$STORM
collaboration.
The facility will be capable of supporting the suite of near detectors
necessary to carry out definitive studies of neutrino-nucleus
scattering.
The PBS therefore identifies the need to develop the specification of
the neutrino-scattering programme and the development of designs for
the suite of detectors required to carry them out.

\subsection{Request for support}
\label{SubSect:PropProg}

We request CERN support to participate in the development of the
$\nu$STORM facility and experimental programme.
In addition, we request CERN support for those tasks where CERN's
particular expertise may be brought to bear to make decisive
contributions to the detailed design of the $\nu$STORM facility.
A number pf work-packages containing evaluations and technical studies
have been defined and are outlined below.
In each case, the evaluation of the necessary manpower needed to
execute the work is an essential early part of the work.
Over the period between April 2013 and June 2013, we request CERN
support to carry out the necessary evaluation so that a more detailed
evaluation of the resources required to carry out the proposed
programme can be presented to the SPSC at its meeting in June 2013.
\begin{itemize}
  \item{\bf Proton beam: SPS extraction, beam lines up to target:} \\
    The $\nu$STORM facility should take advantage of work already
    invested in the CERN North Area "neutrino hub", this means the
    technical evaluations and implementations already performed for
    the CENF and LBNO. 
    Both the 100\,GeV and the 400\,GeV beams extracted from the SPS
    are acceptable for $\nu$STORM. 
    $\nu$STORM would need an additional transport line should a
    new target station be needed; 
  \item{\bf Pion-production target:} \\
    It is likely that an existing target area will need substantial
    investment if it is to be re-used, it would be of interest to
    study a generic, re-usable target station at an early stage in the
    development of the North Area as a neutrino hub. 
    If this is not possible, $\nu$STORM would need to study a new
    target station, however, it would be largely similar to the target
    stations proposed for the CENF and LBNO experiments; 
  \item{\bf Pion transport:} \\
    Pion transport may be different in the CERN implementation to that
    already designed done for FNAL due to site constraints on the the
    topology of $\nu$STORM at CERN. 
    Significant parts of the work that has already done at FNAL can be
    re-used;
  \item{\bf Engineering study of pion-capture magnets:} \\
    The large aperture magnets will have to be studied in detail,
    including the effects of radiation.
    Superconducting magnets in the arcs also need cryogenic
    evaluation and radiation studies;
  \item{\bf Contributions to the design of the muon storage ring:} \\
    The work on a storage ring is ongoing within the $\nu$STORM
    collaboration;
  \item{\bf Contributions to design of storage ring diagnostics:} \\
    Detailed studies of the storage ring are are required to specify
    the instrumentation that is needed.
    Studies of the possibility to use the beam structure from SPS for
    beam instrumentation (how fast is the beam de-grouping) must be
    carried out.
    The influence of electron production from the decay has to be
    studied;
  \item{\bf Evaluation of a possible muon cooling experiment:} \\
    A muon cooling experiment could be set up after the straight
    section that is not used for for neutrino production. 
    A  muon cooling ring could be studied;
  \item{\bf Contributions to the design of the neutrino-scattering
    programme:} \\
    The European Strategy for Particle Physics \cite{Nakada:2013ESG}
    has emphasised the importance of studying the physics of the
    neutrino.
    The next generation of long- and short-baseline, conventional
    neutrino-oscillation experiments rely on the observation of
    electron-neutrino appearance in a muon-neutrino beam.
    To allow such experiments to reach their full potential requires
    that the systematic error related to the neutrino-scattering cross
    sections and modelling of the hadronic final states be minimised.
    As described above, $\nu$STORM is unique in that it is
    capable of delivering the programme that is required.
    CERN has the opportunity to serve the European neutrino community
    which seeks to establish a first-class neutrino programme at CERN
    by contributing to the development of the neutrino-nucleus
    scattering programme at $\nu$STORM.
    We request support from PH Division to provide supervision for a
    CERN Fellow and a research student.
    The latter would be jointly supervised by one of the institutes
    within the European collaboration.
\end{itemize}

%% file: 10A-HiResPhys/10A-HiResPhys.tex
\section{Physics Potential of near-detector suite at \boldmath{$\nu$}STORM}
\label{App:PhysOfHiRes}

We enumerate physics papers that will be engendered with the a suite 
of near detectors proposed for the $\nu$STORM facility.
The topics/papers  are motivated by the published results by NOMAD,
CCFR, NuTeV, MiniBOONE, etc. experiments. 
Criteria for choosing the topics are as follows: 
\begin{enumerate}
  \item Best Measurement: If the topic deals with a Standard Model
    measurement then it should be most precise; 
  \item Most Sensitive Search: If the topic involves a search then it
    should be the most sensitive search; and
  \item New Method: Where 1 and 2 abive are not applicable then the
    topic should include a novel measurement technique.
\end{enumerate}
In all, we have identified over 80 topics. 
The list is not complete. 
For example, it does not include topics involving
detector development, R\&D measurements, or engineering research that
typically are published in journals like NIM, IEEE, etc.  
The list comprising  absolute cross-section measurements, exclusive
and semi-exclusive channels, electroweak physics, perturbative and
non-perturbative QCD, and searches for new physics illustrates the
power of a high resolution, fine-grain-tracker based on the past
experiments.
Over the duration of the project, $\sim 10$ years, the number of
theses/paper will be more than twice as many as the number of topics.

Below we present a salient subset of physics topics. 
\begin{enumerate}
  \item Measurement of the absolute neutrino/anti-neutrino flux using
    neutrino-electron neutral current scattering;
  \item Measurement of the difference in the energy-scale of
    $\overline {\nu}_\mu$- versus $\nu_e$-induced charged-current (CC)
    events;
  \item Exclusive and quasi-exclusive single Pi0 production in
    neutrino- and anti-neutrino-induced neutral current interactions;
  \item Coherent and quasi-exclusive single Pi+ production in
    neutrino-induced charged current interactions; 
  \item Coherent and quasi-exclusive single Pi- production in
    antineutrino-induced charged current interactions;
  \item Proton (neutron) yield in inclusive neutrino and anti-neutrino
    charged current interactions;
  \item The $\nu_e$-$e^-$ and $\overline {\nu}_\mu$-$e^-$ interactions
    and search for lepton number violating process;
  \item Measurement of  neutrino and antineutrino quasi-elastic (QE)
    and resonance  charged current interactions;
  \item Measurement of prompt radiative photon  in muon- and
    electron-neutrino quasi-elastic interactions;
  \item Constraints on the Fermi-motion of the nucleons using the
    2-track topology of neutrino quasi-elastic interactions;
  \item Measurement of the hadronic content of the weak current in
    neutrino- and anti-neutrino CC and NC interactions;
  \item Neutral Current elastic scattering on proton, nu(bar) + p
    $\rightarrow$  nu(bar) + p and measurement of the strange quark
    contribution to the nucleon spin, Delta-S;
  \item Tests of sum-rules in QPM/QCD;
  \item Measurement of nuclear effects on $F_2$ and on $xF_3$ in
    (anti)neutrino scattering from ratios of Ar, Pb, Fe and C targets;
  \item Measurement of  strange mesons and hyperon production in
    (anti-)neutrino charged and neutral current;
  \item Measurement of the $\Lambda$ and $\overline {\Lambda}$
    polarisation in (anti-)neutrino neutral current interactions;
  \item Measurement of backward going protons and pions in neutrino CC
    interactions and constraints on  nuclear processes;
  \item Search for muon-neutrino to electron-neutrino transition and
    the LSND/MiniBOONE anomaly;
  \item Search for muon-antineutrino to electron-antineutrino
    transition  and the LSND/MiniBOONE anomaly; and
  \item Search for heavy neutrinos using its electronic, muonic and
    hadronic decays.
\end{enumerate}